\documentclass[prd,floatfix,onecolumn,amsmath,amssymb,11pt,nofootinbib,secnumarabic]{revtex4}
\usepackage{amsmath,amssymb}
\usepackage{graphicx} 
\usepackage{epstopdf} 
\usepackage{slashed}
\usepackage{wasysym}
\usepackage{subfigure}
\usepackage{array}

\usepackage{hyperref} 
\usepackage[all]{hypcap} 

\newcommand{\newc}{\newcommand}
\newc{\definmath}[2] {\def#1{\ifmmode#2\else$#2$\fi}}

\definmath\gsim{\,\,\rlap{\raise 3pt\hbox{$>$}}{\lower 3pt\hbox{$\sim$}}\,\,}
\definmath\lsim{\,\,\rlap{\raise 3pt\hbox{$<$}}{\lower 3pt\hbox{$\sim$}}\,\,}

\definmath\amin{\mathrm{min}}
\definmath\amax{\mathrm{max}}

\newc{\barr}{\begin{eqnarray}}
\newc{\earr}{\end{eqnarray}}
\newc{\beq}{\begin{equation}}
\newc{\eeq}{\end{equation}}

\newc{\voidcol}{\phantom{m}\begin{rotate}{270}Not
reconstructed\end{rotate}\phantom{mn}}

\definmath\half{{\frac 1 2}}
\definmath\threehalfs{{\frac 3 2}}
\definmath\quarter{{\frac 1 4}}
\definmath\sixth{{\frac 1 6}}
\definmath\third{{\frac 1 3}}
\definmath\twothirds{{\frac 2 3}}
\definmath\fourthirds{{\frac 4 3}}
\definmath{\mPl}{M_\mathrm{Pl}}
\definmath{\invfb}{\mathrm{fb}^{-1}}
\definmath{\nsec}{\mathrm{ns}}

\definmath{\Omegadm}{\Omega_\mathrm{CDM}}
\definmath{\omegadm}{\omega_\mathrm{CDM}}

\definmath{\micron}{{\mu\mathrm{m}}}

\definmath{\silica}{{\mathrm{SiO_2}}}

\definmath{\mnought}{{m_0}}
\definmath{\mhalf}{{m_\frac{1}{2}}}
\definmath{\mthreehalfs}{{m_{3/2}}}
\definmath{\DeltaMChi}{{\Delta M_{\cht_1}}}
\definmath\mtx{{m_{TX}}}
\definmath\mttwo{{m_{T2}}}
\definmath\mttwosq{{m_{T2}^2}}
\definmath\to{\rightarrow}
\newc{\mr}{\mathrm}

\def\slashchar#1{\setbox0=\hbox{$#1$}           
   \dimen0=\wd0                                 
   \setbox1=\hbox{/} \dimen1=\wd1               
   \ifdim\dimen0>\dimen1                        
      \rlap{\hbox to \dimen0{\hfil/\hfil}}#1 
   \else                                        
      \rlap{\hbox to \dimen1{\hfil$#1$\hfil}}/                                    \fi}

\definmath{\etmiss}{\slashchar{E}_T}
\definmath{\pmiss}{\slashchar{p}}
\definmath{\ptmiss}{\slashchar{p}_T}
\definmath{\Ptmiss}{\slashchar{{\bf p}}_T}
\definmath{\pt}{p_T}
\definmath{\qth}{Q_\mr{thr}}

\newc{\appref}[1]{appendix~\ref{#1}}
\newc{\chref}[1]{chapter~\ref{#1}}
\newc{\Chref}[1]{Chapter~\ref{#1}}
\newc{\secref}[1]{\hyperref[#1]{Section~\ref*{#1}}}
\newc{\tabref}[1]{table~\ref{#1}}
\newc{\figref}[1]{fig.~\ref{#1}}
\newc{\figsref}[1]{figs.~\ref{#1}}
\newc{\partref}[1]{part~\ref{#1}}
\newc{\Secref}[1]{Sec.~\ref{#1}}
\newc{\Tabref}[1]{Table~\ref{#1}}
\newc{\Figref}[1]{Fig.~\ref{#1}}
\newc{\Partref}[1]{Part~\ref{#1}}

\definmath{\z}  {\mathrm{Z}^{0}}
\definmath\tbar{{\bar t}}
\definmath\ttbar{{t \tbar}}

\definmath{\cht}{{\tilde{\chi}}}
\definmath{\chgone}{{\cht^+_1}}
\definmath{\chgtwo}{{\cht^+_2}}
\definmath{\chgonem}{{\cht^-_1}}
\definmath{\chgonepm}{\cht^{\pm}_1}
\definmath{\chgall}{\cht^{\pm}_{1,2}}
\definmath{\ntlone}{{{\cht^0_1}}}
\definmath{\ntltwo}{{{\cht^0_2}}}
\definmath{\ntlthree}{\cht^0_3}
\definmath{\ntlfour}{\cht^0_4}
\definmath{\ntlall} {\tilde{\chi}_{1,2,3,4}^{0}}
\definmath{\gluino}{\tilde{g}}

\definmath{\ssul} {{\tilde{u}_{L}}}
\definmath{\ssdl} {{\tilde{d}_{L}}}
\definmath{\sscl} {\tilde{c}_{L}}
\definmath{\sssl} {\tilde{s}_{L}}
\definmath{\sstone} {\tilde{t}_{1}}
\definmath{\ssbone} {\tilde{b}_{1}}
\definmath{\ssur} {\tilde{u}_{R}}
\definmath{\ssdr} {\tilde{d}_{R}}
\definmath{\sscr} {\tilde{c}_{R}}
\definmath{\sssr} {\tilde{s}_{R}}
\definmath{\ssttwo} {\tilde{t}_{2}}
\definmath{\ssbtwo} {\tilde{b}_{2}}
\definmath{\squark} {{\tilde{q}}}
\definmath{\sqr} {\tilde{q}_{R}}
\definmath{\sql} {{\tilde{q}_{L}}}
\definmath{\squark} {{\tilde{q}}}
\definmath{\ssulbr} {\bar{\tilde{u}_{L}}}
\definmath{\ssdlbr} {\bar{\tilde{d}_{L}}}
\definmath{\ssclbr} {\bar{\tilde{c}_{L}}}
\definmath{\ssslbr} {\bar{\tilde{s}_{L}}}
\definmath{\sstonebr} {\bar{\tilde{t}_{1}}}
\definmath{\ssbonebr} {\bar{\tilde{b}_{1}}}

\definmath{\ssurbr} {\bar{\tilde{u}_{R}}}
\definmath{\ssdrbr} {\bar{\tilde{d}_{R}}}
\definmath{\sscrbr} {\bar{\tilde{c}_{R}}}
\definmath{\sssrbr} {\bar{\tilde{s}_{R}}}
\definmath{\ssttwobr} {\bar{\tilde{t}_{2}}}
\definmath{\ssbtwobr} {\bar{\tilde{b}_{2}}}

\definmath{\ssel} {\tilde{e}_{L}}
\definmath{\ssellp} {\tilde{e}_{L}^{+}}
\definmath{\ssellm} {\tilde{e}_{L}^{-}}
\definmath{\ssellpm} {\tilde{e}_{L}^{\pm}}

\definmath{\sser} {\tilde{e}_{R}}
\definmath{\sselrp} {\tilde{e}_{R}^{+}}
\definmath{\sselrm} {\tilde{e}_{R}^{-}}
\definmath{\sselrpm} {\tilde{e}_{R}^{\pm}}

\definmath{\ssmulp} {\tilde{\mu}_{L}^{+}}
\definmath{\ssmulm} {\tilde{\mu}_{L}^{-}}
\definmath{\ssmulpm} {\tilde{\mu}_{L}^{\pm}}

\definmath{\ssmurp} {\tilde{\mu}_{R}^{+}}
\definmath{\ssmurm} {\tilde{\mu}_{R}^{-}}
\definmath{\ssmurpm} {\tilde{\mu}_{R}^{\pm}}

\definmath{\sstauone} {{\tilde{\tau}_{1}}}
\definmath{\sstauonep} {\tilde{\tau}_{1}^{+}}
\definmath{\sstauonem} {\tilde{\tau}_{1}^{-}}
\definmath{\sstauonepm} {\tilde{\tau}_{1}^{\pm}}

\definmath{\sstautwop} {\tilde{\tau}_{2}^{+}}
\definmath{\sstautwom} {\tilde{\tau}_{2}^{-}}
\definmath{\sstautwopm} {\tilde{\tau}_{2}^{\pm}}

\definmath{\sslrpm} {{\tilde{l}_{R}^{\pm}}}
\definmath{\sslr} {{\tilde{l}_{R}}}
\definmath{\ssll} {{\tilde{l}_{L}}}

\definmath{\ssnu} {\tilde{\nu}}
\definmath{\ssnuel} {\tilde{\nu}_{e}}
\definmath{\ssnumul} {\tilde{\nu}_{\mu}}
\definmath{\ssnutl} {\tilde{\nu}_{\tau}}

\definmath{\lqnear} {{l^\mathrm{near}q}}
\definmath{\lqfar} {{l^\mathrm{far}q}}
\definmath{\lqhigh} {{l^\mathrm{high}q}}
\definmath{\lqlow} {l{^\mathrm{low}q}}

\definmath{\lqbnear} {{l^\mathrm{near}\bar{q}}}
\definmath{\lqbfar} {{l^\mathrm{far}\bar{q}}}
\definmath{\lqbhigh} {{l^\mathrm{high}\bar{q}}}
\definmath{\lqblow} {{l^\mathrm{low}\bar{q}}}

\definmath{\lqplus} {{l^+q}}
\definmath{\lqminus} {{l^-q}}

\def\rmax{{{\rm max}}}
\def\rmin{{{\rm min}}}
\def\mttwo{M_{T2}}

\def\llEdge{{$l^+l^-$ edge}}  \def\mll{m_{ll}}

\def\lNear{l_{\rm near}}
\def\lFar{l_{\rm far}}
\def\mlqPlus{m_{q{l^+}}}
\def\mlqMinus{m_{q{l^-}}}
\def\lqNear{{$l^\pm_{\rm near} q$}} \def\mlqNear{m_{l_{\rm near} q}}
\def\lqFar{{$l^\pm_{\rm far} q$}} \def\mlqFar{m_{l_{\rm far} q}}
\def\lqHigh{{${l^\pm q}$ high}}
\def\lqSum{{${l^\pm q}$ sum}}
\def\lqLow{{${l^\pm q}$ low}}
\def\lqEdgeHigh{{\lqHigh-edge}} 
\def\lqEdgeSum{{\lqSum-edge}} 
\def\mlqHigh{m_{l q{\rm(high)}}}
\def\mlqEq{m_{l q{\rm(eq)}}}
\def\lqEdgeLow{{\lqLow-edge}} \def\mlqLow{m_{l q{\rm (low)}}}
\def\lqEdgeNear{{\lqNear\ edge}} \def\mlqNear{m_{l_{\rm near} q}}
\def\lqEdgeFar{{\lqFar\ edge}} \def\mlqFar{m_{l_{\rm far} q}}

\def\xqEdge{{$Xq$ edge}}      \def\mxq{m_{Xq}}

\def\llqEdge{{$l^+l^-q$ edge}}
\def\mllq{m_{llq}}

\def\llqThreshold{{$l^+l^-q$ threshold}}

\def\ifb{{{\rm fb}^{-1}}}

\newc{\sparticle}[1]{{\tilde{{#1}}}}

\def\squark{\sparticle{q}}
\def\slepton{\sparticle{\ell}}

\def\ntlinoOne{{\sparticle{\chi}^0_1}}
\def\ntlinoTwo{{\sparticle{\chi}^0_2}}
\def\gluino{{\sparticle{g}}}

\definmath{\lamp}{\lambda^\prime}
\definmath{\lampp}{\lambda^{\prime \prime}}
\definmath{\rparity}{{R_P}}
\definmath{\rp}{{R_P}}

\newif\iftth
\iftth\else
\newc{\prepareAbbrev}[7]{\newcounter{#5}\newcommand{#1}{\mygloss{#2}{#6}{#4 #7}\ifnum\arabic{#5}=0 {#4 (#3)}\else#3\fi\addtocounter{#5}{1}}}
\fi

\newcommand{\MT}{\ensuremath{M_T}}
\newcommand{\MTTWO}{\ensuremath{M_{T2}}}
\newcommand{\MC}{\ensuremath{M_{C}}}
\newcommand{\MCT}{\ensuremath{M_{CT}}}
\newcommand{\MTTHREE}{\ensuremath{M_{T3}}}
\newcommand{\MTWOC}{\ensuremath{M_{2C}}}
\newcommand{\MTHREEC}{\ensuremath{M_{3C}}}
\newcommand{\MTGEN}{\ensuremath{M_{T\mathrm{Gen}}}}
\newcommand{\MEFF}{\ensuremath{M_{\mathrm{eff}}}}
\newcommand{\ROOTSHAT}{\ensuremath{\hat{s}^{1/2}}}
\newcommand{\ROOTSHATMIN}{\ensuremath{\hat{s}^{1/2}_{\min}}}
\newcommand{\UTM}{\ensuremath{{\bf P}_{UT}}}

\newcommand{\subsubsubsection}[1]{{\textbf{{#1}}.}}
\newcommand{\pref}[1]{{Page~\pageref{#1}}}

\begin{document}
\title{A Review of the Mass Measurement Techniques proposed for the \\ Large Hadron Collider}
\author{Alan J Barr}
\email{a.barr@physics.ox.ac.uk}
\affiliation{Department of Physics, Denys Wilkinson Building, Keble
Road, Oxford OX1 3RH, United Kingdom}
\author{Christopher G Lester}  
\email{lester@hep.phy.cam.ac.uk} 
\affiliation{Department of Physics, Cavendish Laboratory, JJ Thomson
Avenue, Cambridge, CB3 0HE, United Kingdom}
\begin{abstract}
We review the methods which have been proposed for measuring masses of new particles at the Large Hadron Collider paying particular attention to the kinematical techniques suitable for extracting 
mass information when invisible particles are expected.
\end{abstract}
\preprint{CAVENDISH-HEP-10/05}
\maketitle 

\newpage
\tableofcontents
\newpage

\section{Introduction and Scope}

This review principally concerns itself with {\em kinematic} methods
of mass reconstruction, and in particular those that have been
considered for use with hadron colliders, notably the Large Hadron
Collider (LHC).

\par

Specifically, {\em kinematic} methods demand that at
least some particles are sufficiently close to their mass shells that
their energy-momentum Lorentz invariant $p_\mu p^\mu\approx m^2$ can
be used to constrain their masses.  Such methods aim to determine, to bound, or
to otherwise constrain $p$ in order to learn about $m$.  Assuming
momentum and energy conservation, one also can learn about the
four-momenta of -- and hence constrain the masses of -- particles
which are not directly observed experimentally.  Two important
examples are (1) unstable particles which decay and (2) weakly
interacting particles which, though stable, do not interact with the
detector.

\par

It is of course true that when further information 
-- beyond the purely kinematic --  is either known or assumed
that one could use that information too.
With sufficient theoretical and experimental understanding, 
and provided the calculation is tractable, one could obtain maximal information 
about an event by comparing its statistical likelihood under 
different mass (or spin or other) hypotheses.
The ability to numerically marginalise over uncertain information (such as momentum components
of invisible particles) has made such calculations
computationally feasible. This approach -- sometimes called 
the {\em Matrix Element} method -- has been employed at the 
CERN $\mathrm{Sp\bar{p}S}$ (e.g.~\cite{Alitti:1990ch}),
LEP (e.g.~\cite{Acciarri:1999ft}) and the Tevatron (e.g.~\cite{Abazov:2006bd}),
and proposals exist for the LHC (e.g.~\cite{JohanIpmu}).
Such methods are ideal for making precise statements about parameters
when some confidence about the underlying model has been gained.
They are however unwieldy in the early stages of an investigation, 
when there are usually very many interesting hypotheses to test, 
each with a wide range of allowed parameters.
Another non-kinematic method of obtaining information is the investigation
of masses with virtual particles, far from their mass shells.
A well-known example is the constraint on the Higgs Boson mass 
(assuming e.g.~the Standard Model as the underlying theory) from loop contributions to 
electroweak observables \cite{Collaboration:2008ub}.

\par

The main advantage of the kinematic approaches reviewed in this article is that they 
make very few assumptions about the details of the underlying physical model 
(gauge groups, spins etc). 
This means that they can provide rather robust information,
and act as a first step towards understanding the underlying theory.

\begin{table}[p]
\begin{center}
\centering{
\parbox[t]{0.45\linewidth}{
\begin{tabular}{|>{\centering}m{4.3cm}|m{1.75cm}|m{1.25cm}|} \hline
Topology 								& Section(s) 				& Page(s) \\
\hline
\includegraphics[scale=0.5,clip]{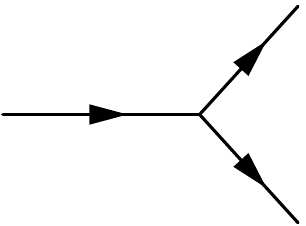}			& \ref{sec:z} 			& \pref{sec:z} 			\\ \hline
\includegraphics[scale=0.5,clip]{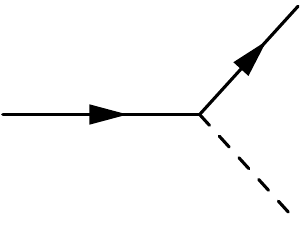}  			& \ref{sec:w}, \ref{sec:mct}	& \pref{sec:w} \pref{sec:mct} 	\\ \hline
\includegraphics[scale=0.5,clip]{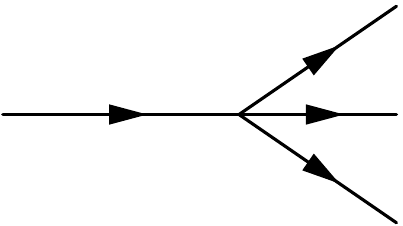}  		& \ref{sec:threebody} 		& \pref{sec:threebody} 		\\ \hline
\includegraphics[scale=0.5,clip]{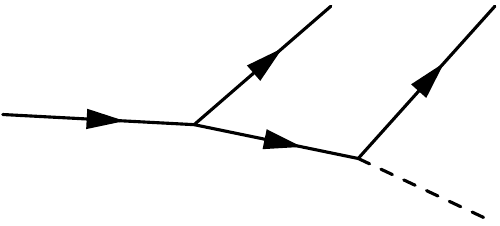}  		& \ref{sec:ll} 			& \pref{sec:ll} 		\\ \hline
\includegraphics[scale=0.5,clip]{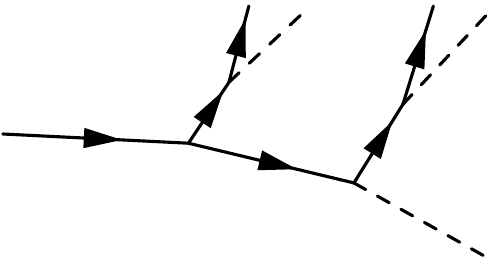} 			& \ref{sec:tautau}	 		& \pref{sec:tautau} 		\\ \hline
\includegraphics[scale=0.5,clip]{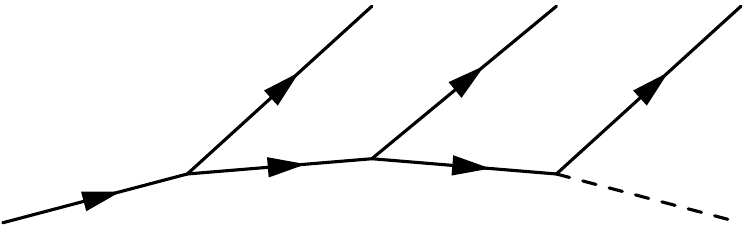} 		& \ref{sec:qll} 			& \pref{sec:qll} 		\\ \hline
\includegraphics[scale=0.5,clip]{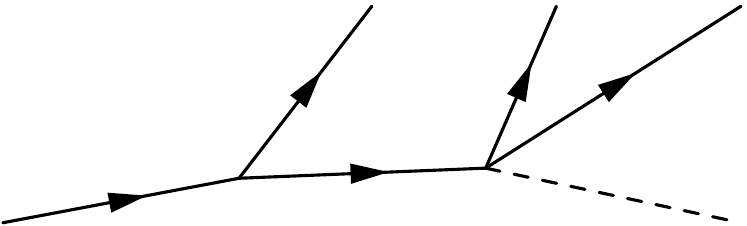}  	& \ref{sec:twothreegetsdiscussed} 	& \pref{sec:twothreegetsdiscussed} \\ \hline
\includegraphics[scale=0.5,clip]{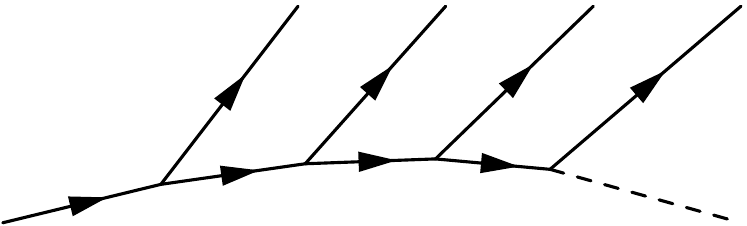}  		& \ref{sec:qqll}		 	& \pref{sec:qqll}  		\\ \hline
\includegraphics[scale=0.5,clip]{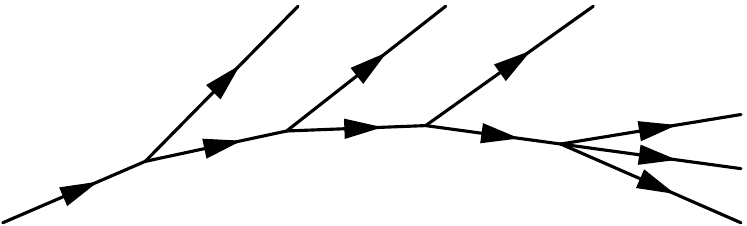}          		& \ref{sec:rpv} 			& \pref{sec:rpv} 		\\ \hline
\includegraphics[scale=0.5,clip]{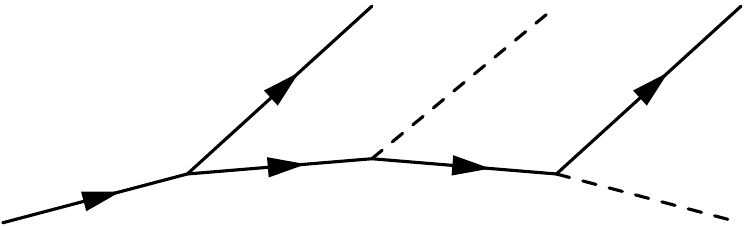}        & \ref{sec:darkmattersandwich} 		& \pref{sec:darkmattersandwich} \\ \hline
\includegraphics[scale=0.5,clip]{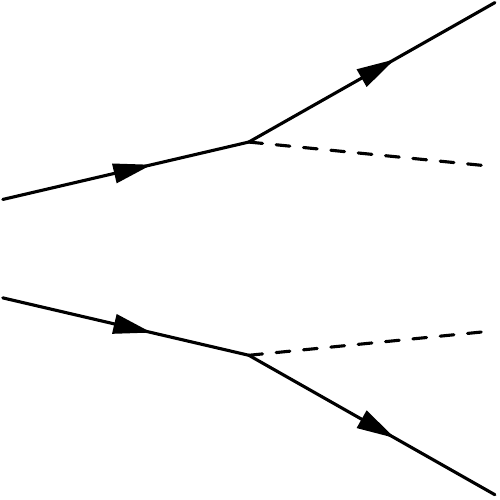}  		& \ref{sec:mttwo}, \ref{sec:alwallme}, \ref{sec:maost} & \pref{sec:mttwo} \pref{sec:alwallme} \pref{sec:maost} \\ \hline
\includegraphics[scale=0.5,clip]{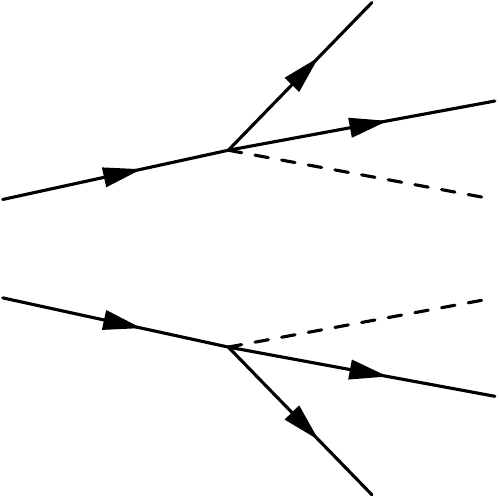}  		& \ref{sec:hybridpair}, \ref{sec:mtwoc} & \pref{sec:hybridpair} \pref{sec:mtwoc} \\ \hline
\end{tabular}}
\hfill 
\parbox[t]{0.45\linewidth}{
\begin{tabular}{|>{\centering}m{4.2cm}|m{1.75cm}|m{1.25cm}|} 
\hline 
Topology 								& Section(s)			 	& Page(s) 			\\ \hline
\includegraphics[scale=0.46,clip]{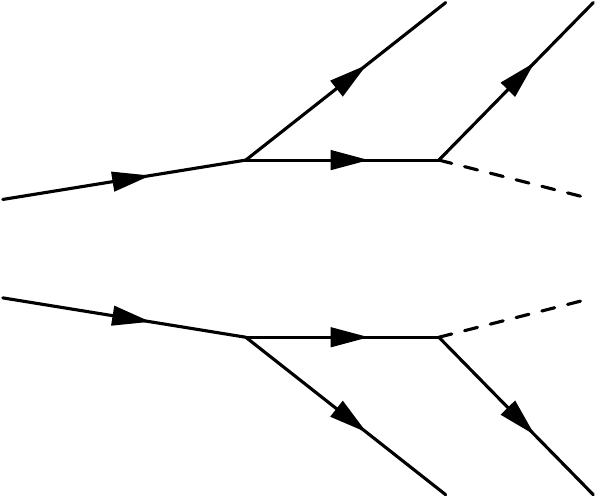}  	& \ref{sec:countsolns}, \ref{sec:hybridpair}, \ref{sec:mthreec} & \pref{sec:countsolns} \pref{sec:hybridpair} \pref{sec:mthreec} \\ \hline
\includegraphics[scale=0.46,clip]{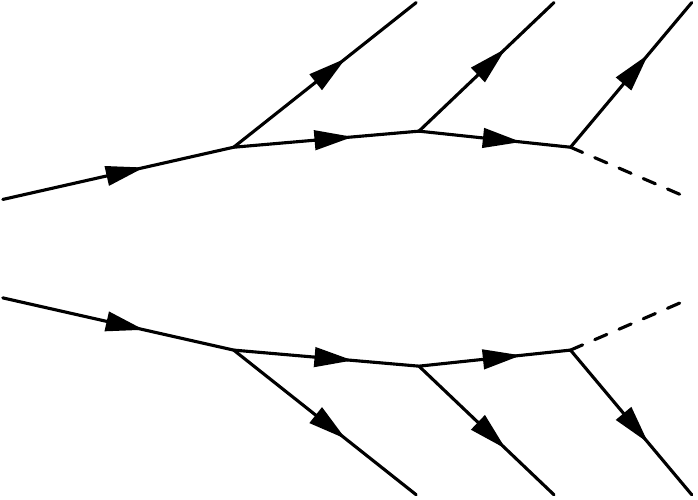}  	& \ref{sec:countsolns} 		& \pref{sec:countsolns} 	\\ \hline
\includegraphics[scale=0.42,clip]{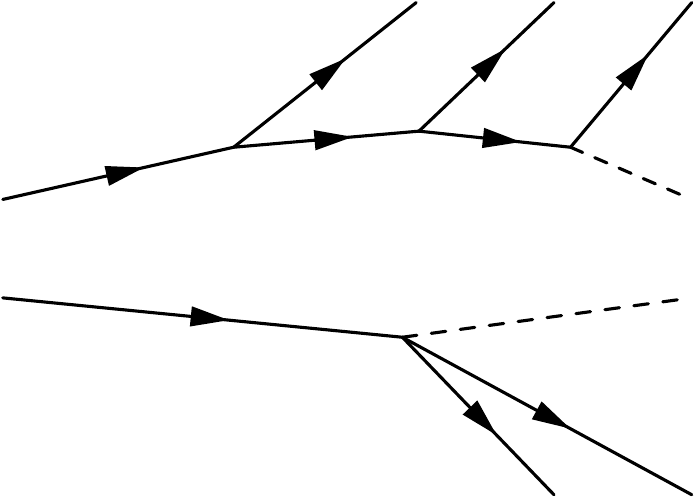}  		& \ref{sec:mttwoasymmetric} 	& \pref{sec:mttwoasymmetric} 	\\ \hline
\includegraphics[scale=0.5,clip]{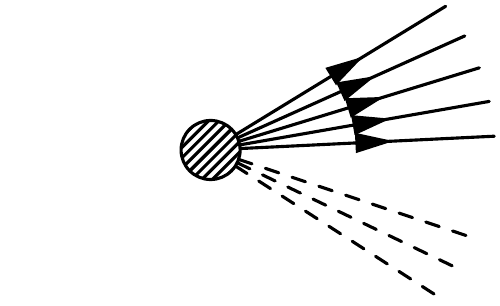}  			& \ref{sec:threshold}, \ref{sec:multipleinvisibles} & \pref{sec:threshold} \pref{sec:multipleinvisibles} \\ \hline
\includegraphics[scale=0.5,clip]{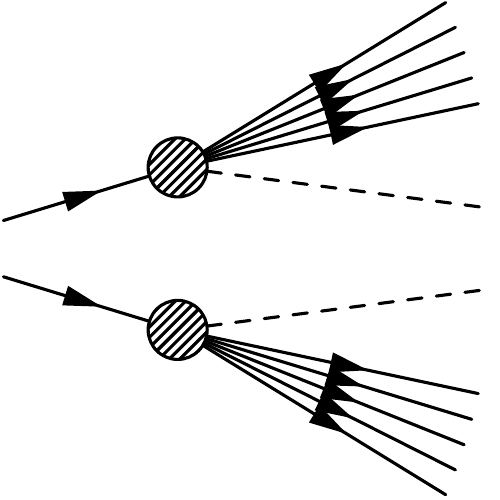}  			& \ref{sec:mtovey} 			& \pref{sec:mtovey} 		\\ \hline
\includegraphics[scale=0.5,clip]{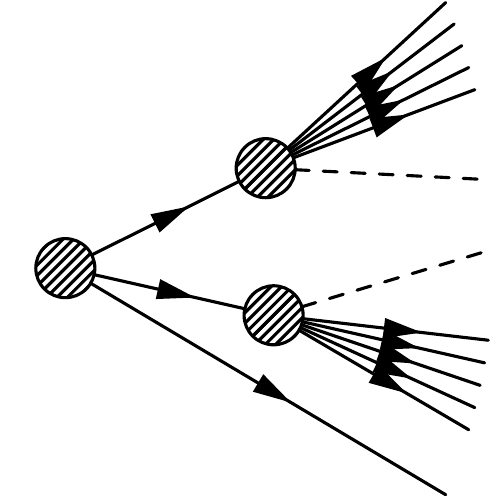}  	& \ref{sec:mtgen} 			& \pref{sec:mtgen} 		\\ \hline
\includegraphics[scale=0.5,clip]{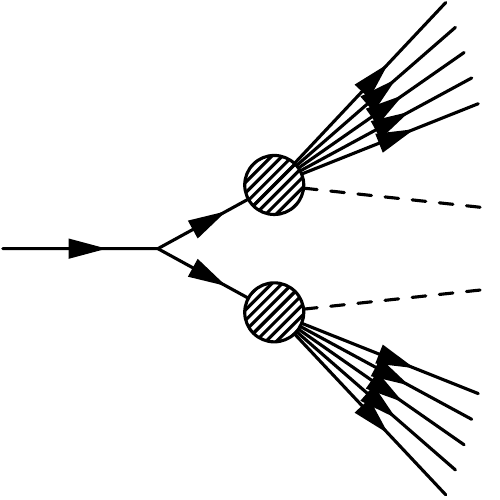} 			& \ref{sec:higgs} 			& \pref{sec:higgs} 		\\ \hline
\includegraphics[scale=0.5,clip]{FunnyDark1.eps} \raisebox{6mm}{$\bigcup$} \includegraphics[scale=0.5,clip]{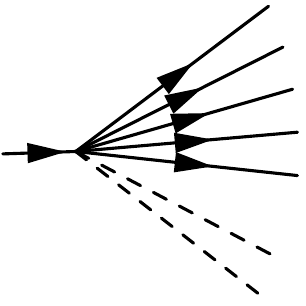}   			& \ref{sec:funnydark}  & \pref{sec:funnydark} \\ \hline
\end{tabular}}
}
\caption{\label{tab:topologies}Cartoons indicating various decay topologies, and relevant
sections of this review.
Dashed lines indicate `invisible' particles which traverse the apparatus undectected.
Blobs indicate decays which may (or may not) have proceeded via one or more on-mass-shell intermediates.
References to sections should be considered indicative rather than exhaustive.
}
\end{center}
\end{table}

\subsection{Outline of the Review}

Many, though not all, mass measurement techniques can be broken down into three phases: (1) the postulation of a hypothesis or hypotheses about the `decay topology' -- by which we mean the sequence of decays which involve the particles whose masses are to be determined, (2) the identification of the most appropriate final-state observables, and (3) the construction of constraints or measurements of the target particle masses, using those observables.

In what remains of the introduction, we say a few more words about these three phases.  In \secref{sec:topologies} we will outline in more detail what we mean by decay topologies and kinematic hypothesis.  Indeed, readers new to the field may find that section a useful starting point from which to determine which mass measurement techniques are relevant to their needs.   We will go on to talk in general terms about the basic observables which we believe are in our remit, and will then comment on the nature of the secondary observables which are derived from them and which are subsequently used to build the mass measurement constraints.  Subsequently we will note some of the issues that present recurrent challenges faced by mass determination methods of all kinds - in particular ambiguities of the final or internal state and of the possibility that multiple decay chains may be present in a single event.

Thereafter, the main part of the review is divided up into sections which, to first order, each cover the mass measurement techniques which are appropriate to a specific decay topology, i.e.~to a specific set of assumptions about the identities of the particles participating in the observed decays, and the kinematical structure of those decays.

Broadly speaking, the review starts with the decay topologies or hypotheses which make the {\em fewest assumptions}  -- for example some assume nothing more than ``that momentum is conserved'' -- and from there the review progresses to the hypotheses which make progressively greater numbers of assumptions.  For example one of the last techniques discussed assumes that the experimenter is able to identify samples of quintuples of events, each of which share the same five-particle topology and particle content (though differing in kinematics), and which together amount to an over-constrained system from which the unknown masses can be determined.

The most significant step-changes in the number of assumptions made by the increasingly complex hypothesis in the main part of the view concern the following: (1) increasing numbers of decay products, (2) increasing numbers of invisible decay products, (3) increasing lengths of ``decay chain'', (4) increasing numbers of decay chains present in each event, (5) assumptions making use of pairs of decay chains related (in single events) by a mutual interaction (recoil), and (6) assumptions requiring increasingly pure samples of events, and (7) assumptions requiring greater control of detector acceptances and efficiencies and background shapes over wide ranges.

In the appendix we gather together some definitions of kinematic variables and useful results.

\subsection{Decay Topologies or Hypotheses}
\label{sec:topologies}

In the literature the shorthand ``topology'' is used to indicate a 
sequence of decays of heavy objects to lighter ones.
The constituents of what is often referred to as the ``final state'' 
may actually have further non-trivial dynamics (e.g.~showering and hadronisation of 
quarks or gluons to jets).
They might be indistinguishable (e.g.~identical leptons) or indeed may be 
unobserved (e.g.~neutrinos).

A list of some topologies and links to corresponding sections of this review
can be found in \autoref{tab:topologies} on \pref{tab:topologies}.
Note that hypotheses of a particular event topology 
may differ in their details yet still provide a correct description of that event.
For example hypothesis (1) might be that particle $A$ decayed to a final state
comprising particles $X$, $Y$ and $Z$, without specifying any details about 
the intermediate mechanism. A refinement, hypothesis (2), might assume 
an explicit form of the decay chain, e.g.~$A\to B X$ followed by the decay
$B\to YZ$. Provided that both hypotheses do indeed correspond to the event observed
it will be possible to extract more information from the more detailed hypothesis (2):
for example the mass of the intermediate particle $B$.

The topologies of \autoref{tab:topologies} are therefore not all mutually exclusive.
There is in fact a tension between the desire to assume more details about topology -- in order
to obtain more information about the event -- and the need, in the presence of 
many competing processes, to propose something general enough that it stands a respectable 
chance of matching the observed event. 

As can be seen from \autoref{tab:topologies}, parts 
(though by no means all) of this review are concerned with topologies
in which two particles are believed to have been produced in the initial state.
This is partially motivated by the expectation that in many models, new particles are odd under a $Z_2$
parity (such as $R$-parity for supersymmetry) under which Standard Model particles are all even.
The lightest such parity-odd particle, if stable for cosmological timescales,
is expected to be weakly interacting, therefore to be unobserved by the apparatus.
Invisible particles (which are not unique to such models) are represented by dashed lines in \autoref{tab:topologies}.

\subsection{Observables and other quantities defined {\em per-event}}

The main information which is obtained from a hadron collider event
is the {\em momentum} and {\em energy} of the observable particles which impinge on the active volume.
Particles which are detected will have their reconstructed momenta and energy smeared by the experimental resolutions, and this smearing will need to be understood in calibration channels and modelled by those interpreting the experiments.
Detailed understanding of the detector response is clearly necessary to perform precision measurements, 
and is also a key component approaches which combine information from different sources (e.g.~different events) in order to over-constrain the kinematics.

Not all particles are necessarily observed -- for example neutrinos and any other weakly-interacting
particles are expected to pass through the apparatus undetected. 
What is more, it is not usually possible to reconstruct particles
with small angles to the beam-pipe
-- for example the hermetic region of the LHC general purpose detectors  \cite{Aad:2008zzm,:2008zzk} is restricted to a fiducial pseudorapidity of $|\eta| \lsim 5$
where $\eta=-\log\tan(\theta/2)$ and $\theta$ is the angle relative 
to one of the beam directions.

The incoming parton momenta are generally not known in hadron-hadron collisions,
so the centre-of-mass energy and the longitudinal boost of the centre-of-mass frame
are not fixed by the initial conditions.
The sum of the momenta of any invisible particles
can be inferred from conservation of momentum.

Except in the very special case of central exclusive production\footnote{For $pp\to ppX$ the centre-of-mass 
four-vector can be fully reconstructed from the outgoing proton momenta
if dedicated detectors are installed at very high $|\eta|$ \cite{Albrow:1005180,Albrow:2008pn}.}
the only information which can be obtained about invisible particles' 
momenta is the sum of the components perpendicular to the beams:
$\sum{\bf p}_T^\mathrm{invisible} \approx \Ptmiss \equiv -\sum{{\bf p}_T^\mathrm{vis}}$ where the second sum is over the
visible transverse momenta of all final-state particles.
The first equality is only approximate since particles at large $|\eta|$ 
will be undetected (though visible in principle), and because of experimental smearing of the ${\bf p}_T^\mathrm{vis}$.

Most heavy particles decay sufficiently rapidly that they do not travel macroscopically measurable distances.
Familiar exceptions include $\tau$ leptons 
and $B$ hadrons 
which can travel macroscopic distances from their production location. 
If decay products can be reconstructed to secondary vertices away from the 
primary interaction point, the additional information can help in particle identification,
and in kinematic reconstruction.
Some examples of using the kinematic information from secondary vertices 
can be found in \secref{sec:secondary}.

In addition to measuring directions, energies and momenta, the majority of hadron collider experiments can make statements about the {\em identities} of the particles they detect by using hardware designed for that purpose.  Muon detectors placed after hadron calorimeters are a prime example of a means by which a species of particle can be identified with a high degree of confidence. Nonetheless, many types of paricles (especially those with similar properties) cannot be identified without dedicated hardware such as \v{C}erenkov detectors \cite{Roberts:1960zz}.

The typical set of observables therefore consists of the four-momenta of a set of 
objects, some of which are individual particles, and some of which may be 
groups of particles (e.g.~hadronic jets). Each of these objects have an identification hypothesis 
or hypotheses with associated probabilities.
Together with a hypothesis about the topology, these observables can be used to make 
inferences about the properties (and particularly for this review, masses) of the particles.

The uses to which our observables are put tend to be restricted by the
invariance of space-time under rotations and Lorentz boosts, or by the
approximate axisymmetry of the LHC detectors, or by the lack of
knowledge of the centre-of-momentum frame of the primary interaction
due to the composite nature of the colliding protons.  As a result,
our primary observables tend to be combined into secondary derived
quantities (which are themselves invariant under general boosts or
axial boosts or general rotations or rotations about the beam axis, or
combinations thereof) and these are in turn used as building blocks or
inputs to more complicated tertiary methods and variables.  The
secondary derived quantities of which we speak include the invariant masses and
transverse masses which, though discussed later, will already be familiar to most readers.  
However, we give advanced warning, that there are sometimes
situations in which it is useful to form derived quantities which lack
some of the usual symmetries of space-time or the detector.  For
example, the first such quantity we will come across is $m_{C}$ (the
``contralinear'' invariant mass) which is decidedly deviant under
Lorentz boosts of any kind, but is nonetheless useful.

\subsection{Constraints and quantities defined {\em per-dataset} and {\em hybrids}}
\label{sec:hybridintro}

As well as having observables and derived quantities which come from individual events, it is natural to expect also observables that are formed from samples containing large numbers of events.  These could be called {\em per-dataset} variables.  One classic example of a class of such variables, which we will discuss in more detail later, are kinematic endpoints.   Typically the position of a kinematic endpoint places a constraint on some relationship of the masses of the particles involved in the decays that generated the endpoint.  A second example would be a mass constraint coming from a fit to the shape of a differential distribution constructed from a large number of events.

Furthermore, one can even conceive of {\em hybrid} variables, by which we mean variables
which mix together pieces of information from {\em per-dataset} and {\em per-event} into something more powerful. The resulting hybrids appear to be defined ``per-event'', but in fact make use of global properties of the dataset as a whole. 
The most common reason for doing
this is the desire to apply, to individual events, one or more
constraints of the type which cannot be deduced from any single event,
but which {\em can} be deduced from the set of all events.

In this review we will try to draw a distinction between {\em per-event},  {\em per-dataset} and {\em hybrid} variables.   We will leave further discussion of the merits and drawbacks of hybrid variables until later (\secref{sec:hybridpair}).

\subsection{Ambiguities}
\label{sec:ambiguities}

It is often the case that final state particles cannot be uniquely
attributed to particular positions in the hypothesised decay chain.
This may simply be due to there being repeated identical particles in
the final state. A second source of ambiguity can arise from initial
state radiation (ISR). Any high-scale process at a hadron collider
will inevitably be accompanied by jets due to ISR, and so mass
measurement techniques, particularly those using jets, need to be
robust with respect to its presence. An extended discussion of complications 
caused by ISR can be found in
Section~\ref{sec:grublingaboutisr} in the context of one particular
event topology.  A third class of ambiguity can result from there
being alternative internal particle assignments that leave the
identities of particles in the final state permuted. For example,
consider the supersymmetric decay chain
$\squark \to \ntltwo q \to \slepton^\mp \ell^\pm q \to \ntlone \ell^\mp\ell^\pm q$.  In this chain
the charge of the intermediate slepton is not known and so
one cannot tell whether the positively-charged lepton originated from the 
decay of the neutralino or from the subsequent decay of the slepton.  
A fourth source of ambiguity can arise from lack of certainty as to whether the decay topology {\em hypothesised} for a given event actually reflects reality.  For example, the decay chain above (in which the slepton is an on-shell resonance) has the same final particle content as a similar chain in which the slepton is much heavier than the $\ntltwo$, thus forcing the $\ntltwo$ to decay by an effective three body decay rather than via two successive two body decays.  

There is substantial variability in the extent to which mass measurement techniques which have been proposed in the literature choose to address the challenges presented by such ambiguities.  Some are developed with these ambiguities in mind from the start, while others do not address them at all and merely hope that ways will be found to address them in the future.  The reader is encouraged to think critically about the assumptions made in each of the techniques reviewed herein, and to consider the ways in which they may or may not be sensitive to unresolved ambiguities.  In particular, it should not be assumed that the presence of a technique in this review guarantees that it can be used in practice, or that any technique can produce a definitive answer that is not strongly dependent on one or more untestable assumptions made at its core.







\subsection{Spins}\label{sec:spins}

The spins of the participating particles and the chiralities of their couplings can play important roles in the dynamics 
of the decays.
In most cases, the effects of the particles spins on experimentally-accessible distributions are small, 
but various analyses have been proposed 
\cite{Richardson:2001df,Barr:2004ze,Goto:2004cpa,Goto:2004cp,Smillie:2005ar,Battaglia:2005zf,Battaglia:2005ma,Datta:2005zs,Meade:2006dw,Athanasiou:2006ef,Wang:2006hk,Smillie:2006cd,Alves:2006kn,Choi:2006mr,Nojiri:2007jm,Kilic:2007zk,Alves:2007xt,Csaki:2007xm,Horsky:2008yi,Burns:2008cp,Cho:2008tj,Gedalia:2009ym,Kramer:2009kp,Ehrenfeld:2009rt}
which indicate sensitivity to spins in a variety of cascade decays.
Angular correlations in variables other than cascade decays have also been studied
\cite{Barr:2005dz,Buckley:2007th,Buckley:2008pp,Alves:2008up,Buckley:2008eb,Boudjema:2009fz}.
A separate review article on the subject of spin determination methods has recently been published \cite{Wang:2008sw} 
and we refer the reader to that article for more details.

\section{Variables for particle production at or near threshold}\label{sec:threshold}

If one wishes to make very few assumptions about the type of interaction, the decay topology, 
and the types or particles involved, then the best one can generally do is 
to construct an observable which (because it is constructed
out of quantities proportional to energy) scales approximately 
as the energy of the centre-of-mass of the collision.

The distribution functions of momenta of partons within protons (``PDFs'') are largely rapidly falling functions 
of the momentum fraction $x$, so above threshold, cross-sections tend to decrease with the
centre-of-mass energy of the parton-parton system, $\ROOTSHAT$.
This means that heavy particles can often be expected to be produced at or near threshold, 
and the energy of the collision can be expected to give a good indication of the mass scale of the particles produced.

A variety of variables sensitive to the overall mass-energy scale have been proposed.
Since the momentum of the parton-parton centre-of-mass generally cannot be known 
when invisible particles have been produced, the majority
are constructed from only those momentum components perpendicular to the beam pipe.

In the context of supersymmetry, the simplest mass-scale measurements are those deriving from ad-hoc variables that have some kind of correlation, even if only approximate, with the masses or mass differences of the primary particles produced in the interaction.  Unfortunately there is little standardisation of nomenclature in this area: variables with the same experimental definition can have more than one name depending on which collaboration uses them (for example, both ATLAS and CMS have at different times defined an identical variable, though the former called their variable $\MEFF$ while the latter named theirs $H_T$) and even worse, neither collaboration has stuck to a single definition of either of these variables for any great length of time.  For example, the earliest \cite{Hinchliffe:1996iu} definitions of $\MEFF$ within ATLAS, which remained in use for more than a decade \cite{Aad:2009wy}, defined the ``effective mass'' variable in terms of the scalar sum of the four highest $p_T$ jets and the missing
transverse momentum as follows:
\begin{equation}\label{eq:meff}
\MEFF = \sum_{i=1,4}|{\bf p}_{T,i}| + |\Ptmiss| .
\end{equation}
More recent ATLAS work \cite{Collaboration:1273174} has re-defined \MEFF\ as the sum of the hightest ``$n$'' $p_T$ jets, where $n$ depends on the analysis channel.  
The peak of the supersymmetric component of the original $\MEFF$ distribution of (\ref{eq:meff}) and \cite{Hinchliffe:1996iu,Aad:2009wy} was found to correlate 
at the $\mathcal{O}(10\%)$ level with 
a characteristic SUSY mass-scale $m_\mathrm{SUSY}\equiv \min(m_\gluino,m_{\ssur})$ 
for models drawn from the 5-parameter constrained Minimal Supersymmetric Standard Model (cMSSM).
A more general MSSM study \cite{Tovey:2000wk}, 
found that the scalar sum over {\em all} jets given by 
\begin{equation}M_\mathrm{est}=\sum_i|{\bf p}_{T,i}| + |\Ptmiss|\end{equation} 
had a peak position which correlates well with a cross-section-weighted SUSY mass scale -- after the latter
was corrected by the ({\em{a priori}} unknown) mass of the lightest supersymmetric particle.

Within CMS, the analogous variable is called $H_T$ whose name, we believe, comes from earlier use at the Tevatron~\cite{PhysRevLett.88.041801}.  In 2006 a single document, the CMS technical design report \cite{cmsphystdr}, defines $H_T$ in two different ways!  The ``jets and missing energy searches'' section defines $H_T$ in terms of the scalar sum of the second, third and fourth (but not the first) jet energies {\em and} the missing transverse momentum according to 
\begin{equation}H_T = E_{T(2)} + E_{T(3)} + E_{T(4)} + |\Ptmiss| \label{eq:ht2a}\end{equation}
where $E_{T(i)}$ is the transverse energy of the $i$th jet, and
\begin{equation}\label{eq:Etprojdef}
E_T = E \sin \theta .\
\end{equation}
In the trigger section of the same document \cite{cmsphystdr} $H_T$ is defined differently as the scalar sum of the $E_T$ values of {\em all} jets, {\em excluding} the missing transverse momentum:
\begin{equation}H_T = E_{T(1)} + E_{T(2)} + E_{T(3)} + \cdots .\label{eq:ht2b}\end{equation}
In later CMS work \cite{cms:CMS-PAS-SUS-10-001} the definition of $H_T$ has changed for a third time, and is now the scalar sum of the transverse {\em momenta} of all jets:
\begin{equation}H_T = p_{T(1)} + p_{T(2)} + p_{T(3)} + \cdots .\label{eq:ht2c}\end{equation}

\par

Regardless of the specific definition used, the implicit assumption behind variables such as \MEFF\ and $H_T$ is that in a hadron collider particles tend to be produced
near threshold. Any particles produced exactly at at rest, and which decay in a semi-invisible two-body decay, produce visible daughters with transverse momenta $p_T = |{\bf p}_T|$ less than the two-body decay momentum,
\begin{equation}\label{eq:ptlimit}
p_T \leq p^* = \frac{\lambda^\half(m_A,m_B,m_C)}{2m_A}
\end{equation}
where $m_A$ is the mass of the parent, $m_B$ and $m_C$ are the masses of the two daughters and
\[
\lambda(a,b,c) = \left(a^2 - (b+c)^2 \right)  \left(a^2-(b-c)^2\right) .\
\]
The inequality in \eqref{eq:ptlimit} becomes an equality in the case where there is no momentum along the direction of the beam pipe. Thus the transverse momentum distribution contains information about the mass scale of any heavy particles produced, though the above inequality will be smeared by detector resolution, by production of heavy states well above threshold and by recoil of the parent against initial state radiation.

As discussed above, the centre-of-mass energy of the (parton-parton) collision  $\ROOTSHAT$ is 
sensitive to the mass scale of heavy particles even if few details about their decay topology are known.
When invisible particles are produced, there is insufficient information to reconstruct
$\ROOTSHAT$ for any particular event, but it will be bounded from below by the observable\label{sec:shatminfirstmentionsec}
\[\ROOTSHATMIN = (E^2-P_Z^2)^\half + (\ptmiss^2 + M^2_\mathrm{invis})^\half\]
where $M_{\mathrm{_{invis}}}$ is the sum of the mass of all invisible
particles thought to have been produced \cite{Konar:2008ei}.  It has been noted
\cite{Papaefstathiou:2009hp} that though $\ROOTSHATMIN$ and other
similar variables are very heavily modified by initial state
radiation, the amount of modification is nonetheless calculable.  This
is all we shall say about \ROOTSHAT\ for the moment, however we will
return to \ROOTSHATMIN\ in more detail in \secref{sec:rootshatmininpairs} where
we discuss how it might concretely be used to measure masses of
pair-produced events.

We note in passing that \cite{Konar:2008ei}
does not confine itself to suggesting that \ROOTSHATMIN\ be used only
to place strict bounds on \ROOTSHAT, or to place constraints on
particle masses in the manner described in
\secref{sec:rootshatmininpairs}. On the contrary, \cite{Konar:2008ei}
advances a number of quite different potential uses for \ROOTSHATMIN,
some not even related to mass measurement, which are not discussed further in this review.

\section{Variables for single cascade decay chains \label{sec:single}}

\subsection{Decays to two visible particles (``two-body visible'')}\label{sec:z}

\begin{figure}[t]
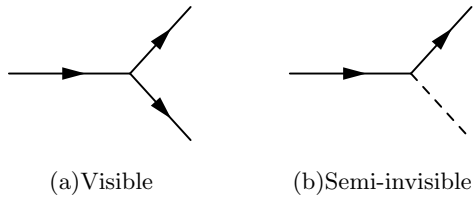

\begin{center}
\subfigure[Visible]{\includegraphics[width=2.5cm]{Zdecay.pdf}}
\hspace{1cm}
\subfigure[Semi-invisible]{\includegraphics[width=2.5cm]{Wdecay.pdf}}
\caption{\label{fig:atobc} Two very simple decay topologies.}
\end{center}
\end{figure}

\begin{figure}[t]
\begin{center}
\subfigure[Dilepton invariant mass]{\includegraphics[width=7cm,height=6.5cm,clip]{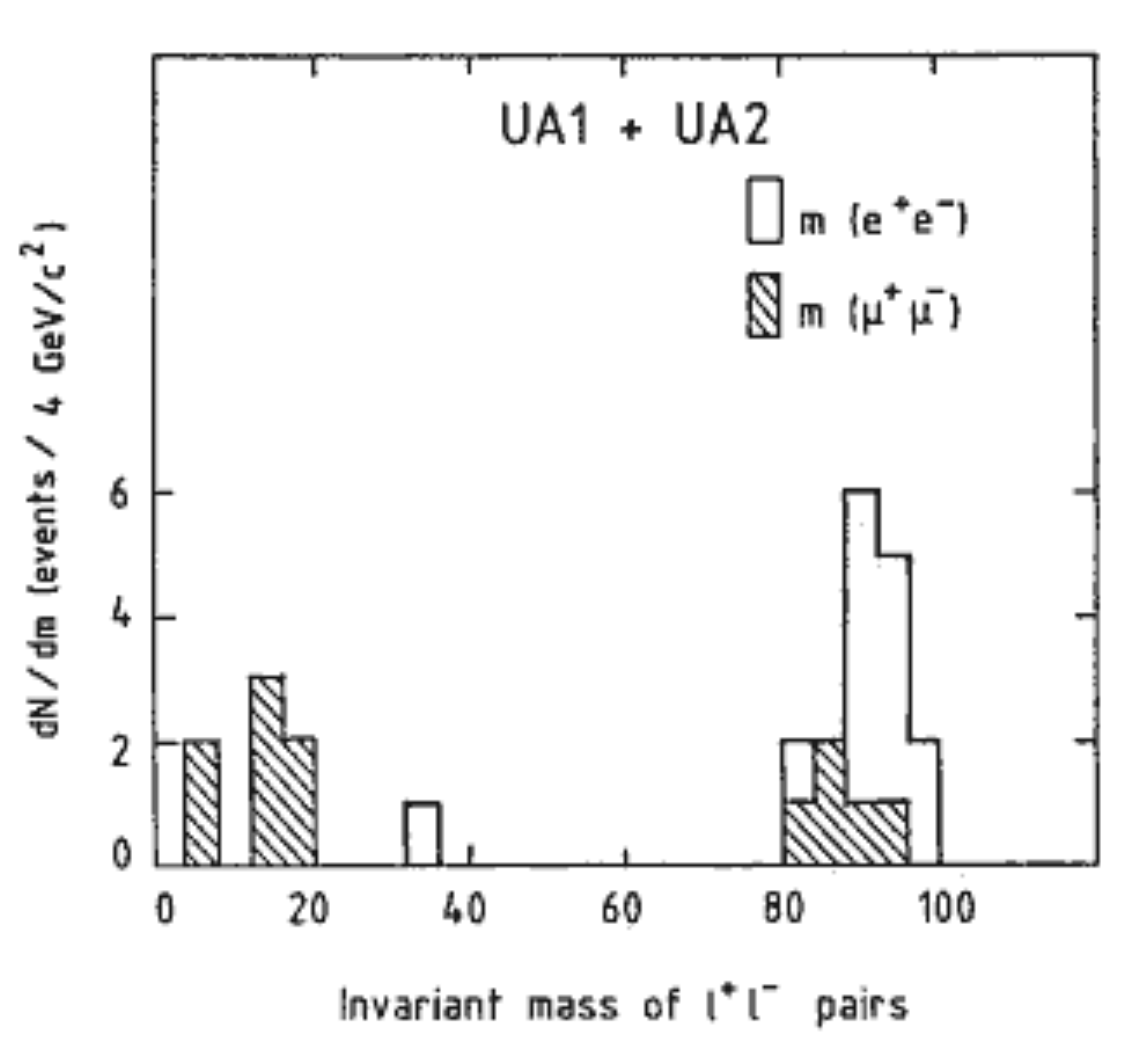}} 
\hspace{5mm}
\subfigure[$W$ boson transverse mass]{\includegraphics[width=7cm,height=6.5cm,clip]{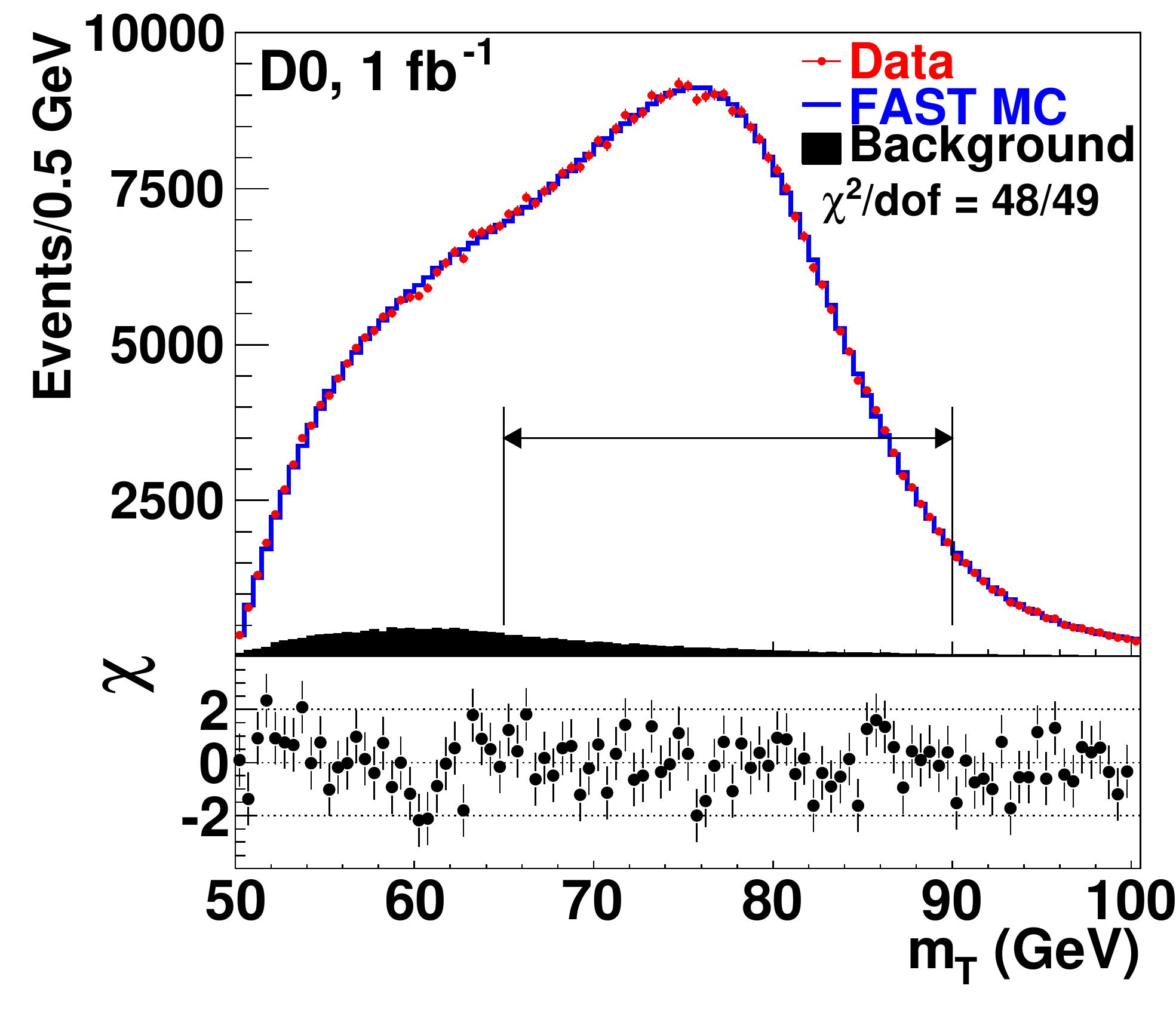}}
\caption{\label{fig:wandzplots}
(a)~Dilepton invariant mass distribution for the process $p\bar{p}\rightarrow Z/\gamma \rightarrow \ell^+\ell^-$.
From~\cite{RevModPhys.57.699}.
(b)~Transverse mass distribution for $p\bar{p} \rightarrow W \rightarrow e\nu$.
The $W$ boson mass is determined from a fit to the range indicated with the 
double-headed horizontal arrow.
From~\cite{Abazov:2009cp}.
}
\end{center}
\end{figure}

The simplest examples of kinematic mass reconstruction, 
e.g in the case of $Z\to e^+e^-$, are familiar. The decay topology can be written $A\to B C$ (\autoref{fig:atobc}a) where capital letters are used to label particles, and
corresponding lower case letters represent their four-momenta.
The parent ($Z$ boson) mass can be obtained from the straightforward 
construction of the ``invariant mass'' from the square of the sum of the visible four-momenta: 
\begin{equation}\label{eq:minv} m_{bc}^2 = (b+c)^2 .\ \end{equation}
One therefore obtains an event-by-event estimator of the $Z$ boson mass,
and can form a distribution which may be calibrated by comparison to calculations and/or Monte Carlo simulations (\autoref{fig:wandzplots}a).

\subsection{Decays to a visible and an invisible particle (``two-body semi-invisible'')}\label{sec:w}
A more interesting case, because the final state contains missing information,
can be found by considering leptonic $W$ boson decay (\autoref{fig:atobc}b).
For $W\to\ell\nu$, the topology is again $A\to BC$,
but the neutrino is essentially invisible.
Henceforth we will
denote invisible particles with a slash; writing this now as $A\to B
\slashed C$.  Although the three-momentum of the neutrino is not
observed, its transverse momentum $ \slashed{\bf c}_T$ may typically
be inferred from energy momentum conservation in the transverse plane
if there are no other invisible particles in the event.
For each event there is some range of values of $m_W$ which are consistent 
with the observables $b$, $\slashed{\bf c}_T$, and the known mass of
the lepton $m_B$ and the (negligible) mass of the neutrino $m_\slashed
C$.
The boundary of the allowed domain is conveniently \label{sec:mtasboundary} found by the explicit construction of the 
{\em transverse mass}, \label{sec:mtgetsdefined} $\MT$ \cite{NeervenVermaserenGaemersMT,Arnison:1983rp,Banner:1983jy,PhysRevLett.50.1738,PhysRevD.36.295}:
\begin{equation}\label{eq:mtdef}
\MT^2 \equiv m_B^2 + m_{\slashed C}^2 + 2 \left( e_b e_\slashed c - {\bf b}_T \cdot \slashed{\bf c}_T \right) .
\end{equation}
The (lower case) ``transverse energy'' quantities $e$ for each particle are defined by
\begin{equation}\label{eq:et_noz}
e^2 = {m^2 + {\bf p}^2_T} .\
\end{equation}
These $e$ are equal to the $E_T$ quantities (also denoted ``transverse energy'') defined in \eqref{eq:Etprojdef} in the massless limit.
That the function in \autoref{eq:mtdef} gives the largest value of $m_W$ consistent with the observations is noted in \cite{Cheng:2008hk,Barr:2009jv}.
While the results of hypothesising incorrect values for the mass of one of the daughter 
particles are of great interest -- and are explored further in \secref{sec:kinks} --
one can also obtain a simple but equally important result when the {\em correct} values of 
the daughter particles masses are assumed.
For the true values of $m_B$ and $m_\slashed{C}$
and in the approximation where the widths are narrow and experimental resolutions small, the inequality 
\begin{equation}
\MT \leq m_A
\end{equation}
is satisfied by construction, with equality when the relative rapidity of the
daughter particles vanishes.
Therefore a histogram of values of $\MT$, for many events with the same topology, 
should populate some regions (corresponding to allowed values of $m_W$)
but not other regions, corresponding to disallowed values of $m_W$.
The mass could then be determined from the boundary of the populated region -- the 
{\em kinematic endpoint} or {\em edge}.
In practice, background events, finite-width effects and experimental
resolutions smear the edge, so precise determinations of the $W^\pm$ mass
using this method need to model such effects (see \cite{Arnison:1983rp,Banner:1983jy} and others subsequently including the example in \figref{fig:wandzplots}b).



\subsection{Fully visible three-body decays}\label{sec:threebody}

\begin{figure}[t]
\begin{center}
\includegraphics[width=3.5cm]{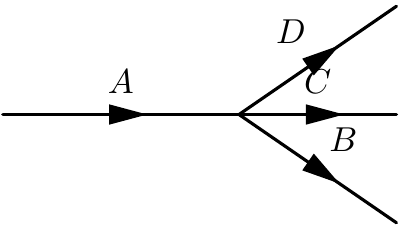}
\caption{\label{fig:threebodyvisiblelabelled} A single particle ``A'' decaying to three visible particles ``B'', ``C'' and ``D''.}
\end{center}
\end{figure}

Techniques for analysing three-body decays of the type shown in Figure~\ref{fig:threebodyvisiblelabelled}, i.e.~where all three daughters are visible,
can be most conveniently analysed using the tried-and-tested method of the {\it Dalitz plot} 
\cite{Dalitz:1953cp,Amsler:2008zzb}.
This plot projects the momenta onto a surface (usually $\{m_{BC}^2,m_{BD}^2\}$)
which is uniformly populated for a three-body decay with a constant matrix element. 
Intermediate resonances can be observed as bands in these plots for particular values of invariant mass.
Angular momentum multipoles can be determined from the rank of the spherical tensor needed to 
reproduce the observed angular distributions.

Attempts to reproduce the desirable features of the Dalitz plot when invisible particles are 
unobserved are revisited in \secref{sec:maost}.

\subsection{The dilepton edge: two successive two-body decays}\label{sec:ll}

\begin{figure}
\begin{center}
\includegraphics[width=4cm,clip]{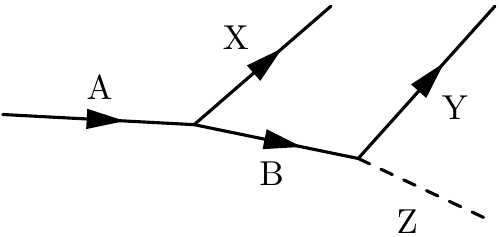}
\caption{\label{fig:dilepton}
The ``dilepton'' decay topology. The particle labelled $Z$ is assumed to be unobserved by the detector.
}
\end{center}
\end{figure}

\begin{figure}
\begin{center}
\includegraphics[width=0.4\linewidth]{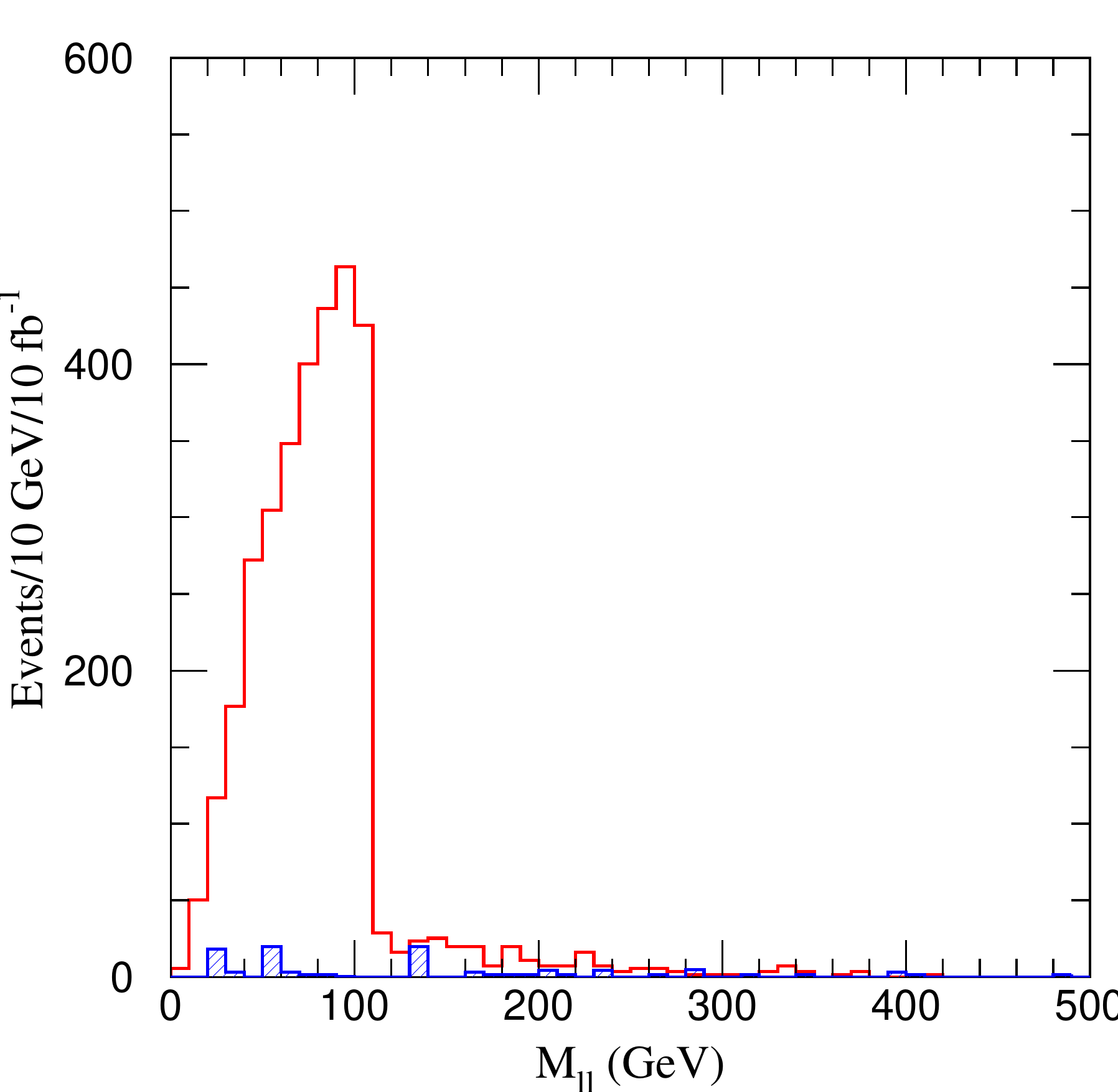}
\caption{\label{fig:mll_9609373}
An example ``dilepton'' distribution (taken from \cite{Paige:1996nx}) for the topology shown in \autoref{fig:dilepton}.  In this example, the kinematic endpoint is at approximately 100~GeV.
}
\end{center}
\end{figure}

An example of a hypothesis used for the partial reconstruction of one part of an event
is the topology shown in \autoref{fig:dilepton}.
This is sometimes called the ``dilepton'' topology, since it was first studied in the context of the LHC
\cite{Paige:1996nx} for the case of the supersymmetric decay 
$\ntlinoTwo \to q l^\pm \slepton^\mp \to q l^\pm l^\mp \ntlinoOne$.
The kinematics are most easily studied in the rest-frame of particle $B$ (the slepton in the example above)
in which if the masses are fixed, the sizes of the momenta of the final state particles $X$, $Y$ and $Z$
are fixed. The invariant mass of the visible system, $m_{XY}$, then depends only on the 
angle $\theta$ between $X$ and $Y$. In the limit of small masses of $X$ and $Y$ 
(which is approximately true for the dilepton case), 
the density of states is proportional to $m_{XY}$ up to a maximum at 
\begin{equation}(m_{XY}^\mathrm{max})^2=\frac{(m_A^2-m_B^2)(m_B^2-m_Z^2)}{m_B^2}\label{eq:dileptonconstraint}\end{equation} when $\theta=\pi$.
Plotting a distribution of $m_{XY}$ one therefore obtains a triangular distribution, such as the one shown in \autoref{fig:mll_9609373}.  The maximum endpoint of this distribution
can be measured, giving one constraint on the three variables, $m_A$, $m_B$, and $m_Z$.
It is worth noting in this context that while the endpoint of the sequential two-body decay 
\eqref{eq:dileptonconstraint} 
constrains differences in squared masses, 
the equivalent single step three-body decay $A\rightarrow X Y Z$
would have an endpoint at $m_{XY}^{\max}=m_A-m_Z$, so would constrain the difference in unsquared masses.

Examples of applications include 
sensitivity for multiple kinematic endpoints from competing decay chains \cite{atlasphystdr},
calculations of the $m_{\ell\ell}$ distribution shapes \cite{Nojiri:1999ki,Gjelsten:2004ki,Birkedal:2005cm,Smillie:2005ar,Athanasiou:2006ef},
tests of lepton universality \cite{Hinchliffe:2000np,Goto:2004cpa,Allanach:2008ib}, 
and an examination of pairs of such dilepton chains \cite{Bian:2006xx,Bisset:2008hm}.

If individual lepton flavour numbers are assumed to be conserved 
then in the dilepton case the signal can be expected to be found in opposite-sign same-flavour
(OSSF) pairs ($e^+e^-$ and $\mu^+\mu^-$). Backgrounds from e.g.~$t\bar{t}$ will not 
have lepton flavour correlations, and so an estimate of the OSSF background distribution 
(resulting from such flavour-uncorrelated sources) 
can be obtained from the opposite-sign, different flavour (OSDF) $e^\pm\mu^\mp$ distribution \cite{atlasphystdr}.

\begin{figure}
\begin{center}
\includegraphics[scale=0.8,clip]{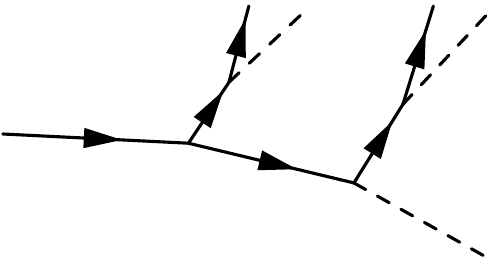}
\caption{\label{fig:ditau}
The ``ditau'' decay topology. 
}
\end{center}
\end{figure}
\phantomsection
\label{sec:tautau}
The di-tau invariant mass was investigated in \cite{Hinchliffe:1999zc}. This last case
is not strictly an example of the topology of \autoref{fig:dilepton} since each tau
decay also generates invisible particles (neutrinos), so the appropriate topology is that of \autoref{fig:ditau}.
More about chains with multiple invisible particles can be found in \secref{sec:multipleinvisibles}.
Helicity effects in tau distributions are discussed in \cite{Choi:2006mt,Mawatari:2007mr,Godbole:2008it}.

\subsection{Constraints from the $qll$-like chain
\label{sec:qll}}

If the ``dilepton'' topology of \secref{sec:ll} is extended by one two-body
decay, we reach a chain having three successive two-body decays,
resulting in a final state consisting of three visible (frequently but
not always light) particles, and one (frequently but not always
massive) invisible particle.
\par\begin{center}
\includegraphics[scale=0.5]{Mode1Unlabeled.pdf}
\end{center}\par

The most frequently considered context in which this topology is used
is the decay of $\squark \to q
\ntlinoTwo \to q l^\pm \slepton^\mp \to q l^\pm l^\mp \ntlinoOne$
\begin{center}
\includegraphics[scale=0.8]{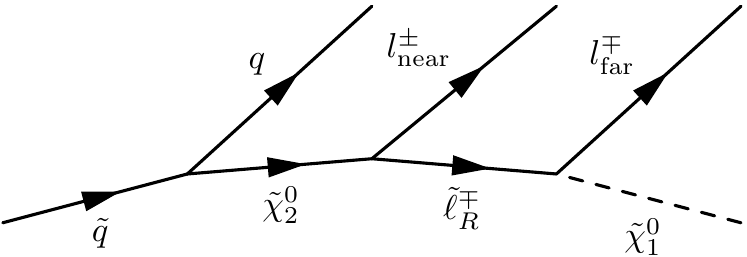}
\end{center}
which has led to this chain being known as the ``$qll$-chain''.
In fact the $qll$ case is really a special one in the sense that it assumes
particular identities of particles, 
and hence admits only particular possibilities for ambiguities.
This chain was first suggested as a means of measuring sparticle masses in
\cite{Hinchliffe:1996iu,Bachacou:1999zb}.  These early works proposed
that, following on from the di-lepton edge technique described above,
other one-dimensional invariant mass distributions be plotted
involving the quark (or rather jet) momenta in addition to the momenta
of the leptons.  As before, relativistic kinematics impose an upper
limit on any particular invariant mass distribution, and the position
of any particular upper limit (or more generally {\em kinematic
end-point}, or in some cases just ``{\em end-point}'') may be
established as a function of the masses of the particles involved in
the chain. As always, these kinematic end-point positions are
valid only if the events are from the topology considered, and will be
smeared by detector resolution effects.  Events from ``backgrounds''
may have almost any invariant mass.

Conventionally, the lepton produced first in the decay of the heavier neutralino is called the
``near lepton'' (near to the quark) and is notated $\lNear$, while the
lepton produced second in the decay of the slepton is called the ``far
lepton'' and is notated $\lFar$ \cite{Lester:2001zx,Allanach:2000kt}.
For the concrete case of the $qll$
it is not possible in a single event (in isolation from any other
information) to determine which observed lepton is $\lNear$ and which
is $\lFar$ and thus it is not possible to construct an invariant mass
distribution consisting of exclusively of the combination
$\mlqNear$ (or that of $\mlqFar$).  The early $qll$-chain studies
\cite{Hinchliffe:1996iu,Bachacou:1999zb,phystdr} elected to put to one
side the issue of this ambiguity (c.f.\ general discussion in
section~\autoref{sec:ambiguities}).  Subsequent attempts at addressing the
issue of this ambiguity established the need to build mass constraints
out of kinematic end-points of distributions which were truly
``observable''.  For example, the first such attempts
\cite{Lester:2001zx,Allanach:2000kt} proposed that the distributions
of $\mll$, $\mllq$, $\mlqHigh \equiv \max\{\mlqPlus,\mlqMinus\}$ and
$\mlqLow \equiv \min\{\mlqPlus,\mlqMinus\}$ be used.
along with other
variables (discussed later) to measure the corresponding kinematic
endpoints $\mll^\rmax$, $\mllq^\rmax$, $\mlqHigh^\rmax $ and
$\mlqLow^\rmax$.  
From a kinematic perspective, though not from a spin
perspective (see \secref{sec:spins}), there is no point in using
$\mlqPlus$ and $\mlqMinus$ in place of $\mlqHigh$ and $\mlqLow$: even
though either pairing is ``observable'', the Majorana nature of the
neutralino makes the two distributions identical.

Note that many of the invariant mass combinations
that can be constructed are not independent of the others.  For
example, in the limit of massless visible particles,
$\mll^2+\mlqLow^2+\mlqHigh^2 = \mll^2+\mlqPlus^2+\mlqMinus^2 =
\mllq^2$.  In some but not all cases, this can lead to the kinematic
end-points themselves being related.  For example $(\mll^\rmax)^2 +
(\mlqHigh^\rmax)^2 = (\mllq^\rmax)^2$ over {\em some but not all} parts of
mass-space \cite{Gjelsten:2004ki}.

There is definitely a clear benefit to be derived from critically
(re-)examining the choices of one-dimensional distributions used to
constrain the $qll$-chain for
there is no reason to believe that the ``traditional'' choices of
endpoint \cite{Lester:2001zx,Allanach:2000kt} are optimal in any sense
-- indeed it is very unlikely that the ``traditional'' choices are
optimal by any definition as a measure of optimality was never part of
their design.  For example, \cite{Matchev:2009iw} point out that it
may be preferable to look for two endpoints (i.e.~the \lqEdgeNear\
and the \lqEdgeFar) in the ``union'' distribution of
$\mlqMinus\cup\mlqPlus$ rather than to split this distribution into
the \lqHigh\ and \lqLow\ components, as the resulting inversion space
has only a twofold rather than a threefold ambiguity.  Similarly, it can be advantageous to look for maxima in linear combinations of invariant masses.  For example \cite{Matchev:2009iw} investigates the properties of kinematic endpoints of distributions of $(\mlqMinus^{2\alpha}+\mlqPlus^{2 \alpha})^{\frac{1}{\alpha}}$ for different values of $\alpha$ and finds merit in the particular case $\mlqMinus+\mlqPlus$ (in this regard note the kinematic end-point of the ``\lqEdgeSum'' in \autoref{tab:obs} on \pref{tab:obs}).

One consequence of moving from technically unobservable distributions
(like that of $\mlqNear$) to observable distributions (like that of
$\mlqHigh$) can be that the locations of the kinematic end-points may
become {\em piecewise-smooth} functions of the unknown masses
\cite{Lester:2001zx,Allanach:2000kt,Gjelsten:2004ki,Gjelsten:2005aw}.  Furthermore,
such invariant mass distributions can evolve non-trivial shapes, and
can acquire undesirable features (so called ``feet'') near the end
points which might in some cases make end-point measurement prone to
large systematic errors \cite{Gjelsten:2004ki,Gjelsten:2005aw,Miller:2005zp}.  Local
non-differentiability of end-point position need not, in itself, be a
problem for mass determination (note that piecewise-smooth functions
like $|x|+2x$ can have well defined inverses) however it can be a
visible symptom of a separate issue which is of concern in certain
cases: {\em ambiguity in end-point inversion}, discussed below.
Accordingly there has been some recent interest in alternative
observable distributions for
which end-point positions are smooth functions of the masses
\cite{Matchev:2009iw}. 

\subsubsubsection{Ambiguity in end-point ``inversion''}\label{sec:inversion}
Very often one finds oneself in the unfortunate position of having too few
observables to constrain all the parameters of a model.  On other
occasions one may find oneself with a much larger number of independent
measurements, sufficient to over-constrain a model.  In this fortunate
position, one potentially has the power to rule out a model, or else
to give strong constrains on the parameters of a model which is
consistent.

Occasionally one may find oneself in the very special situation in
which the number of independent observables or measurement happens to
match exactly the number of free parameters (e.g.~masses) in the
model.  In such situations it can be very hard to resist the
temptation to search for analytic or closed-form ``inversions'' :
i.e.~solutions for the parameters (e.g.~masses) in terms of the
observables or measurements (e.g.~the position of the end-points).
Many such ``inversions'' have been published published for different
sets of observables for the $qll$-chain
\cite{Gjelsten:2004ki,Miller:2005zp,Burns:2009zi,Matchev:2009iw,Costanzo:2009mq}.  For
some sets of end-point measurements the inversion process
may yield a single set of consistent
masses -- hopefully the correct ones -- while for some other sets of
end-point methods there may be more than one set of consistent masses
(of which one is hopefully correct while the others are spurious).

In fact the previous statement applies to the idealised situation in which
detector resolution is perfect.  
While the performance of some ``inversions'' degrade rapidly in response to even small amounts of experimental smearing/resolution, others are much more tolerant.  It seems that there is much scope for  future work to determine which inversions are best suited to experimental application and which are not.

For more detailed discussion see \cite{Burns:2009zi} and
\cite{Matchev:2009iw}.  In particular these papers pose the further
question: ``Can one find sets of distributions whose end-points always
yield the smallest number of spurious solutions?'' and in answering
this yield entirely new sets of invariant mass distributions for the
$qll$-chain.

There are benefits, clearly, in widening our understanding of what
features in data drive our mass constraints.  Looking at endpoint
inversion formulae (and minimal sets of invertible endpoints) is one
way that can be accomplished.  Nevertheless, it should be remembered
that the issue of analytic invertibility {\em alone} must not drive the choice of
variables used.  Frequently there will be other more important issues to
address which might include: (1) which end-points are easiest to
observe (dependent on slope and shape near the end-point; relative
numbers of signal and background events near the edge; the degree to
which the background shape and size may be independently predicted);
(2) which are least smeared by detector resolutions; and (3) which are
least sensitive to cuts and acceptance or things which can affect
systematic uncertainties.  Furthermore, it seems very likely that the
best measurements will be made by putting together the largest
possible number of pieces of (sometimes overlapping) evidence in a
joint numerical fit, rather than by inverting a set of equations for a
particular set of constraints at the expense of other observables.

The reader who is not convinced that there is much work yet to be done
in identifying better (or at least additional) means of constraining
masses in the $qll$-chain would be well advised to review the
cautionary tale of the hitherto undiscussed lower kinematic endpoint
known as the \llqThreshold.  Most of the ``traditional'' sets of
endpoints
\cite{Lester:2001zx,Allanach:2000kt,Gjelsten:2004ki,Miller:2005zp} as
well as some of the new proposals \cite{Burns:2009zi} rely to a lesser
or indeed greater extent on the \llqThreshold\ proposed first in
\cite{phystdr}. This is the lower end-point of the $\mllq$ distribution
under the additional constraint that $\mll < \mll^\rmax / \sqrt(2)$.
This lower end-point is notorious\footnote{A discussion of the experimental drawbacks of the \llqThreshold\ distribution may be found in Section 1.2 of \cite{Matchev:2009iw}.}
for having experimental systematic errors associated with its
measurement (in part due to the shape \cite{Lester:2006yw} at turn-on
being concave) which are in some cases much larger than those required
to make use of the constraint it provides.  Such
an end-point may turn out to be just the sort of measurement that
looks good on paper but turns out to be under poor experimental
control.

\subsubsubsection{Moving away from one-dimensional constraints}
It is often suggested, particularly by the experimental community, that one-dimensional distributions of variables like $m_{ll}$, $m_{qqll}$ and the other
Lorentz invariants discussed above offer the simplest, and probably
the most easily measurable distributions from which to extract
information about the masses of the parents.  But is this suggestion correct?  It is certainly being challenged.  These 
one-dimensional invariant mass distributions can all be thought of as
``projections'' of the higher-dimensional space in which the
measurements live, onto a single dimension.  The full
three-dimensional shape of the $qll$-chain has been noted in
\cite{Costanzo:2009mq}, and there are many promising proposals to use
fits to structures in observables of two (and higher) dimensions in
order to gain information from correlations that are not otherwise
available in one-dimensional distributions \cite{Bian:2006xx,Bisset:2008hm,Matchev:2009iw,Costanzo:2009mq,Burns:2009zi}.  In principle there is a lot more information available in these higher-dimensional distributions -- but whether that information is easier or harder to extract than that from the one-dimensional distributions will depend to a large extent on the relative degree to which the systematic uncertainties can be understood/controlled by the experimental collaborations in the two cases.

\subsection{Constraints from the $qqll$-like chain}\label{sec:qqll}

\begin{figure}[t]
\begin{center}
\includegraphics[scale=0.6,clip]{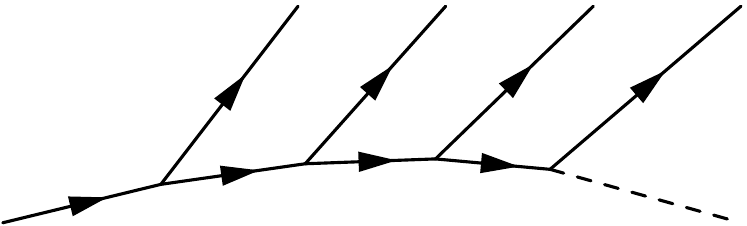}
\caption{\label{fig:gluino}
The ``gluino'' decay chain.
}
\end{center}
\end{figure}

Adding a further two-body decay to the ``$qll$'' chain produces the topology shown in \autoref{fig:gluino},
which sometimes called the ``qqll'' or ``gluino'' chain since the most studied example has been 
$\gluino \to \bar{q} \squark \to \bar{q} q \ntlinoTwo \to 
\bar{q} q l^\pm \slepton^\mp \to \bar{q} q l^\pm l^\mp \ntlinoOne$.
Many of the kinematic endpoints for this longer chain
can be found in the results of section~\autoref{sec:ll} and section~\autoref{sec:qll} (or relabellings thereof).
The new endpoints, including the maximum of the four-body $qqll$ distribution 
have been calculated using massless approximation for the 
visible particles \cite{Gjelsten:2005aw} assuming all particles on the backbone are on mass-shell.\footnote{Contrast how little has been written \cite{freedmanlester} about the case where some particles on the backbone of the $qqll$-chain are {\em off} mass-shell.} These can depend on any of the other five masses in the problem
($\gluino$, $\squark$, $\ntlinoTwo$, $\slepton$ and $\ntlinoOne$).
The same chain has been used to put constraints on the spin of the
gluino \cite{Alves:2006df}.   When dealing with chains of this length containing many jets in the final state, most studies have found it necessary to have addional information about the jets (for example bottom-quark tags) to reduce ambiguities due to combinatorics and ISR {\it etc}.

\subsection{Other chains containing successive two-body decays}\label{sec:darkmattersandwich}

\begin{figure}[t]
\begin{center}
\includegraphics[scale=0.6,clip]{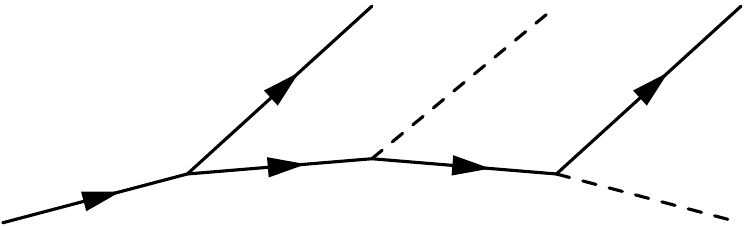}
\caption{\label{fig:darkmattersandwich}
The ``dark matter sandwich'' decay chain.
}
\end{center}
\end{figure}

One interesting chain which has been studied (to the best of our knowledge) only in \cite{Agashe:2010gt} is the ``dark matter sandwich'' topology of Figure~\ref{fig:darkmattersandwich}.  What makes this chain different from $qll$-like chains is that the missing invisible particle (the dark matter particle) emerges at the mid-point of the decay chain, rather like the filling in a sandwich.  In the context in which it was investigated, the particles on the back bone were allowed arbitrary masses, while the two visible ejecta were treated as massless.  As such the invariant mass distribution of the two visible particles depends on five unknown masses (the four backbone masses and the sandwiched invisible particle's mass).  Formulae for the position of the kinematic endpoint and the differential shape of the invariant mass distribution of the invariant mass of the pair of visible ejecta may be found in \cite{Agashe:2010gt}.

These distributions are notable for having kinematic {\em cusps} -- places in the differential distribution where two curves with different slopes come together. Further discussion on such cusps can be found in Sections~\ref{sec:higgs} and \ref{sec:robotics}.

\subsection{When backbone sparticles are off mass-shell in multi-step decay chains}

It is worth asking whether we would be able to tell if
an observed ensemble of similar multi-particle final states is likely to have come from 
events containing a long series of 2-body decays,
or whether it might (for example) have originated from a somewhat shorter 
decay chain with more particles emitted at a smaller number of vertices.
This question can be rephrased as asking whether the narrow-width approximation
is valid for all the intermediate states.

\begin{figure}[t]
\begin{center}
\includegraphics[scale=0.6,clip]{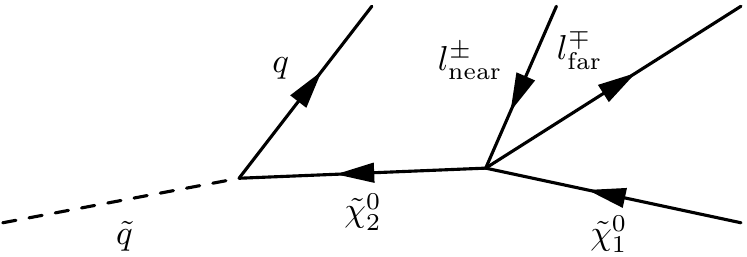}
\caption{\label{fig:twothreelabeled}
The $qll$ decay chain with a terminal three-body decay.
}
\end{center}
\end{figure}

In the context of the discussion above,
it may be noted that not all sparticles on the ``backbone'' need be on
their mass-shell.  
It is possible to imagine scenarios in which (for
example) the sleptons are heavier than the second-lightest neutralino.
The second-lightest neutralino would then not be able to decay via an
on-shell slepton, though a three body decay via a highly virtual
off-shell slepton might still \label{sec:twothreegetsdiscussed} be
possible, illustrated in \autoref{fig:twothreelabeled}.  The observed
final state particle content (a jet and two opposite sign sane flavour
leptons) offers no clue as to whether the decay topology has a virtual
or an on-shell slepton in the backbone.  This is a problem, because
the positions of the kinematic endpoints of the usual invariant mass
distributions are entirely different functions of the masses of the
sparticles involved.  Consequently it might be possible, if events
coming from the on-shell scenario were analysed using the off-shell
hypothesis (or {\em vice versa}) to obtain entirely spurious mass
measurements.  Fortunately, the way that the event-space is populated
(i.e.~the shapes of the distributions
\cite{Gjelsten:2004ki,Gjelsten:2005aw,Miller:2005zp,Smillie:2005ar,Athanasiou:2006ef,Lester:2006yw,Burns:2008cp,Matchev:2009ad,Konar:2009wn,Burns:2009zi,Agashe:2010gt})
and the relationships between the positions of the kinematic endpoints
\cite{Gjelsten:2004ki,Gjelsten:2005aw,Miller:2005zp,Lester:2006cf,Burns:2009zi}, betray clues as to the nature of the topology
and can permit the type of the decays (two-body versus
three-body or similar) to be determined correctly under favourable circumstances.

\par

The $qqll$-chain is as susceptible as the $qll$-chain to ambiguities
introduced from not knowing which (if any) of the particles on the
backbone are on or off mass shell.  It is possibly the case that the
only work which has considered the $qqll$-chain with off-mass-shell particles on the backbone is an incomplete undergraduate project
\cite{freedmanlester}.

\par

\begin{figure}[t]
\begin{center}
\includegraphics[scale=0.6,clip]{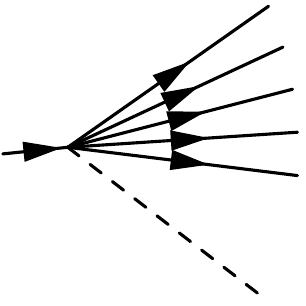} \ \raisebox{6mm}{$\bigcup$} \ 
\includegraphics[scale=0.6,clip]{FunnyDark2.pdf}
\caption{\label{fig:funnydark}
Occasionally there are models which can produce {\em both} of the above topologies in the decays of a single type of particle.  For such models, variables have been proposed which constrain the unknown masses. See the end of Section~\ref{sec:funnydark}.
}
\end{center}
\end{figure}

We round off this section \label{sec:funnydark} by noting an observation of \cite{Agashe:2010gt} regarding a particular class of models considered therein in which a single massive particle could decay via multi-body decays into two or more visible particles and {\em either one or two invisibles particles of identical and unknown  mass}. These two topologies are shown in  Figure~\ref{fig:funnydark}.  It was noted therein that the distribution of the invariant mass of the visible system would show a double endpoint structure -- one endpoint being the difference between the mass of the parent and the invisible daughter: $(m_{\mathrm{vis}}^{\max{}})_1 = m_A - m_C$, and the other being $(m_{\mathrm{vis}}^{\max{}})_2 = m_A - 2 m_C$.  Were such a double endpoint structure to be observed, and were one prepared to hypothesise the underlying structure of Figure~\ref{fig:funnydark}, the authors of \cite{Agashe:2010gt} propose that one use the linear combination of endpoints $2 (m_{\mathrm{vis}}^{\max{}})_1 - (m_{\mathrm{vis}}^{\max{}})_2$ to measure the parent mass $m_A$, while the combination $ (m_{\mathrm{vis}}^{\max{}})_1 - (m_{\mathrm{vis}}^{\max{}})_2$ would measure the invisible particles' mass $m_C$.

\subsection{Directly reconstructible}\label{sec:direct}

Most of the discussion until this point has involved final states
for which at least one of the daughter particles is expected to go undetected.
When the particle(s) of interest decay to a set of daughters all of which
are visible, then determining the mass of the parent(s) 
should generally more straightforward (at least in principle).
A simple example of the fully-visible case, $A\to B C$ (\autoref{fig:atobc}a) was discussed in 
\secref{sec:z}, and the three-body case $A\to B C D$ in \secref{sec:threebody}.

However, even when all the particles are visible the kinematical reconstruction 
is not necessarily trivial. For example 
it is often difficult to assign the visible particles to the appropriate decay,
particularly if there is a large number of final-state objects or much initial state radiation. 
Other ambiguities can arise when attempting to associate 
final state hadronic jets to particular types of decay,
since jets are themselves composite objects.
Some examples of papers considering these more difficult cases are surveyed in what follows.


\subsubsection{Combinatorial complications.}\label{sec:rpv}
Even when all of the final state particles can be identified, the task 
of reconstructing the masses of the parent particles can be far from trivial.
In events with many objects (jets, leptons, \ldots) in the final state
the attempt to associate such objects to particular parents
involves considering a factorially large number of different possible combinations.
Though one can attempt to resolve some of these ambiguities in particular cases by
appealing to e.g.~lepton number conservation~\cite{phystdr,Barr:2008ba,Serna:2008zk} 
or by looking for correlated kinematic features~\cite{Bian:2006xx,Bisset:2008hm},
often there is no alternative but to assume that all kinematic combinations are possible.

\begin{figure}[t]
\begin{center}
\includegraphics[scale=0.6]{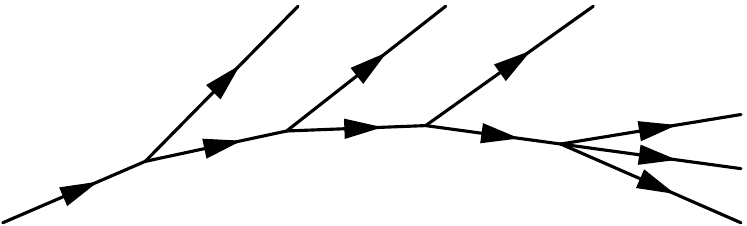}
\caption{The topology explored in \cite{Allanach:2001xz}.\label{fig:rpv}}
\end{center}
\end{figure}

A particularly difficult case -- because very many jets are expected in the final state -- is 
baryon-number violating, $R$-parity violating Supersymmetry \cite{Allanach:2001xz,Butterworth:2009qa}.
If the baryon-number-violating couplings are small, then the Supersymmetric 
decay chain proceeds as in the $R$-parity conserving case, but then each of the 
two lightest supersymmetric particles decays into three (different-flavour) quarks:
i.e.~$\ntlone\to qqq$.

One can attempt to reconstruct the $\ntlone$ mass from three-jet invariant mass
combinations. However in a hadron collider one expects 
(as well as the six jets from the two \ntlone\ decays) further jets from cascade decays,
so the combinatorial background from wrong jet combinations can be very significant.
The first attempts to reconstruct such complex topologies
made use of leptons from the cascade decays (\autoref{fig:rpv}) to simultaneously 
form invariant mass combinations for several heavy particles \cite{Allanach:2001xz}.

It has been shown that for one can reconstruct the heavy particle masses 
in such cases without relying on the existence of leptons in the cascade decays. 
That analysis made use of more sophisticated jet algorithms to 
determine the scale at which a single merged jet from the 3-quark system 
(from each \ntlone\ decay) 
can be resolved into sub-jets \cite{Butterworth:2009qa}.

Similar sub-jet analyses have been proposed in reconstruction
of other boosted heavy objects, with recent examples including searches for boosted Higgs bosons decaying to $b\bar{b}$ when produced in association with a vector boson~\cite{Butterworth:2008iy} or in association with  $t\bar{t}$~\cite{Plehn:2009rk}. Anther example of using jet substructure to improve the mass resolution (and hence signal to background discrimination) of heavy objects can be found in the context of highly boosted top quarks~\cite{ATL-PHYS-PUB-2009-081}. 


\subsubsection{Mass from velocity of metastable particles}\label{sec:longlived}

When charged massive stable particles traverse the detector
their mass can by determined from simultaneous 
measurements of momentum and velocity ($\beta$).
The momentum measurement is usually obtained in the same way as for a muon
-- i.e.~be determining the bending radius of the particle in an externally applied 
magnetic field.
The particle's velocity can be found from precision timing information,
or from measurements of energy loss ($\frac{dE}{dx}$) or from a combination of 
both methods.
When the mass of the metastable particle has been determined,
the full 4-vectors of all instances of that particle can be determined event-by-event.
This allows the mass of its parents/ancestors to be reconstructed 
by forming invariant masses along appropriate cascade decay chains.

LHC-related studies have considered the case of heavy leptons 
\cite{Nisati:1997gb,atl-muon-99-006,Kazana:687147,Ambrosanio:2000ik,Ambrosanio:2000zu,Allanach:2001sd,Ellis:2006vu,Aad:2009wy,Raklev:2009mg}
and bound states of heavy coloured objects (so-called $R$-hadrons) \cite{Kilian:2004uj,Hewett:2004nw,Kraan:2005ji,atl-phys-pub-2006-005,Aad:2009wy}.
Slow-moving particles present particular experimental difficulties
because the delay in reaching the outer parts of the detectors (the muon chambers)
means they risk being identified with the wrong bunch crossing. 
The experimental issues associated with triggering and reconstructing such particles 
have been addressed and are understood \cite{atl-phys-pub-2005-022,Tarem:2009zz}.
For more details on searches and measurements of massive stable particles 
we refer the reader to a recent review paper dedicated to that topic \cite{Fairbairn:2006gg}.

\subsection{Using spatial as well as momentum information}\label{sec:secondary}

If invisible long-lived particles decay within the detector then 
the location of the decay vertex in space 
can be used to provide constraints on the kinematics.
Examples of models predicting such displaced vertices include
bilinear \cite{deCampos:2007bn} or baryon-violating \cite{Allanach:2001if} $R$-parity violating supersymmetry,
and anomaly-mediated supersymmetry \cite{Barr:2002ex}.

A demonstration of how the position of the secondary vertex can be combined with 
direct kinematic information has been given in the context of a 
gauge-mediated supersymmetry breaking model \cite{Kawagoe:2003jv}.
Cascades terminating in the decay $\ntlone \to \gamma \tilde{G}$ were considered,
which (provided they occur within the tracking volume)
produce photons which detectable in the calorimeter but which do 
not point back to the primary interaction point.
The position, arrival time and momentum direction of the photons
are used to determine the photon momentum, 
allowing the (invisible) gravitino momentum to be completely determined.
Knowing both the photon and the gravitino momentum, the kinematics of the 
rest of the decay chain can also be determined.

\subsection{Multiply branched trees}\label{sec:higgs}

\begin{figure}\begin{center}
\subfigure[Multi-branched topology]{\includegraphics[width=3.5cm,clip]{Cusps.pdf}}
\hspace{8mm}
\subfigure[Cusps in invariant mass distributions]{\includegraphics[width=9cm,clip]{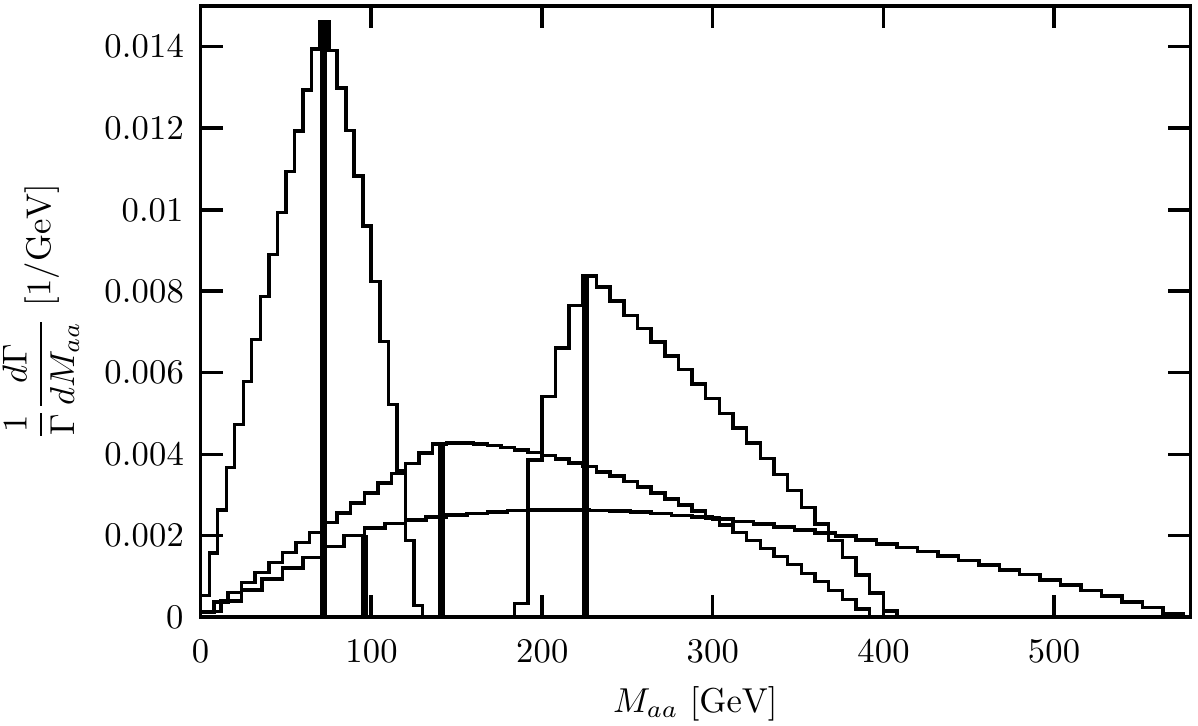}}
\caption{\label{fig:cusps}
(a) A branched decay chain of the sort considered in \secref{sec:higgs}.
(b) Examples of kinematic cusps. Adapted from \cite{Han:2009ss}.
}
\end{center}\end{figure}

One way to extract mass information from multi-branched graphs 
(such as that shown in \figref{fig:cusps}a) is by treating them with the same methods 
as single decays of unknown internal structure to $n$-body final states 
(i.e.~ignoring the existence of on-shell intermediate particles).
For example one get a good measurement of the Higgs boson mass by  constructing 
a transverse mass variable $\MT$ for the decay $H\rightarrow W^+W^-\rightarrow \ell^+\nu\ell^-\bar{\nu}$ 
treating it just like a single four-body decay \cite{Barr:2009mx}.
This method has some merits, but does not make use of the full kinematic information available in such topologies.

Multi-branched trees can have multiple on-shell constraints and so 
can contain a rich spectrum of possible Lorentz invariants.  
Plotting correlations between appropriate combinations can dilute backgrounds,
thereby improving the measurability of kinematic endpoints \cite{Huang:2008qd,Huang:2008ae}.

These decay topologies have also bring to light other interesting features. In particular one can observe in projected variables kinematical {\em cusps} \cite{Han:2009ss}. These features are places in the differential distribution where two curves with different slopes come together, as shown in \autoref{fig:cusps}b. Such cusps are a general feature of kinematic distributions 
not just multi-branched trees. 
For example one can find a \cite{Agashe:2010gt} for a cusp in topology described in \secref{sec:darkmattersandwich}.
The source of these cusps, as well as the other singularity structures -- endpoints and thresholds -- are discussed further in \secref{sec:robotics}. 

Kinematic cusps could well be as useful for extracting mass information as endpoints. In fact because the differential distributions are 
typically populated by large numbers of events near these cusps, uncertainties in cusp positions might well be smaller than endpoint positions.

In these multi-branched decays -- as elsewhere -- it is possible to use the extra kinematic constraints to select events in which all final state-momenta are well-constrained despite the presence of invisible particles \cite{Choi:2009hn} (see also \secref{sec:maost}).

\subsection{The contransverse mass}\label{sec:mct}
The invariant mass \eqref{eq:minv} has the property that it is not modified under any operation which transforms all of the particles with the {\em same} boost $p_i\mapsto \Lambda^{\vec{\beta}} p_i$.
It is also possible to construct variables which are invariant when {\em different} boosts are applied to their constituent particles. In particular, one can construct a variable which is constructed from the sum of two arbitrary Lorentz vectors $a$ and $b$ and which is invariant under {\em equal and opposite} boosts of those vectors 
\begin{eqnarray}\label{eq:backtoback}
b &\mapsto& \Lambda^{\vec{\beta}} b \\ c &\mapsto& \Lambda^{-\vec{\beta}} c .\
\end{eqnarray}
A variable which satisfies this back-to-back boost invariance condition was defined in \cite{Tovey:2008ui},
\[
\MC^2 = m_B^2 + m_C^2 + 2 ( E_b E_c + {\bf b} \cdot{\bf c} )
\]
where the bold quantities again represent the Euclidean three-vector momenta.
Note the plus sign before the dot product which distinguishes \MC\ from the invariant mass \eqref{eq:minv}.
In the limit when $m_B$ and $m_C$ are negligible, and the visible particles have originated 
from the decay $A\rightarrow BC $, one can see that $\MC = \sqrt{4E_bE_c} = 2p^*$
where $p^*$ was defined in \eqref{eq:ptlimit}.

Because of our ignorance (in a hadron collider) of the $z$-momentum of the initial state it is useful to define the related quantity constructed from purely transverse quantities
\begin{equation}\label{eq:mctdef}
\MCT^2  = m_B^2 + m_C^2 + 2 \left( e_b e_c + {\bf b}_T \cdot{\bf c}_T \right)
\end{equation}
where $e$ is defined in \eqref{eq:et_noz}.
This quantity is known as the {\em contransverse mass}.

Neither \MC\ nor \MCT\ have found much application for {\em single} two-body decays, but 
they have interesting invariance properties when {\em pairs} of identical semi-invisible decays (see \secref{sec:mtovey}) are produced back-to-back in the transverse plane.
The resulting contra-linear invariance properties are of interest because back-to-back configurations with extremal \MCT\ can be generated from the 
threshold-production (`at rest') configuration
by applying the transformations \eqref{eq:backtoback} to the respective parent particles.

\section{Variables for pairs of cascade decay chains}\label{sec:pairdecays}

\begin{figure}[t]
\begin{center}
\subfigure[\ ]{\includegraphics[scale=0.4]{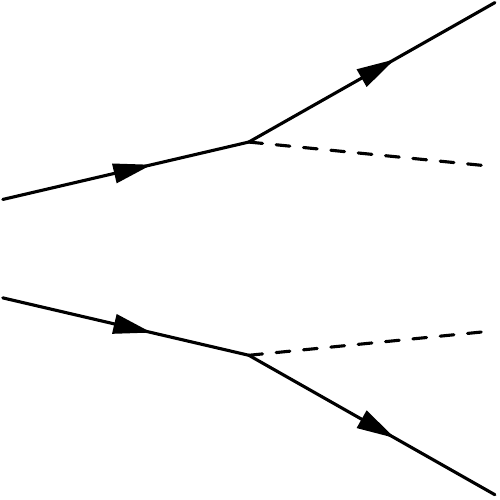}}\hspace{3mm}
\subfigure[\ ]{\includegraphics[scale=0.4]{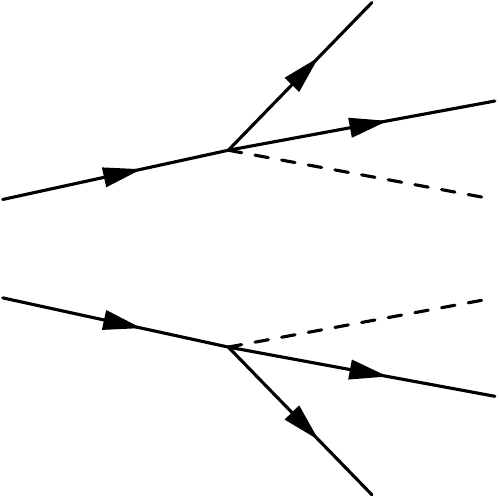}}\hspace{3mm}
\subfigure[\ ]{\includegraphics[scale=0.4]{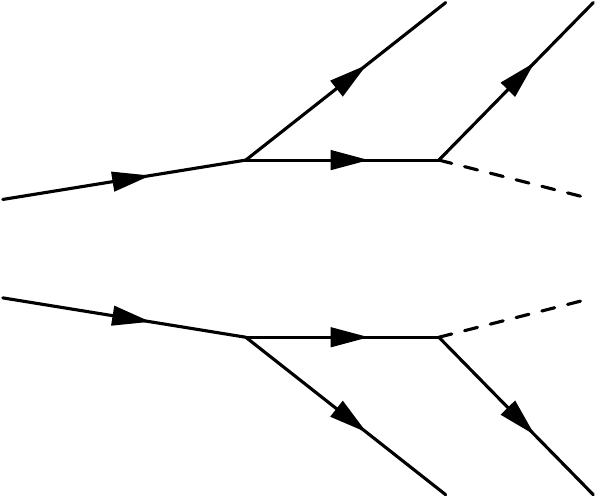}}\hspace{3mm}
\subfigure[\ ]{\includegraphics[scale=0.4]{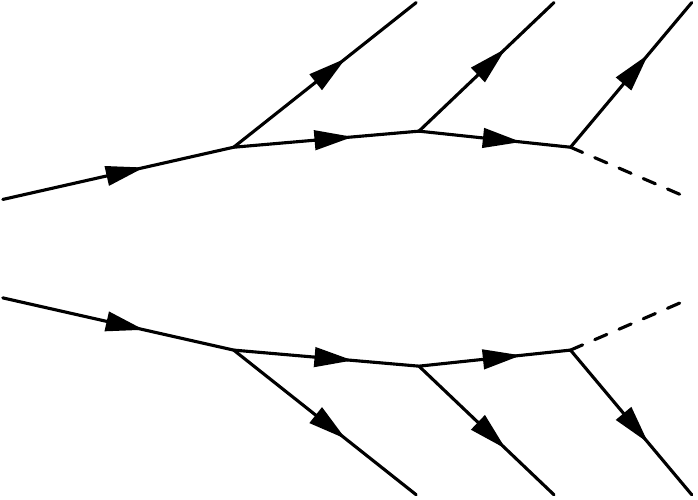}}\hspace{3mm}
\subfigure[\ ]{\includegraphics[scale=0.4]{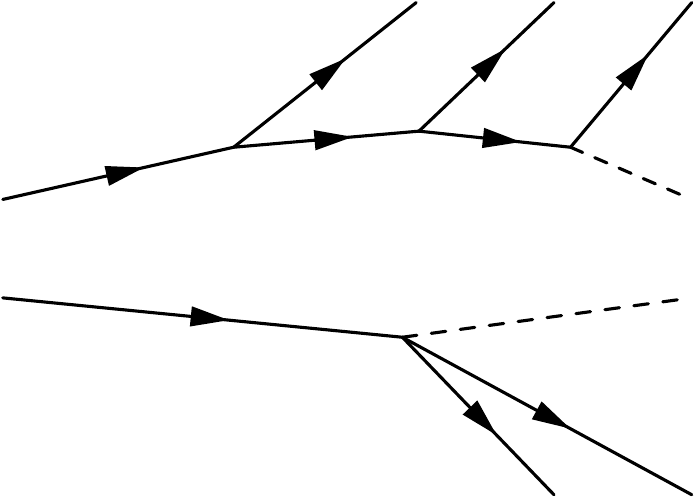}}\hspace{3mm}
\subfigure[\ ]{\includegraphics[scale=0.4]{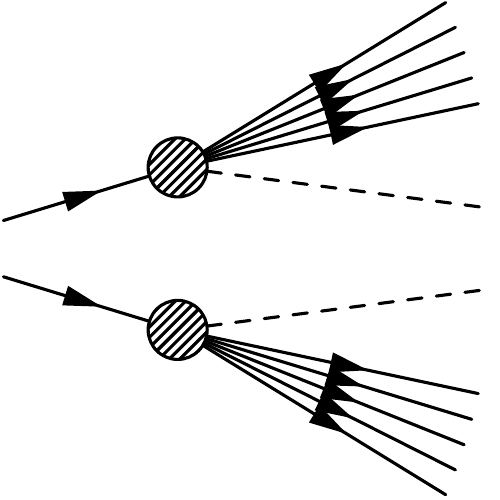}}
\caption{Examples of the ``dual-sided'' event topologies discussed in \secref{sec:pairdecays}.}
\label{fig:mt2sortsoftopologies}
\end{center}
\end{figure}

We have seen that the transverse mass
$\MT$ (\autoref{eq:mtdef}) is useful in situations involving $A\to B
\slashed C$ where $\slashed C$ is the {\em only} invisible particle in the event.  This
begs the question: ``What comparable tools can be employed in
situations where there are {\em two identical invisible particles} in
each event -- such as might arise in models with 
stable or meta-stable weakly interacting particles whose creation is
protected by a multiplicative quantum number?''  ($R$-parity
conserving supersymmetry \cite{Dimopoulos:1981zb} and universal extra-dimensional models \cite{Cheng:2002ab} being
just two examples of such models.) 

Techniques for extracting mass information from {\em pairs} of cascade decays are described in 
this section.
First we introduce the sorts of event-topology which are relevant to this question 
and the notation convention we will use when describing them.

\par
\label{sec:doublesideddefined}
\par

The topologies of interest in this section (shown in \autoref{fig:mt2sortsoftopologies})
share the common feature that each event is composed of two ``sides'' -- where each
``side'' consists of a decay chain which terminates in an invisible
particle and one or more visible particles.  For obvious reasons we
call such events ``double sided''.  The sides need not be identical
(\autoref{fig:mt2sortsoftopologies}(e) provides an asymmetric example,
and we will discuss asymmetric examples in more detail in section
\autoref{sec:mttwoasymmetric}) though topologies with identical chains
on each side have historically been the most studied.

\par

To distinguish the sides of the particles when discussing events of these
types, we used un-primed indices for the particles on one side, and
primed indices for particles on the other.  Where possible we use
letters nearer the beginning of the alphabet for the most senior
parent particles and letters nearer the end of the alphabet for the
most junior daughter particles.  For example, the simplest double sided
topology (\autoref{fig:mt2sortsoftopologies}(a)) might be denoted
$(A\to B \slashed C) + (A' \to B' \slashed C')$.

If a pair of particles is produced in the collision, and then each of these
goes on to decay, there are both additional constraints and additional complications
compared to the single decay case. New combinatorial ambiguities arise,
since it is no longer generally possible to associate a particular visible particle
with one or other of these decay chains. In addition there are constraints which 
link information between the two cascades -- for example the missing transverse momentum
is usually assumed to be equal to the sum of the momenta of any invisible particles
from {\em both} decay chains.

\begin{figure}[t]
\begin{center}
\includegraphics[width=5cm,clip]{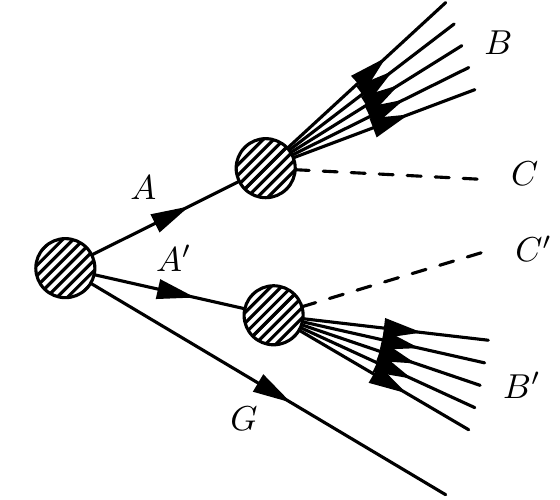}
\caption{Generic \MTTWO\ configuration.\label{fig:mt2generallabelledmt2configuration}
The sets labelled $B$ and $B^\prime$ may correspond to 
individual particles, or groups of visibles. In the latter case
the visible could result from internal cascade decays within the `blobs'.
$G$ labels `upstream' particles, as defined in the text.
}
\end{center}
\end{figure}

In this section we describe the simplest non-trivial example of a pair of decay chains -- 
that being an identical pair of single-step decays, with each decay producing one visible and one invisible daughter (\autoref{fig:mt2generallabelledmt2configuration}). We examine the kinematic 
constraints for that case, and then go on to examine more complicated topologies 
including multi-step cascade decays and non-identical chains.

\subsection{Identical semi-invisible pair decays: \MTTWO}
\label{sec:mttwo}

We already saw in \secref{sec:mtgetsdefined} that the {\bf trans}verse
mass could be applied in circumstances where there is a single mother
particle (frequently the particle whose mass we hope to bound)
decaying in one or more steps ultimately into a single invisible
particle (whose mass we may not know) and one or more visible
particles.

\par

The \MTTWO\ variable \cite{Lester:1999tx} (also known as the {\bf
strans}verse mass)
\footnote{The nick-name ``stransverse mass'' 
arose as a shortened form of ``supersymmetric transverse mass''
as \MTTWO\ was originally applied most frequently to supersymmetric
events in cases where the transverse mass was no longer usable.}

is the analogue of the transverse mass which can be applied in the
situations where there are not one but {\em two} parent particles,
each undergoing decays to a single invisible particle (whose mass we
may not know) and one or more visible particles.  The most general
topology of this type may be seen in \autoref{fig:mt2generallabelledmt2configuration} while specific
examples may be seen in \autoref{fig:mt2sortsoftopologies}.

\par

The usual definition of \MTTWO\ in this case written for the general
case shown in \autoref{fig:mt2generallabelledmt2configuration} casts
the variable as a function of six things. The first four are
straight-forward, being the invariant masses ($m_B$ and $m_{B'}$) and
the transverse
momenta (${\bf b}_T$ and ${\bf b}'_T$) of the visible
final state particles, or collections thereof, on each side of
the event. 

The fifth input is the observed missing transverse momentum
in the event, often denoted ${\slashed {\bf p}}_T$.
If $G$ in \autoref{fig:mt2generallabelledmt2configuration} is taken to represent
the totality of all other visible momenta in the event regardless of source, then
${\slashed {\bf p}}_T$ is equivalent to $-({\bf g}_T+{\bf b}_T+{\bf
  b}'_T)$.  Whether or not ${\slashed {\bf p}}_T$ is ``useful'' is
dependent on how closely it resembles ${\slashed{\bf
    c}}_T+{\slashed{\bf c}'}_T$, which depends on how many other
invisible particles there are in the event and on the detector
reconstruction resolution for ${\bf g}_T$, ${\bf b}_T$ and ${\bf
  b}'_T$. 

The sixth and final input is a pair of hypothesised masses
for the invisible particles ($\tilde m_{\slashed C}$ and $\tilde
m_{\slashed{C}'}$).  To distinguish the real from hypothesised masses,
the latter have been given a tilde.  
In principle these two hypothesised masses could be taken to be
different from each other
(see \secref{sec:mttwoasymmetric})
however in practice most studies take
them to be identical.  When both hypothesised masses are taken to be
identical that common value is often denoted by $\chi$.
In these terms, the usual definition of \MTTWO\ is as follows:\footnote{
Computer libraries that can evaluate \MTTWO\ may be found in
\cite{oxbridgeStransverseMassLibrary} and in
\cite{zenuhanStransverseMassLibrary} The library of
\cite{zenuhanStransverseMassLibrary} can only compute \MTTWO\ using the
bisection algorithm of \cite{Cheng:2008hk}, but it is very simple to
use and is not dependent on external packages.  
It is also distributed as part of the {\tt WIMPMASS} library \cite{ucdWimpmassLibrary}.
The library of \cite{oxbridgeStransverseMassLibrary} contains algorithms for a larger
number of variables (including \MTGEN, \MTWOC, etc, as well as a copy
of the algorithm in \cite{zenuhanStransverseMassLibrary,Cheng:2008hk})
but depends on the external {\tt Minuit2} library \cite{minuit2Library,James:1975dr}.}
\begin{equation}\label{eq:mttwodef}
\mttwo(m_B, m_{B'}, {\bf b}_T, {\bf b}'_T, {\slashed {\bf p}}_T ; \chi) \equiv \\ 
     \min_{ \slashed{\bf c}_T + \slashed{\bf c}'_T = \ptmiss }
     \left\{
     \max { \left( \MT, \MT' \right) }
       \right\} .\
\end{equation} 
where $\MT$ is the transverse mass constructed from $m_B$,
$\tilde{m}_{\slashed C}(=\chi)$, ${\bf b}_T$ and $\slashed{\bf c}_T$, while
$\MT'$ is the transverse mass constructed from $m_{B'}$, $\tilde{m}_{\slashed
{C}'}(=\chi$), ${\bf b}'_T$ and $\slashed{\bf c}'_T$, and where the
minimisation is over all hypothesised transverse momenta $\slashed{\bf
c}_T$ and $\slashed{\bf c}'_T$ for the invisible particles which sum
to the observed missing transverse momentum.  In
\autoref{eq:mttwodef} the dependence on $\chi$ (or equivalently
on $\tilde{m}_{\slashed C}$ and $\tilde{m}_{\slashed {C}'}$ in the case that they
differ) has been separated from the dependence on the other inputs by
a semi-colon to emphasise that the quantities to the left of the
semi-colon are observables, while $\chi$ to the right is instead a
parameter.  \MTTWO\ might thus be better described  not as an
observable in the usual sense, but rather as an ``observable
function'' -- in this case a function of $\chi$.

\par

\par

There are many parallels between the stransverse and the transverse
mass.  Most importantly (as was first mentioned in \autoref{sec:mtasboundary})
the transverse mass can be
viewed in two different but equivalent ways:
{\em either} as an event-by-event lower bound on the mass of the
parent particle (in terms of a mass hypotheses for the invisible
particle), {\em or} as a curve delineating the boundary between the
regions of the two-dimensional space of the unknown parent and daughter masses
which are -- or are not -- consistent with a particular event.
The same two interpretations are valid for the stransverse mass:

\par

In the first interpretation, most frequently used in the case that particles $A$ and
$A'$ (though not necessarily $\slashed C$ and $\slashed {C'}$) have
the same mass,
the stransverse mass can be viewed as
providing an event-by-event lower bound for $m_A$ in terms of a
hypothesis (i.e.~$\chi$) for the mass of the invisible
particles.  It may be shown \cite{Lester:1999tx,Barr:2003rg} that it
is possible to saturate this bound with certain kinematic
configurations.  A typical usage pattern therefore would be to plot a
histogram of $\mttwo(\chi)$, over all events, with the intention of identifying a
clear kinematic end-point in that distribution located at $m_A$ -- at
least for the case where $\chi$ is chosen to be equal to the true
value $m_{\slashed  C}$.   This technique has been used by the CDF
collaboration to measure the top quark mass in the dilepton channel \cite{cdftop} and has been suggested for the same use at the LHC \cite{Cho:2008cu}. The freedom to re-evaluate \MTTWO\ at
different values of $\chi$ corresponds to the need to obtain different
bounds on $m_A$ under differing assumptions about the mass of the
invisible particles that $A$ and $A'$ decayed into.

\begin{figure}[t]\begin{center} \includegraphics[width=4.5cm]{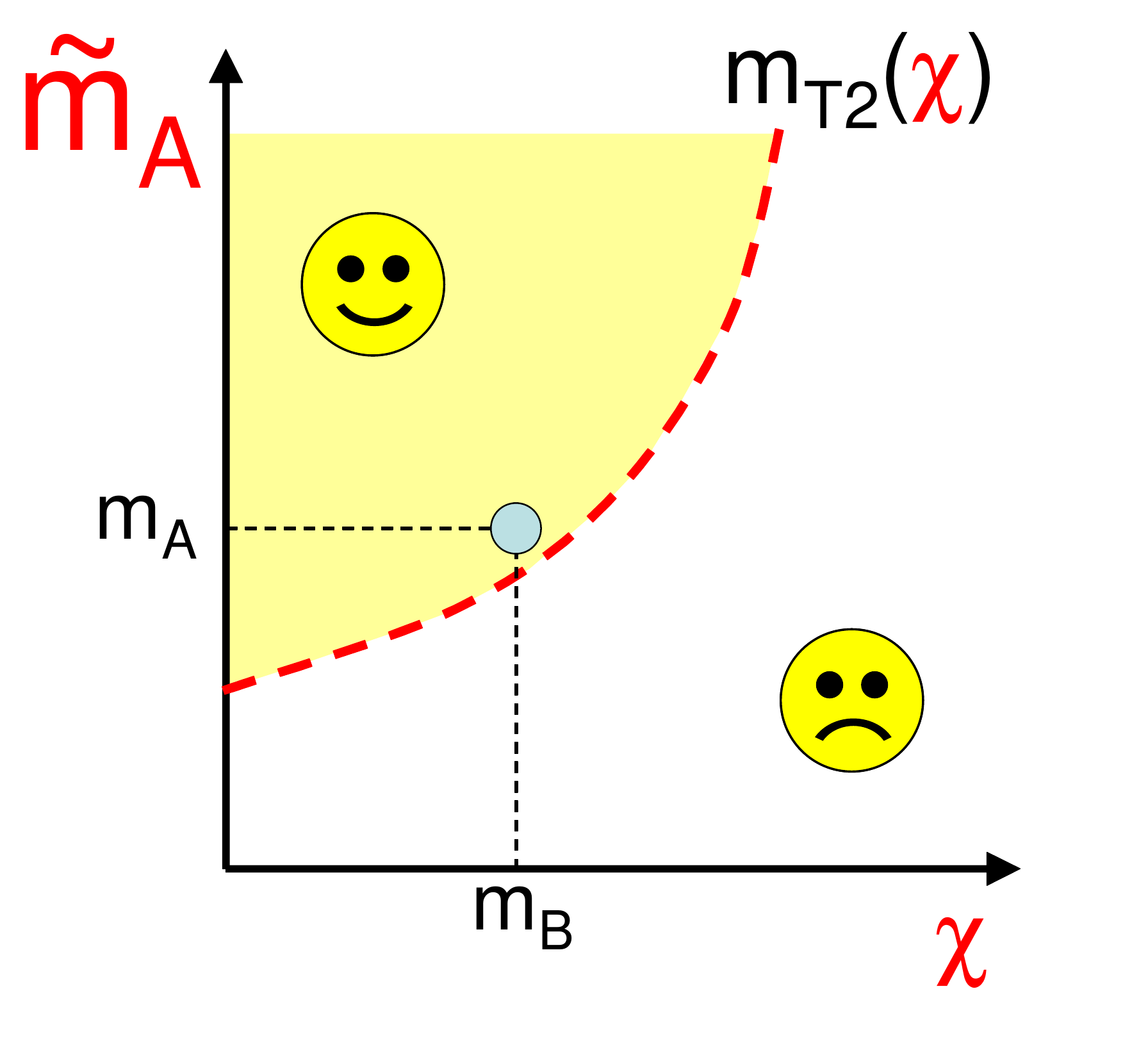}
\caption{\label{fig:mt2asboundary}
The nature of the \MTTWO\ constraint from a single event.
The region above the dashed line (marked {\LARGE \smiley}) is consistent with the constraints, 
while that below and to the right of the line (marked {\LARGE \frownie}) is inconsistent.
Similar regions can be drawn for a single decay chain 
where the regions are bounded by \MT.
}\end{center}\end{figure}\par

The second interpretation of \MTTWO\ is that it (or more specifically
the functional form of the curve $\mttwo(\chi)$) describes, for each
event, the boundary between the region of (parent, daughter)-mass
space that is consistent with that event and the region that is
inconsistent with that event in the manner indicated in
\autoref{fig:mt2asboundary}. The first explicit proof of this property
was recorded in \cite{Cheng:2008hk} and similar ideas have been expressed elsewhere \cite{Baumgart:2006pa}. Viewing \MTTWO\ as a ``boundary
of a consistent region of mass space'' is a powerful idea, not only 
because it provides a different way of understanding \MTTWO, but also
because  it allows us to see that the transverse mass and even the
ordinary invariant mass could similarly have been defined as such
boundaries. Indeed, once it has been seen that the transverse
mass $\MT$ could have been defined as the boundary of an allowed
region, the proof that the stransverse mass is such a boundary follows
immediately from its usual min max definition~\eqref{eq:mttwodef}.
The interpretation of \MTTWO\ as a boundary also shows us that generalisations of \MTTWO\ 
-- for example to situations with dissimilar parent masses --  would ideally be 
constructed so that they give the boundary of the consistent region of an
``extended'' mass-space with a higher dimensionality.

\par

\subsection{Dependence of \MTTWO\ on the WIMP mass(es)}\label{sec:kinks}

\begin{figure}
\begin{center}
\includegraphics[width=10cm]{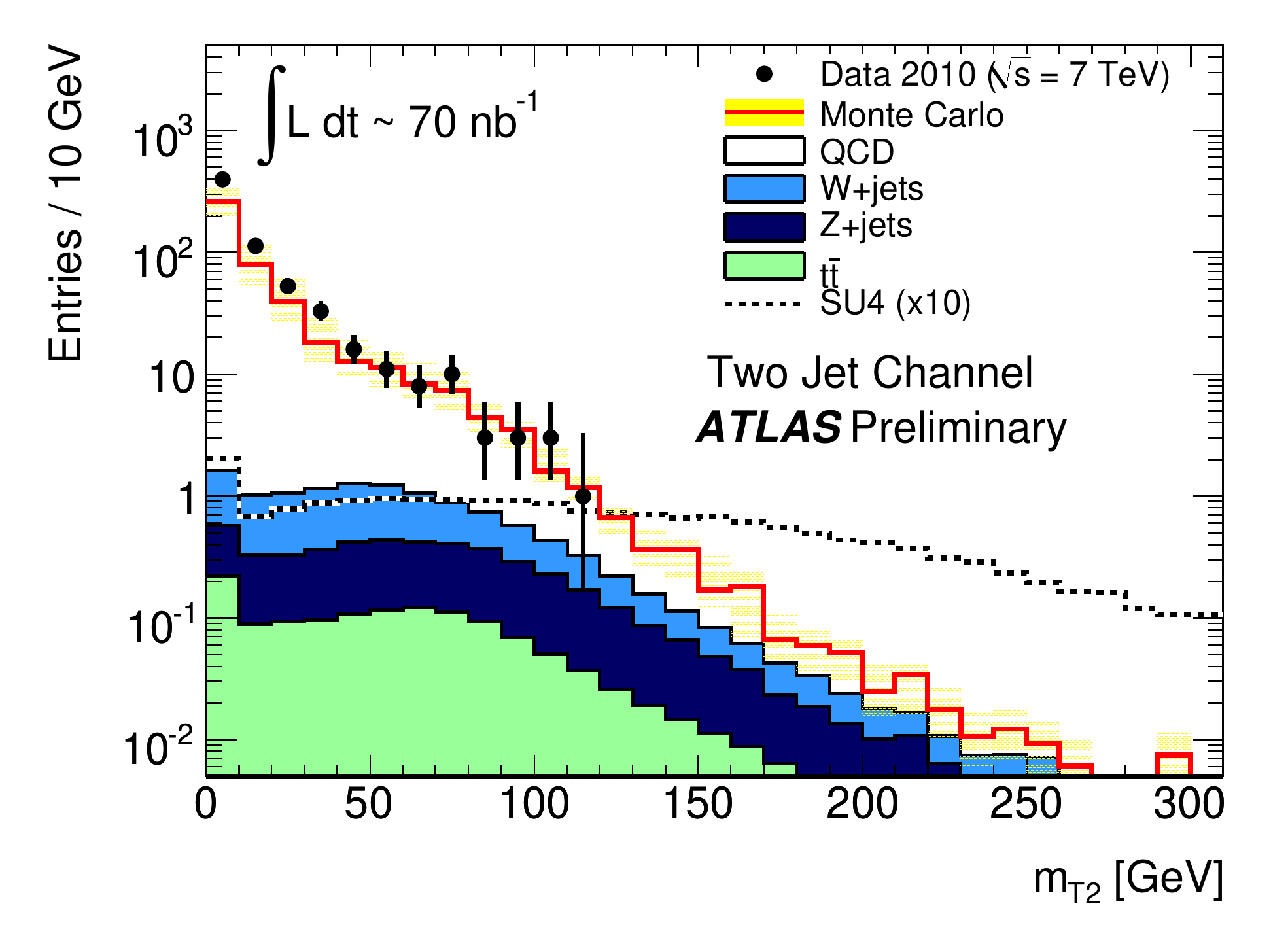}
\caption{This figure, reproduced from \cite{Collaboration:1273174}, shows the preliminary leading dijet \MTTWO\ distribution for the first 70 inverse nanobarns of ATLAS data. \label{fig:mt2atlasichep}  The dotted line shows the shape of a potential SUSY signal in a model with a large amount of strongly interacting sparticle production.  Note that the QCD dijet background constitutes the majority of the events passing cuts, but that as it lacks a mass scale the majority of those events fall at very low \MTTWO\ values.  This contrasts with the behavour of top-pair and potential SUSY events which have high mass scales and occur at high \MTTWO\ values.}
\end{center}
\end{figure}

Different approaches can be made to the problem of the dependence of \MTTWO\ on the {\em a priori} unknown parameter $\chi$, the hypothesis for the mass of the invisible particles. If one is using \MTTWO\ as a bound on the mass of the parent particle, one possibility would be to take the most conservative value -- i.e.~to set $\chi=0$. Since 
\[
\MTTWO(\chi=0) < \MTTWO(\chi>0) \leq m_A
\]
using a trial value $\chi=0$ will return a value which is certainly less than $m_A$, the mass of the parent. This conservative approach has been shown to be useful when using \MTTWO\ as a tool to distinguish events which are not consistent with particular Standard Model decay topologies \cite{Lytken:2003rw,Barr:2005dz,Barr:2009wu}, because the invisible particles of the Standard Model -- the neutrinos -- do indeed have very small masses and so satisfy $\chi\approx0$.
The problem with assuming $\chi=0$ is that for $m_\slashed{C}\ne0$ the bound is not saturated; while $\MTTWO(\chi=0) < m_A$ for all events, there are no events for which   $\MTTWO(\chi=0)$ approaches $m_A$, so one cannot use the end-point of the  $\MTTWO(\chi=0)$ as a measurement of $m_A$.  The first example of \MTTWO\ being used in LHC data to separate expected standard model backgrounds from potential signals from supersymmetry is shown in Figure~\ref{fig:mt2atlasichep}.

To measure masses we want to use the property that if the {\em correct} hypothesis is made for the mass of the invisible particle, then $\MTTWO$ returns a value $\leq m_A$, {\em with equality } for some state configurations.
The dependence of the \MTTWO\ distribution on the unknown mass of the invisible daughter particle $\chi$ is therefore important. For a distribution of interest to depend upon an unknown parameter might be seen as a disadvantage. But it is possible to turn this argument on its head; the fact that the distribution of $\MTTWO(\chi)$ depends on $\chi$ might allow us to simultaneously extract {\em both} the mass of the parent and the mass of the invisible daughter.

\begin{figure}\begin{center}\includegraphics[width=8cm]{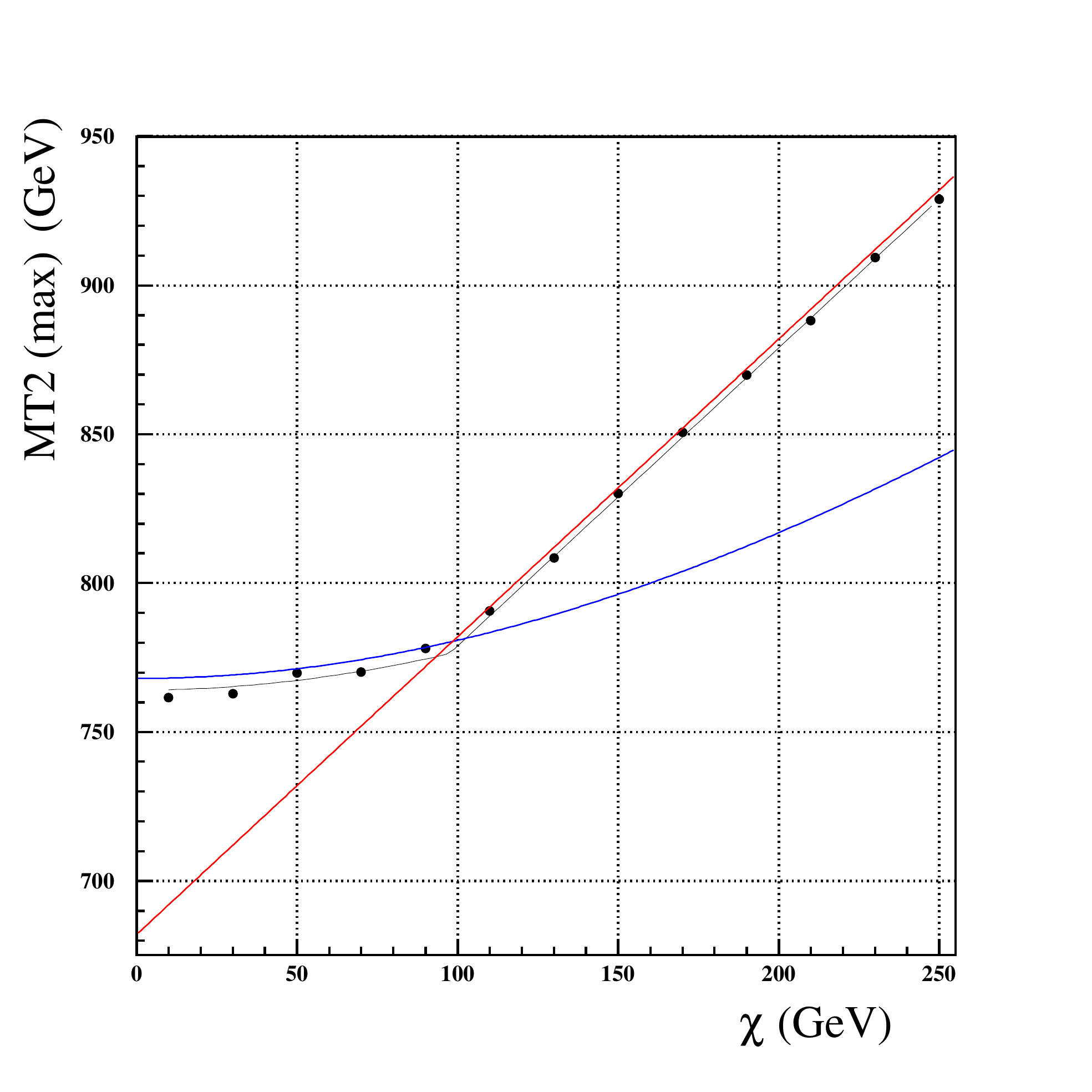}
\caption{\label{fig:kinkplot}
The points show how a measured upper bound of the \MTTWO\ distribution for pairs of three-body decays 
($\tilde{g}\rightarrow q \bar{q} \tilde{\chi}_1^0$) 
depends on the {\it a prior} unknown mass of the invisible particle.
The straighter red (more curved blue) line shows the configuration which is maximal 
for $\chi$ greater than (less than) $m_\slashed{C}$.
The area above and to the left of {\em both} curves gives the 
domain of allowed values of $(m_{\tilde\chi_1^0},m_{\tilde{g}})$.
Notice the change in gradient in the envelope curve near (97,780);
the coordinate of the position of this kink corresponds to the masses of the neutralino and the gluino 
used in the simulation. Adapted from \cite{Cho:2007qv}.
}\end{center}\end{figure}

To see how the dependence of $\MTTWO(\chi)$ on $\chi$ can be made to help us, consider the envelope of the maximum of the curves $\MTTWO(\chi)$ over all events.
The individual bounding curves from different events will generally have different shapes, but all must share the property (by construction) that
\begin{equation}\label{eq:mttwoprop}
\MTTWO(\chi=m_\slashed{C}) \leq m_A ,
\end{equation}
with the equality satisfied for some set of kinematic configurations.
In general different events will be maximal for $\chi < m_\slashed{C}$ and for $\chi > m_\slashed{C}$.
However the property \eqref{eq:mttwoprop} means that the bounding curves just 
above and below $\chi = m_\slashed{C}$ must both pass through the point $(m_\slashed{C},\,m_A)$.
If events with different slopes are maximal for $\chi < m_\slashed{C}$ and $\chi > m_\slashed{C}$,
then the overall envelope function $\max_{\rm events}\MTTWO(\chi)$ 
will be continuous but non-differentiable at the point $(m_\slashed{C},m_A)$: 
\[
\left[\frac{d}{d\chi}\max_{\mathrm{events}}\MTTWO(\chi)\right]_{\chi=m_\slashed{C}-}\ne
\left[\frac{d}{d\chi}\max_{\mathrm{events}}\MTTWO(\chi)\right]_{\chi=m_\slashed{C}+} .
\]
This feature was first spotted in simulations of pairs of three-body gluino decays $\tilde{g}\rightarrow q\bar{q}\ntlone$ \cite{Cho:2007qv} (see also \autoref{fig:kinkplot})
but has also been explored for simpler and more complex topologies \cite{Gripaios:2007is,Barr:2007hy,Cho:2007dh}. 

The existence of this `kink' in the $\MTTWO$ endpoint makes it tempting to infer that it will be straightforward to extract both the parent and the invisible-daughter masses. However for a substantial change in the gradient $\frac{d}{d\chi}\max_{\mathrm{events}}\MTTWO(\chi)$ at $(m_\slashed{C},m_A)$ there must be contributions from events with substantially different properties. 
Pairs of two-body decays in which the sum of the parents' transverse momenta is zero, and which 
have fixed $m_A$ will not produce kinks because the kinematics are so constrained that the gradients at $\chi=m_\slashed{C}\pm$ have to be equal ~\cite{Barr:2007hy}..

The event-by-event changes which lead to measurable `kinks' in the end-point function can come from the two different sources below~\cite{Cho:2007qv,Gripaios:2007is,Barr:2007hy}.
\begin{itemize}
\item
Different values of $m_B$, the invariant mass of the visible-particle subsystem,
will lead to different boundary curves in the space $(m_\slashed{C},\,m_A)$.
This mechanism is the dominant source of the kink seen in \figref{fig:kinkplot}, 
where the $q\bar{q}$ invariant mass changes significantly between events.
Other topologies in which the visible system is a composite 
constructed from the sum of two or more visible particles will share this behaviour.
\item
When one allows the two-parent center-of-mass to be boosted, the extremal boundary curves 
correspond to configurations with arbitrarily large parent momenta. 
The bounding curves for $\chi < m_\slashed{C}$ ($\chi > m_\slashed{C}$)
come from events in which the invisible particle is emitted 
parallel to (anti-parallel to) the boost direction. 
Systems with finite boosts have correspondingly less-pronounced kinks.

The two-parent center-of-mass frame can be expected to have a small transverse boost 
unless the parents were themselves created from previous decays, or there was large initial state radiation. The negative sum of the transverse momenta of the parents -- i.e.~the momentum against which those parents are recoiling -- is often known as the `upstream' momentum, \UTM\ and is represented in 
\autoref{fig:mt2generallabelledmt2configuration} by the label $G$. 
\end{itemize}

Since only a small number of signal events are expected to contribute near the kinematic endpoint, it might be considered difficult to extract information from the $\chi$ dependence of $\max_{\mathrm{events}}\MTTWO(\chi)$ without having a very good knowledge of the backgrounds.
Nevertheless attempts to measure the invisible particle mass from the kink (or variables characterising the location of that kink) have shown some promise.
The position of the kink of the first sort (coming from a variable-mass visible particle system) was successfully captured in simulations \cite{Cho:2007qv}. 

A method for exploiting the second (boost-generated) kink in the case where the invariant mass of the visible particle system is fixed has been proposed in \cite{Konar:2009wn} and further explored in \cite{Cohen:2010wv}. 
The central observation of that method is that one can construct one-dimensional analogues of \MTTWO\ using only the components of the visible momenta parallel to (perpendicular to) the upstream momentum direction. 
Because the perpendicular analogue $M_{T2\perp}$ has a distribution with an endpoint (and indeed a shape) which is independent of \UTM\ it can be used as a `control' sample against which the  $M_{T2\parallel}$ distribution (constructed from components parallel to \UTM) can be compared. 
For a pair of two-body decays the experimental problems in understanding the behaviour of the parallel and the perpendicular endpoints are likely to be considerable. For that case, the fraction of near-extremal events will be small, the backgrounds important, and the systematic uncertainties in fitting the shapes are likely to be significant. It is not yet clear whether the method could be practical in such cases.

We note that while these `kinks' have been most frequently studied for topologies containing a pair of decay chains, the same effects also generate kinks in single decays or decay chains \cite{Cho:2007qv,Gripaios:2007is,Barr:2007hy}, or in asymmetric decay chains (for which see \secref{sec:mttwoasymmetric}).

The mathematical structure of these `kinks' is discussed further in \secref{sec:robotics}.

\subsection{Decomposing \MTTWO\ with respect to upstream momenta: ($M_{T2\perp}$ and $M_{T2\parallel}$)}\label{sec:mttwodecomposed}
It can be useful to decompose \MTTWO\ into ``components''\footnote{We use the term ``components'' figuratively since \MTTWO\ is not a vector.  Strictly it is the transverse momenta which are {\em inputs} to \MTTWO\ which are resolved into components.} which are perpendicular or parallel to the upstream transverse momentum.  These components are called, respectively, $M_{T2\perp}$ and $M_{T2\parallel}$ \cite{Konar:2009wn}.  One advantage of this decomposition is that the component {\em perpendicular} to the upstream transverse momentum, $M_{T2\perp}$, has no dependence on the magnitude of the recoil supplied {\em by} the upstream transverse momentum for any value of the trial mass $\chi$.\footnote{Recall that this was not true for \MTTWO, and in fact the dependence of \MTTWO\ at unphysical values of $\chi$ was a necessary ingredient for forming one of the two types of \MTTWO\ ``kink'' (see Section \ref{sec:kinks}).}   A second advantage of this decomposition is that under a reasonable set of circumstances, the shape of the differential distribution of $M_{T2\perp}$ becomes fully determined (i.e.~it does not depend on unknown parameters such as the unknown centre of mass energy) \cite{Konar:2009wn}.  This existence of this universal shape might therefore make it possible to fit the distribution more accurately and make it possible to extract masses by a secondary step \cite{Konar:2009wn}. The kinematic endpoints of the $M_{T2\perp}$ and $M_{T2\parallel}$ distributions may be found in the appendix in equations (\ref{eq:mt2perpmax}) and (\ref{eq:mt2paramax}).  Compare the related properties of  $M_{CT\perp}$ and $M_{CT\parallel}$ discussed in Section~\ref{sec:mctdecomposed}.

\subsection{Identical chains of decays}\label{sec:countsolns}

If one is willing to assume that the visible particles originate from two identical two-step decays of identical sparticles, i.e.
\par
\begin{center}\includegraphics[scale=0.6,clip]{Mt2TwoSequential.pdf}\end{center}
\par\noindent
then several \MTTWO\ variables can be calculated for each event \cite{Serna:2008zk,Burns:2008va} (first suggested in the context of the variable $\MCT$ in \cite{Tovey:2008ui}). Using the endpoints of three versions of \MTTWO, changing what is interpreted as visible transverse momentum, missing transverse momentum and upstream momentum one can, in combination with the dilepton endpoint, identify the correct masses (assuming perfect resolution and no combinatoric ambiguity). In \cite{Burns:2008va} a similar approach to forming \MTTWO\ subsystems is proposed, and the origin of kinks in the maxima of the various distributions is explored. Several methods of extracting particle masses using multiple \MTTWO\ distributions are introduced, including a hybrid method that uses the dilepton endpoint. The kink analysis is discussed in more detail in this review in section~\autoref{sec:kinks} (\pref{sec:kinks}), and the hybrid method in section~\autoref{sec:hybridpair} (\pref{sec:hybridpair}).

An alternative to finding limits is to make hypotheses about the
particle masses, and then for each mass hypothesis to count the number
of events for which there are real positive solutions for the energies
of the unseen particles (`consistent events')
\cite{Cheng:2007xv,Cohen:2010wv}. A region is formed in
three-dimensional mass space which is consistent with all events, and
this tends to a minimum volume as the number of events approaches
infinity. The correct masses correspond to a point at the region's
tip.  By looking for the point at which the number of consistent
events is a maximum, one can obtain an estimate for all three masses.
A computer library called {\tt WIMPMASS} \cite{ucdWimpmassLibrary} that
facilitates implementation of the techniques of \cite{Cheng:2007xv}
may be downloaded.

Events containing slightly longer pairs of identical chains
are amenable to other treatments.  For example, under the hypothesis that the invisible particles are massless (relevant in many GMSB models with gravitinos in the final state) events containing a pair of identical decay chains of the form \par
\begin{center}\includegraphics[scale=0.6,clip]{Mt2ThreeSequential.pdf}\end{center}
\par\noindent
are fully reconstructible even though there are two unseen particles in the final state.  Such a reconstruction is demonstrated in the appendix of \cite{Hinchliffe:1998ys} using a GMSB-like scenario as an example with two copies of the chain $\ntltwo \to \slepton^\mp \ell^\pm \to \ntlone \ell^\mp\ell^\pm \to \tilde G \gamma \ell^\mp\ell^\pm$i.

If one drops the assumption that the unobserved final state particles are massless, then there are too few constraints from a single event to reconstruct the event.  However we will return to this double-chain later in \secref{sec:MULTIPLE} where we will see that there are a number of techniques that would permit the masses to be recovered if one is prepared to consider {\em more than one event simultaneously}.

\subsection{Pair decays with small upstream momentum: $\MCT$}\label{sec:mtovey}

Identical pairs of semi-invisible decays in which the parents had zero upstream momentum 
(i.e.~\autoref{fig:mt2generallabelledmt2configuration} with recoil momentum $g$=0)
have interesting properties if the visible daughters are used as inputs to 
the \MCT\ variable \cite{Tovey:2008ui}.
The definition of \MCT\ used for pair decays of the above type uses only the momenta of the two visible decay products (or systems of products) and is as follows:
\begin{equation}
\MCT^2 = m_B^2 + m_{B^\prime}^2 + 2 \left( e_b e_{b^\prime} + {{\bf b}_T}\cdot{{\bf b}^\prime _T} \right)\label{eq:mctdefinedforpairs}
\end{equation}
where $e_b$ is once again defined as in Equation \eqref{eq:et_noz}. 
Take note of the subtle difference between the definition of \MCT\ in equation \eqref{eq:mctdefinedforpairs} for pair decays, and the definition of \MCT\ for in equation \eqref{eq:mctdef}.  The definition in \eqref{eq:mctdefinedforpairs}\footnote{Historically this was how \MCT\ was first defined in \cite{Tovey:2008ui}.} uses only half of the final state momenta (namely those of the two systems $B$ and $B^\prime$ in Figure~\ref{fig:mt2generallabelledmt2configuration} which are visible!) while the definition in equation \eqref{eq:mctdef}\footnote{So far as we are aware, active use of \MCT\ for single particle systems does not seem to have been encouraged prior to \cite{Cho:2009ve}.} uses {\em all} the final state momenta of a single two-body decay.

\MCT\ enjoys the property that it is invariant under equal and opposite boosts in the transverse plane
of the primed and un-primed systems.
This insensitivity is a welcome feature as equal-mass parent particles produced in hadron collisions will have 
(in the absence of initial-state radiation) back-to-back transverse boosts, and the magnitude of those boosts will be unknown and unmeasurable if there are invisible daughters.

It has been shown \cite{Serna:2008zk} that $\MCT$ is equal to \MTTWO\
in the special case where $\chi=0$, the visible particles are massless, and the
upstream transverse momentum is zero.
Since it has also been shown \cite{Cheng:2008hk} that \MTTWO\ delineates the boundary 
between allowed and disallowed regions in mass space, 
we can see that $\MCT$ has the same bounding property in mass space under these conditions.

Although $\MCT$ does not quite describe the boundary of the allowed region of 
mass space when the event contains non-zero upstream momentum it is
nevertheless bounded above by a value $\MCT^{\max}$ which is calculable for any boost, so a 
``boost-corrected'' \MCT\ can be used to recover a good determination of the masses in sequential decays \cite{Polesello:2009rn}.
The combination of masses determined by this maximum value of this contransverse mass 
is (in the limit where $m_B=m_{B^\prime}=0$, and \UTM=0)
\[
\MCT^{\max} = \frac{m_A^2-m_\slashed{C}^2}{m_A} = 2p^*.
\]
To within a factor of two, this endpoint is therefore telling us the momentum of the daughters in the rest frame of the parent. This simple dependence of $\MCT^{\max}$ on the unknown parameter $m_\slashed{C}$ may make distributions of $\MCT$ very convenient for later interpretation, since the endpoint can be measured for one hypothesised value of ${m}_\slashed{C}$ (for example 0), and later reinterpreted for other trial invisible-particle masses.
This can be compared to the behaviour of \MTTWO\ which (though it exactly describes the bound in mass space for any trial mass $\tilde{m}_\slashed{C}$ and under arbitrary boosts of the parents) has a non-trivial (and boost-modified) dependence on the invisible particle trial mass, $\chi$.
Because of the ease of reinterpreting $\MCT$ for different $\tilde{m}_\slashed{C}$, one could suggest that even in the case of arbitrarily boosted parents (for which the variable does not quite describe the boundary of a domain in mass space) the $\MCT$ distribution could still be the most suitable choice for presenting endpoints kinematics relevant to mass measurements \cite{Polesello:2009rn}.

Two one-dimensional decompositions of \MCT\ were proposed in \cite{Matchev:2009ad} and named $M_{CT\perp}$ and $M_{CT\parallel}$. \label{sec:mctdecomposed}
These were constructed from the components of the visible momenta in the directions respectively perpendicular to and parallel to the upstream transverse momentum two-vector. Since the projection of a vector onto a plane is not changed by any boost which is perpendicular to that plane, the distributions of $M_{CT\perp}$ is unmodified by 
the magnitude of the boost which the visible systems acquire from recoil against the upstream transverse momentum. Indeed the $M_{CT\perp}$ distribution (or at least that part of it which has $M_{CT\perp}>0$) has a universal shape in the absence of spin correlations \cite{Matchev:2009ad}. This distribution does not depend on, for example, $\hat{s}$ or the longitudinal boost of the parents, which it is claimed should make it much easier to fit in order to extract the kinematic limit $M_{CT\perp}^{\max}$. The maximum of the $M_{CT\parallel}$ distribution does depend on the boost, but does so in a simple calculable way \cite{Matchev:2009ad}. A summary of some of these results may be found in the Appendix in equations~(\ref{eq:repromctparamaxresult}) and (\ref{eq:repromctperpmaxresult}).  Compare also the related properties of  $M_{T2\perp}$ and $M_{T2\parallel}$ discussed in Section~\ref{sec:mttwodecomposed}.

In \cite{Cho:2009ve} a variant has proposed which considers separately the $\MCT$ for each side of a pair-decay event, and then uses a \MTTWO-like construction but now using $\MCT$ rather than $\MT$ on each branch
\[
M_{CT2} =
     \min_{ \slashed{\bf c}_T + \slashed{\bf c}^\prime_T = \ptmiss }
     \left\{
     \max { \left( \MCT, \MCT' \right) }
       \right\} .\
\]
With a judicious choice of $\tilde{m}_C$, the resultant variable has a Jacobian which increases the density of events near the kinematic endpoint $M^{\max}_{CT2}$. The shape of the distributions for typical Standard Model backgrounds were not investigated in \cite{Cho:2009ve}, but provided that the backgrounds near the peaked endpoint structures can be reduced~\cite{Barr:2010ii} then this variable could increase the observability of kinematic end-points and the precision with which they might be determined.

\subsection{Multiple invisible daughters per chain}\label{sec:multipleinvisibles}

The generalisation of \MTTWO\ to cases with more than one invisible particle in each decay chain have been considered in \cite{Barr:2002ex,Chang:2009dh,Barr:2009mx}. The transverse mass (and hence the stransverse mass \MTTWO) remains bounded below by a minimum value $m_<$ and above by $m_>$ ($=m_A$ for $\chi=m_\slashed{C}$). In the case of a $n$-body decay to a set of visible particles and a set of invisible particles, $A\rightarrow B+C+ \ldots + \slashed{X}+\slashed{Y}+ \dots$ the minimum value of is simply the sum of the masses of the daughters $\sum m_B + m_C + \ldots + m_\slashed{X} + m_\slashed{Y} + \ldots$. With larger numbers of invisible particles produced, the fraction of states near $\MT=m_A$ (or $\MTTWO=m_A$ for the two-chain case) is reduced. In such cases the end-point might only be inferred from a measurement of the shape of the distribution. Sequential decays producing invisibles at each step further restrict the range of $\MT$ (and hence \MTTWO). 

In \cite{Chang:2009dh} it is recognised that while chains containing multiple light invisible particles 
will generally have different kinematic properties to those containing smaller numbers of heavier invisible particles, nevertheless there are cases (such as the decay of a neutralino to multiple neutrinos $\tilde{\chi}\rightarrow \nu\ldots\nu$) for which the presence of multiple invisibles would be very difficult to infer.

\subsection{Non-identical decay chains}\label{sec:mttwoasymmetric}

If the two decay chains do not contain identical mass particles along their 
length then the results above need some modification.
In some cases it may possible to find parts of the chains (particular decays) which are identical
and to apply the identical-chain methods of the previous sections
to those subsystems \cite{Nojiri:2008vq}.

The generalisation of the above methods to two-chain processes with non-identical masses
has been considered in \cite{Barr:2009jv,Konar:2009qr}. 
For example in \cite{Barr:2009jv} it was shown that in a pair decay with 
different-mass parents one can hypothesise a value for the ratio of the parents masses $m_A/m_{A'}$, 
and produce a distribution which is sensitive to their product $m_A m_{A'}$.
In principle one can also determine the correct value of the input ratio $m_A/m_{A'}$ from a kink structure 
in the endpoint of this distribution in a manner
reminiscent of, but different to that discussed in \secref{sec:kinks} (when changes due to a 
different input parameter -- the hypothesised mass $\chi$ of the invisible daughters -- were considered). 

A similar method can be employed to measure the mass of non-identical mass invisible 
particles: by using either the inverse of the transverse mass $(\MT^{-1})$ \cite{Barr:2009jv}
or by constructing a variant of \MTTWO\ with two different invisible particle masses 
\cite{Barr:2009jv,Konar:2009qr}.
With either variant, ridges or creases in the domain of consistent masses are found.
These crease structures -- which are generalisations of the `kinks' discussed in \secref{sec:kinks} --  intersect at the special point in the parameter space
for which the assumed masses (or relationships between masses) were correctly hypothesised.
In \cite{Konar:2009qr} the latter method is studied in detail,
the \UTM\ dependence of the modified \MTTWO\ distribution determined,
and the configurations of the bounding events described.

\subsection{Inclusive pair-decay variables}\label{sec:mtgen}

For double sided event topologies 
in which one (or both) of the two equal mass parent particles generates a
large number of visible
particles in its decay (e.g.\ as suggested in
\autoref{fig:mt2generallabelledmt2configuration}) it is reasonable
to ask the question ``Could one, in a real detector, tell
which of the observed/reconstructed final state particles or calorimetric energy
deposits belong to $B$ (i.e.\ to one side) and which belong to $B'$
(i.e.~to the other side)?''    In some particular cases
(principally those in which $A$ and $A'$ are guaranteed to be produced
with large and opposite boosts) the answer to the above question might
be ``yes'', as the constituents of $B$ and $B'$ might be found in
opposite hemispheres.  This leads to so-called ``hemispheric
\MTTWO'' techniques \cite{Nojiri:2008hy},
previously investigated in 
\cite{Barnett:1993ea,Baer:1995nq,cmsphystdr}.
The remainder of the time the answer to the above question is likely
to be ``no'', in which case there are a few alternative approaches.
One such approach, which is specifically targeted at analysing these
pair decays, is to define the inclusive variable \MTGEN\
\cite{Lester:2007fq}.  After we have discussed \MTGEN, we will discuss
an alternative approach involving \ROOTSHATMIN.

It may be shown that \MTGEN, canonically defined \cite{Lester:2007fq}
as the smallest value of \MTTWO\ obtained after trying all possible
partitions of the visible momenta (excluding those visible momenta
deriving from $G$ in
Figure~\ref{fig:mt2generallabelledmt2configuration} in cases where it
is possible to determine which they are) between the two sides, again
has an interpretation as the boundary. The equivalence
between these two different definitions of \MTGEN\ was shown in
\cite{Barr:2009jv} using insights from \cite{Cheng:2008hk}.  In this
case \MTGEN\ is the boundary of the region of parent/invisible
mass-space which is consistent with the hypothesis that the visible
momenta were, in some unknown order, derived from $A$ or $A'$ in
association with two invisible particles.  The \MTGEN\ endpoint thus
provides a constraint which tells you the parent particle mass as a
function of a hypothesised daughter particle masses, in just the same
way that \MTTWO\ does.  In short, \MTGEN\ is the natural
generalisation of \MTTWO\ to events in which you believe there is a
pair of parent particles but whose momenta you cannot be sure of
uniquely assigning to their respective parents.

During the process of scanning all possible partitions of the visible
momenta into two groups (one for each side of the event) \MTGEN\ has
to combine observed particle {\em four}-momenta together into resultant
{\em transverse}-momenta for input to \MTTWO.
There are a variety of ways in which this combination can
be done, each of which has merits and demerits.  
\MTGEN\ is not
unique in needing to combine four-momenta to produce
transverse-momenta, but we will only discuss this issue in the context
of \MTGEN. 
Options for determining the 1+2 vector for the visible system (all of which have
the same space-like components ${\bf p}_T$ but differ in their
time-like component) include \cite{MatchevUnpublished}: (a) summing
the Lorentz four vectors $(E,{\bf p})$ then ignoring the final $p_z$;
(b) summing the Lorentz 1+2 vectors $(e,{\bf p}_T)$; (c) projecting
the total energy onto the transverse plane (using $E_T=|{\bf E}|
\sin\theta$) after summing the constituent four-vectors; (d) as for
(c) but projecting each constituent before summing.
Which of these combinations is used for combining momenta has
important additional consequences for the dependence of these
variables on initial state radiation and/or multiple parton
interactions as will be discussed shortly.

A second and alternative approach to analysing pair-decays inclusively
involves returning to \ROOTSHATMIN\ \cite{Konar:2008ei}, which the
reader will recall has already featured in this review in
Section~\ref{sec:shatminfirstmentionsec} where it was discussed as a global
event variable.  All that is needed to re-use \ROOTSHATMIN\ in the
context of the pair-decay hypothesis it is to revisit how it should be
interpreted. For example, \cite{Konar:2008ei} noted that when
sparticle-pair decays are generated {\em without initial state
radiation and without multiple parton interactions}, then there is a
strong correlation between (a) the position of the peak of the
\ROOTSHATMIN\ distribution and (b) the sum of the masses of the two
parent particles.  \label{sec:rootshatmininpairs} One might hope to use this correlation to measure
those parent particle masses.\footnote{\ROOTSHATMIN\ shares with
\MTTWO, \MTGEN\ and similar variables the property that it measures
parent masses for a given hypothesised mass of the invisible
daughters.  So when we say ``this correlation can be used to the
masses of the two parent particles'' we mean for suitable hypothesised
invisible particle masses.}  It is noted in \cite{Konar:2008ei} that
this correlation has a status closer to a ``conjecture'' or a ``rule
of thumb'' than a ``necessity'', since it results from a fortuitous
(and somewhat process dependent) cancellation between two effects with
quite different origin \cite{Konar:2008ei}. The first effect is
that the true value of \ROOTSHAT\ exceeds the value it would take for
threshold production by an unknown and almost certainly different
amount in each event (i.e.~$\ROOTSHAT > 2 m_{\mathrm{parent}}$) with
the mean offset carrying some production-process dependence.  The
second effect is that the bound on \ROOTSHAT\ provided by
\ROOTSHATMIN\ seems to be lower than \ROOTSHAT\ by approximately the
same amount as the excess in the first effect.  In consequence we have
a somewhat process dependent coincidence, with two quite different
effects approximately cancelling out, leading to the correlation
between the peak of the \ROOTSHATMIN\ distribution and 
$2 m_{\mathrm{parent}}$.\footnote{Note the difference here between
\ROOTSHATMIN\ and \MTGEN.  The {\em construction} of the \MTGEN\
variable makes direct use of the pairwise hypothesis, whereas the
construction of \ROOTSHATMIN\ does not.  However both can have the
pairwise hypothesis applied to their {\em interpretation}, and
consequently both can potentially say useful things about pairwise
events.  Whereas an \MTGEN\ measurement places a {\em direct and
calculable bound} on the parent masses in each event (subject to the
pairwise hypothesis), a \ROOTSHATMIN\ measurement of those parent
masses is reliant on a correlation based on a fortuitous cancellation
between quite different and process dependent effects).  This does not
mean that either variable is better or worse than they other. The two
variables are simply different.  Which you should use depends on which
assumptions you do or do not wish to make and at what level you wish
to make them.  For example, if you are prepared to make a strong
pairwise hypothesis about the pairwise nature of all the events in
your sample, then you should perhaps use \MTGEN\ since it can make
more use of that hypothesis.  On the other hand, it might be
advantageous to avoid making the pairwise assumption at construction,
so that you can interpret the same set of events in many ways.  In
which case \ROOTSHATMIN\ may be more useful.}

The above correlation was demonstrated in simulations {\em without
initial state radiation}.  It is noted in
\cite{Papaefstathiou:2009hp,Konar:2008ei} that the effect of inclusion
of initial state radiation (ISR) and multiple parton interactions
(MPI) can be to substantially increase the measured values of
\ROOTSHAT\ and to substantially modify the shape and position of the
peak of the \ROOTSHATMIN\ distribution by large factors (order 2).
This means that, in the presence of ISR, the correlation spotted in
\label{sec:grublingaboutisr} \cite{Konar:2008ei} requires either
accurate modelling of such effects, or alternatively requires care to
reduce the effect of ISR (by for example restricting attention to
momenta in the small $|\eta|$ region).

Sensitivity to initial state radiation and multiple parton interactions 
is by no means unique to \ROOTSHATMIN. 
For example, \MTGEN\ 
distributions also can gain undesirable high tails when ISR or other
momenta not deriving from the parents are ``mistakenly'' identified as
the products of the parents and used as inputs. 
Indeed inclusive variables
tend, by construction, to be sensitive to all objects in the final
state, and so will, to a greater extent than more selective variables,
tend to contain contributions from initial state QCD radiation as well
as decay products from heavy object decays. Inclusive variables which
combine four momenta using options (a) and (c) above are often
particularly sensitive to ISR, {\em etc}, since these momentum
combination schemes are more sensitive to the appearance of a small
number of momenta at high rapidity.  For all such variables, the
contributions from ISR and MPI will need to be well modelled before
the variables can be used in precision constraints.

\par

It has been shown that QCD radiation and multiple parton interactions
can play an important role in modifying inclusive distributions
\cite{Papaefstathiou:2009hp,Konar:2008ei,Plehn:2008ae}.  In
\cite{Papaefstathiou:2009hp} it is even argued such dependence should
be regarded as beneficial, since it allows us to test not only our
understanding of QCD, but also our understanding of the link between
the scale of the QCD radiation and the mass scale of the particles
that were involved in its generation.  In this sense, accurate
modelling of the sensitivity of inclusive variables to this radiation
might in itself lead to indirect mass determinations or constraints.

One final approach to controlling the effect of ISR is to attempt to
find ways of ``removing'' it from the analysis. Such techniques
presuppose that the decay chain of interest can be well
reconstructed from only a {\em subset} of the jets in the final state.
In such a case, this information can be used to discard some jets
(under the assumption that they had a high probability of coming from
initial-state radiation) and improve the mass reconstruction
\cite{Alwall:2009zu}.

\subsection{Hybrid variables}\label{sec:hybridpair}

The concept of {\em hybrid} (as opposed to {\em per-event} and {\em per-dataset} variables) was introduced in \secref{sec:hybridintro}.   What use might hybrid variables be for pair decays?
No single event containing a $qll$-chain of
\secref{sec:qll} can generate the constraint on the slepton and
neutralino masses seen in \autoref{eq:dileptonconstraint} --
however such a constraint can emerge from the ``dilepton edge'' by
consideration of the sample of events as a whole.  In a scenario where
it is foreseen that such a constraint could be obtained from a large
sample of events, one might profit from constructing a variable which
``re-interprets'' each event in the light of that constraint.  Ideally
this process injects valuable additional information into each event,
and the resulting hybrid variables are more powerful than their
non-hybrid relatives.  This supposes, however, that the information
injected is ``good'' or ``relevant'' to the event into which it was
injected, which need not be true.  Successful application of hybrid
variables is thus limited by the degree of homogeneity in the samples
of events to which they might be applied, the extent to which the
events in those samples satisfy the ``injection hypothesis'', and the
degree to which both these requirements could be verified.

We will now look at two hybrid variables that apply the dilepton edge constraint to pair decays, \MTWOC\ and \MTHREEC.

\label{sec:mtwoc}\label{sec:mthreec}

\begin{figure}[t]
\begin{center}
\subfigure[\MTWOC]{\includegraphics[width=3.5cm]{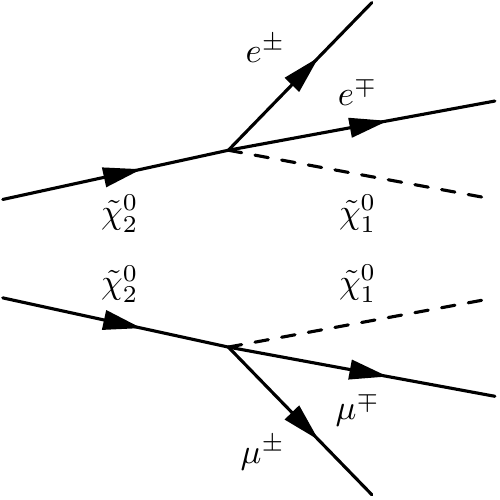}}
\hspace{1cm}
\subfigure[\MTHREEC]{\includegraphics[width=4.5cm]{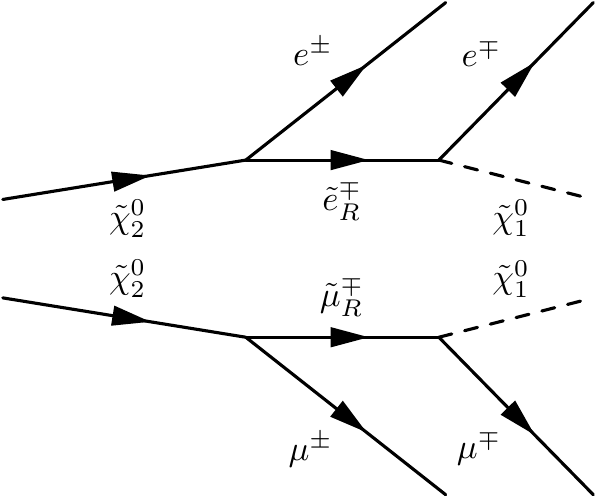}}
\caption{Event topologies required (a) by \MTWOC\ and (b) by \MTHREEC. \label{fig:picofm2candm3ctopologies}}
\end{center}
\end{figure}

\begin{figure}[t]
\begin{center}
\includegraphics[scale=0.25]{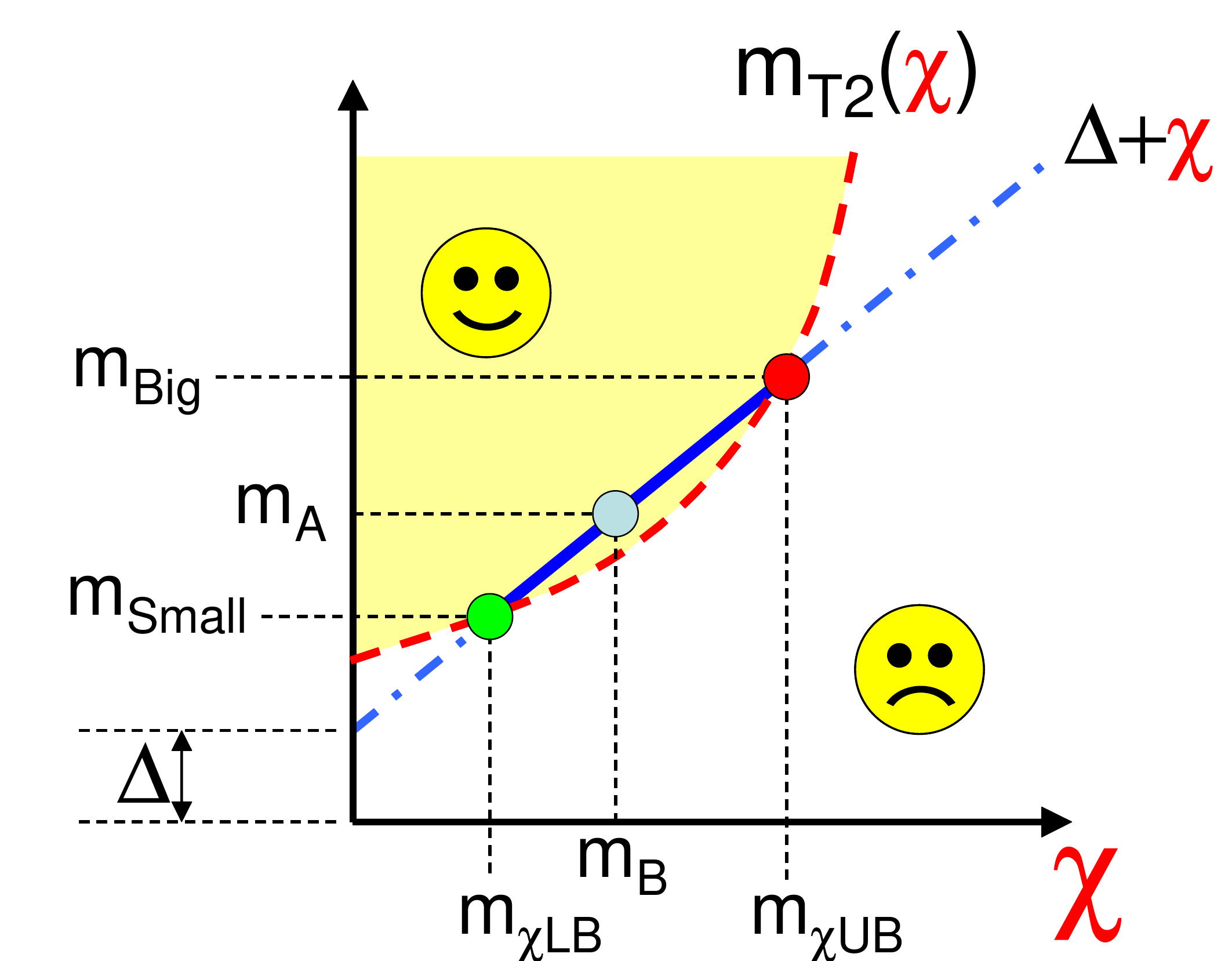}
\includegraphics[scale=0.25]{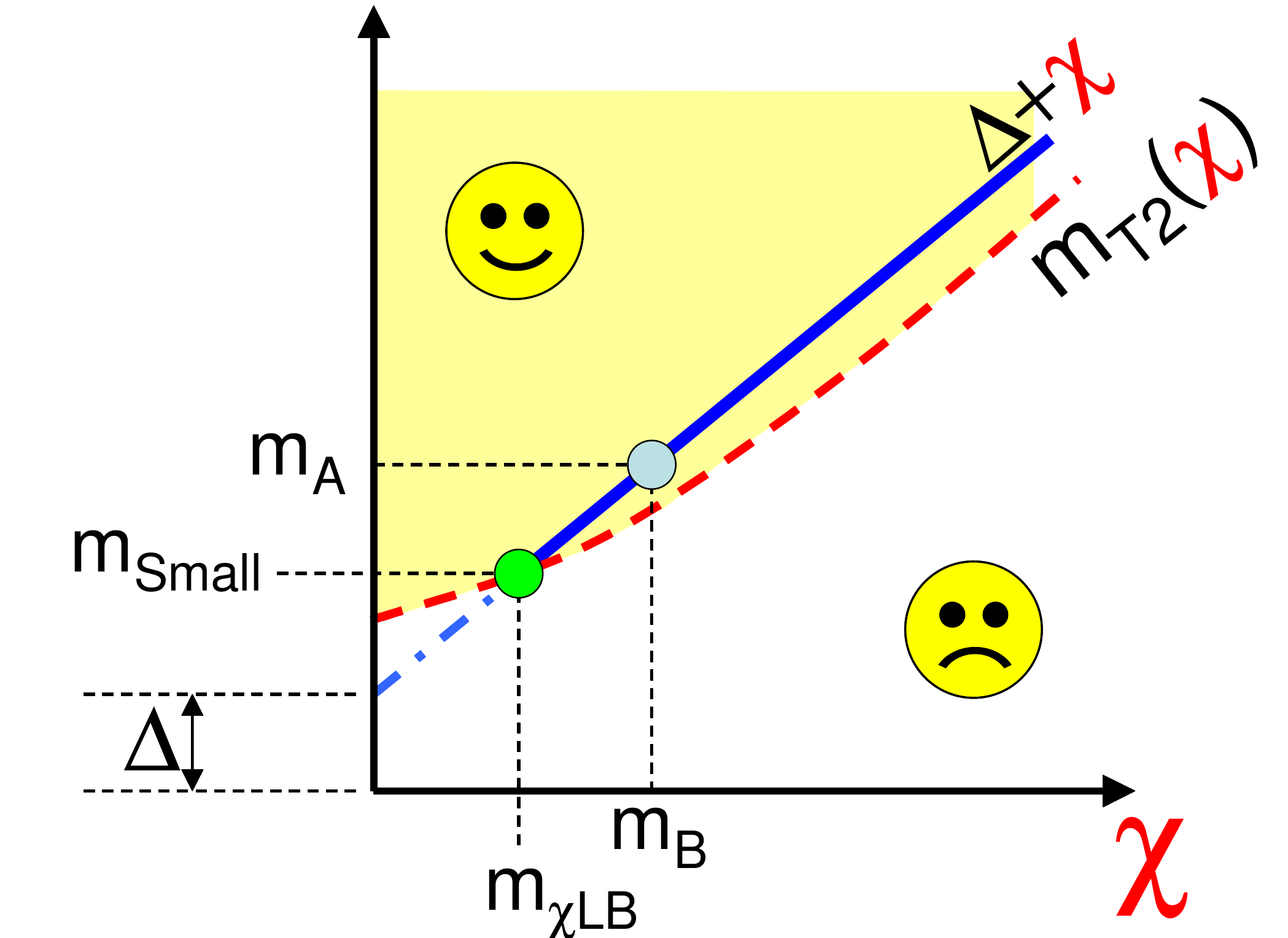}\\
\qquad\includegraphics[scale=0.25]{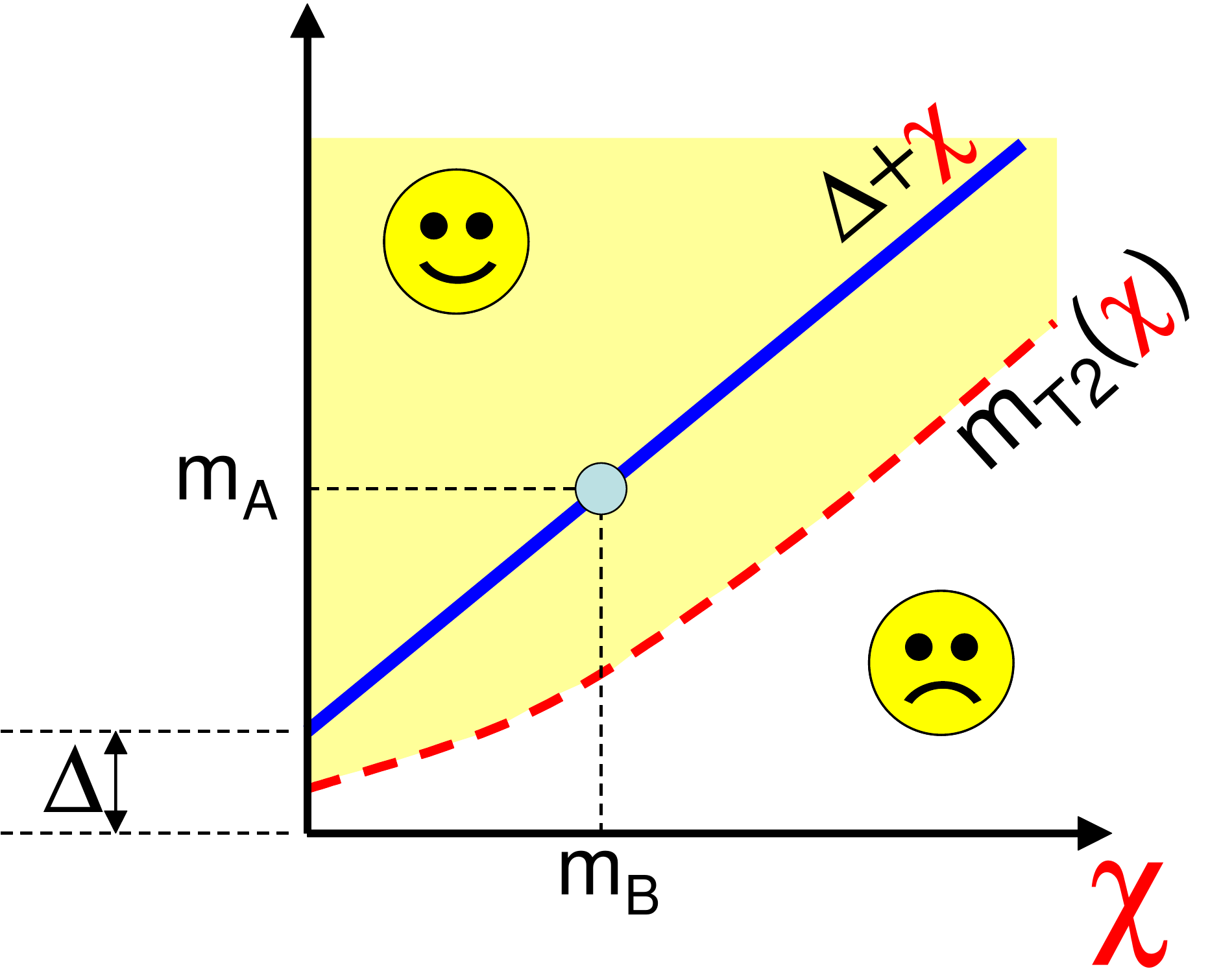}
\qquad\includegraphics[scale=0.25]{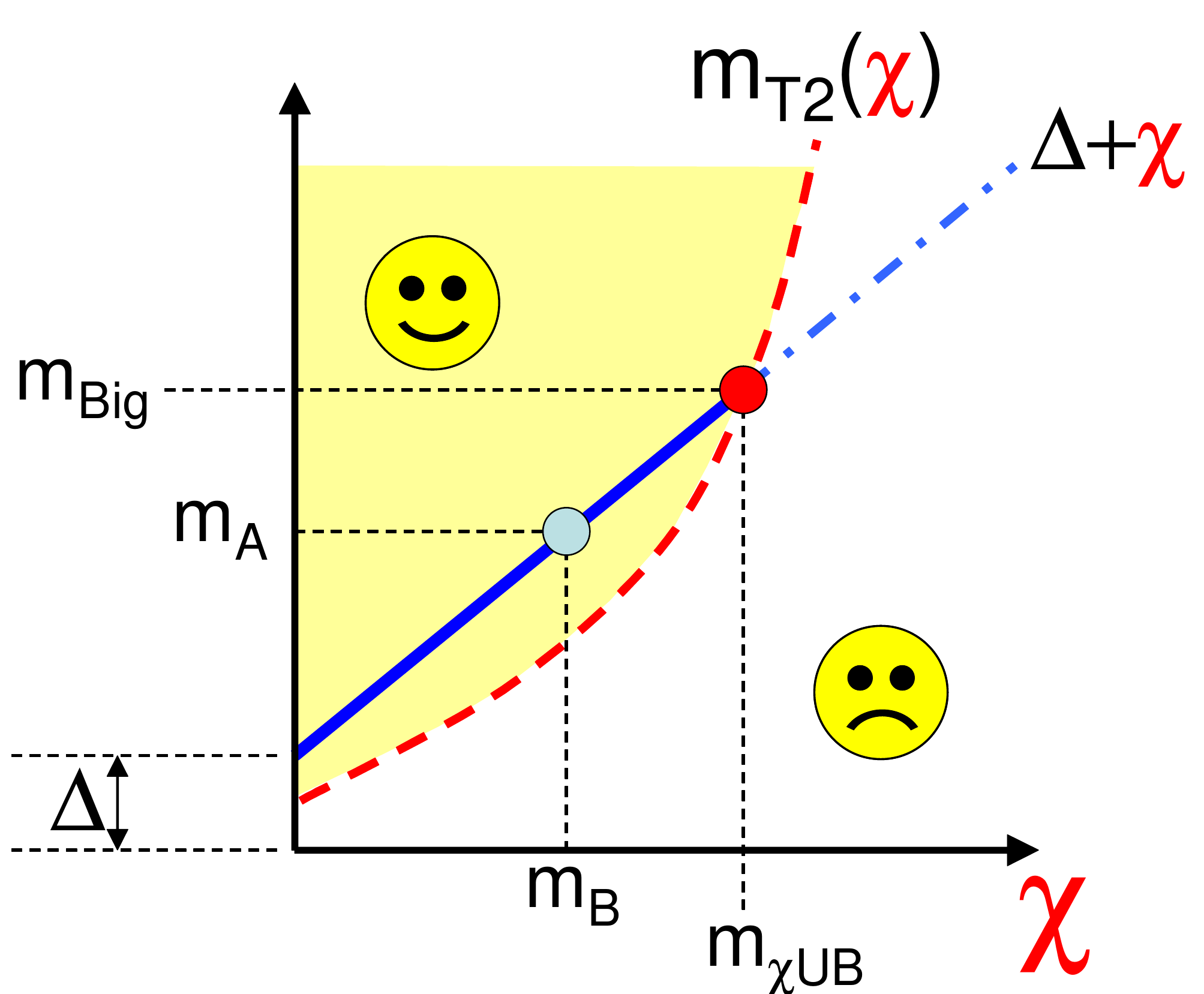}
\caption{The four potential combinations of constraint coming from
  \MTTWO\ and the dilepton edge which together contribute to \MTWOC\ are shown. 
  The mass difference $\Delta=m_A-m_B$ is assumed to have been
  measured independently, e.g.~from the endpoint of the dilepton invariant mass distribution 
  in three-body decays $A\rightarrow B \ell^+\ell^-$.
  Knowledge of $\Delta$ constrains the $(m_A,\,m_B)$ solution to lie on the diagonal line. 
  The constraint from the pair decay kinematics gives the separate $\MTTWO$ constraint,
  for which the solution must lie above and to the left of the dashed line.
  The intersection of the shaded region with the diagonal line
  then gives an event-by-event allowed solution space.
  \label{fig:casesforMTWOC}}
\end{center}
\end{figure}

The hybrid variables \MTWOC\ \cite{Ross:2007rm,Barr:2008ba} and
\MTHREEC\ \cite{Barr:2008hv} use as their hybrid ``ingredient'' the
mass relationships obtained from the ``dilepton edge'' of the
$qll$-chain.  Using this ingredient, together with \MTTWO,
these variables reconstruct event-by-event lower and/or upper bounds
on the LSP mass.

\par

The only important difference between \MTWOC\ and \MTHREEC\ is the
topology to which each variable is applied.  Both have the same final
state, but the internals are different.  The topologies for each
variable are shown side-by-side in
\autoref{fig:picofm2candm3ctopologies}.  \MTWOC\ assumes a pair of
identical particles, decaying identically to one LSP and two SM
particles. \MTHREEC\ requires two sequential two-body decays per
branch, with an intermediate real particle.  In both cases, the
``daughter'' particles are commonly taken to be the lightest
neutralino, and the ``parent'' particles are taken to be the second
lightest neutralino.  The slepton, if it is on shell, is treated as
neither a parent nor a daughter but as an internal particle.

\par

It is easy to see conceptually how \MTWOC\ generates an upper and/or a
lower bound on the LSP mass for each event using the constraint coming
from the dilepton edge.  We have already seen in
\autoref{fig:mt2asboundary} that, in the absence of any additional
``hybrid'' information, the nature of the ``ordinary'' \MTTWO\
constraint is to identify the region of (parent,daughter)-mass space
which is compatible with the event.  This allowed region was shaded
and labelled with a {\LARGE \smiley} in \autoref{fig:mt2asboundary}.
With only the information from \MTTWO, all we know is that the
true parent and daughter mass combination is {\em somewhere} in this
region.  If we now introduce the hybrid constraint from the
dilepton edge on the same set of axes we get one of the cases shown in
\autoref{fig:casesforMTWOC}.  The constraint from the dilepton edge is
always the straight blue (dot-dash or solid) line passing through the
point at which the parent and daughter particles take the correct,
though unknown, masses, with 
\[
\Delta = m_A - m_B = m_{\ell\ell}^{\max}\ .
\]
If we apply the \MTTWO\ constraint and the
dilepton edge constraint {\em simultaneously}, we can see that the
point where the true parent and daughter masses lie {\em must be found
somewhere on the part of the straight line which overlaps the shaded
\MTTWO\ region}.  This smaller allowed region is indicated by the
solid (rather than dot-dashed) sections of the straight blue lines
in \autoref{fig:casesforMTWOC}.  It will be noted that, depending on the
shape of the \MTTWO\ boundary, the allowed region of joint constraint
may be either finite or unbounded in extent.  If finite,
there is always an upper bound on the neutralino masses (both parent
and daughter), and sometimes a lower bound too.  If unbounded in
extent, there may be only a lower bound on the neutralino masses, or
if unlucky, no bound at all.  Which situation one finds depends very
much on the kinematics of the individual events, and depends in
particular on the net transverse momentum in the lab-frame of the pair
of parent neutralinos (\UTM).  Where a lower bound and/or upper bound for the
mass of the lightest neutralino exists, it is called $M_{2C,LB}$
and/or $M_{2C,UB}$ as appropriate. 

Analogous definitions and results apply to the case of \MTHREEC,
except that in this case the sequence of two-body decays means
that the dilepton edge provides a constraint \eqref{eq:dileptonconstraint} 
on the differences of squared masses of participating particles
rather than the mass difference $\Delta$.

Alternatively, instead of generating hybrid variables {\it per se}, one can instead talk of hybrid techniques, where again one combines constraints (such as those from a dilepton edge) with what would otherwise be under-constrained event topologies to reconstruct masses.  A fine example of this includes \cite{Nojiri:2007pq} which combines kinematic edges with events of the type shown in \autoref{fig:mt2sortsoftopologies}(d).  

\subsection{Going beyond pairs of decay chains}
It is straightforward to generalise the kinematical boundary method to 
chains involving more than two decay chains. For example one can define an analogue of 
the \MTTWO\ variable \eqref{eq:mttwodef} suitable for the case where three parents are produced, and each 
decays to a system of visible particles and invisible particles.
One can construct the best bound possible, without knowing the splitting of the invisible momentum between chains, but subject to the constraint that the momenta of the invisibles should sum to the total missing momentum \cite{MatchevUnpublished,Barr:2009wu}
\[
\MTTHREE
\equiv \min_{ \slashed{\bf c}_T + \slashed{\bf c}^\prime_T + \slashed{\bf c}^{\prime\prime}_T= \ptmiss }
     \left\{
     \max { \left( \MT, \MT', \MT'' \right) }
       \right\} .\
\]
There are obvious generalisations to more than three cascades, but such constructions have not received much attention in the literature.

\subsection{Singularity variables}\label{sec:robotics}

It has been noted \cite{Kim:2009si} that all of the kinematic endpoints identified in \secref{sec:single} and \ref{sec:pairdecays} (e.g.~$m_{ll}$, \MT, \MTTWO, \MCT, \ldots)
as well as the cusps described in \secref{sec:darkmattersandwich} and \ref{sec:higgs} are points where the projection from 
momentum space into the variable of choice becomes singular {\em provided that the correct hypothesis has been made for the relevant masses}.
It is shown in \cite{Kim:2009si} that one can systematically identify all the singularities. It is also possible to construct normalised variables (singularity coordinates) locally perpendicular to all such singularity structures in a procedural manner which, though somewhat involved, is well defined. These variables can be constructed for different trial masses and the most singular behaviour sought.

\section{Variables based on suppositions relating to multiple events\label{sec:MULTIPLE}}

We have, by now, seen many examples of events or topologies in which it is not possible to determine the full kinematics of a single event in isolation from any others.  The problem has been that a typical event can be expected to contain far more unknowns (e.g.~the components of the momenta of the invisible particles, and the masses of the unseen internal resonances) than can be constrained by the available observations.   However, although each event {\em in isolation} may contain too little information to allow kinematics and unknown masses to be fully determined, it is sometimes the case that {\em one or more events taken together with some joint assumptions} can overcome this hurdle.  In practice, this requires three conditions to be satisfied: (1) that you have a sample of events in which a sufficiently large fraction can be expected to share a common topology and particle content, (2) that the ``unknowns'' may be divided into those which are ``shared'' among all events (e.g.~model parameters such as unknown masses) and those which are ``independent'' (e.g.~kinematical variables such as the four momenta of the LSPs in each event), and (3) that the number of independent (kinematical) unknowns per event is {\em smaller} than the number of constraints that may be applied to each event by observation and/or hypothesis.   If these three conditions are satisfied, then when a sufficiently large sample of events is considered as a whole, the joint system should become over-constrained and it should be possible to determine not only the unknown masses of the participating particles, but also information about the momentum components of the unseen particles in each event. 

Some methods taking this approach look quite literally at pairs (or triples or quintuples) of events -- indeed however many are necessary to get an over or exact constraint -- and attempt thereby to use each such pair (or \ldots) to gain multiple independent measurements of model parameters \cite{Nojiri:2003tu,Kawagoe:2004rz,Cheng:2008mg,Cheng:2009fw}.  We will discuss some of these in \secref{sec:pairwiseEvents}. Other studies shy away from such an approach, preferring to derive a measurement from the sample of events ``as a whole'' rather than from pairings.  In this latter category come not only all matrix element methods and model dependent fits (of which there are too many to produce a definite list\footnote{Nevertheless, it is worth mentioning \cite{Alwall:2009sv} as an example of a matrix element method providing an alternative to the purely kinematic methods of \secref{sec:mttwo} for measuring particle masses in pair production.}) but also methods that form approximate goodness of fit minimisations.  We will mention one such method \cite{Webber:2009vm} in \secref{sec:webberfitsec}. 

There is still much debate and little consensus as to whether it is
better to work with event ``pairs'' (etc) or to work with the sample
of events in its entirety.  There are arguments and proponents on both
sides - and it is hoped that both methods will be tested on the LHC data.

\subsection{Methods looking at small groups of events}
\label{sec:massrelation}
\label{secondMcElrath}
\label{sec:pairwiseEvents}

One of the first studies to attempt to extract masses by looking at small groups of events that individually would be unconstrained was the ``mass relation method'' \cite{Nojiri:2003tu,Kawagoe:2004rz}.  This considered long decay chains of the form
\par
\begin{center}\includegraphics[scale=0.6,clip]{GluinoChain.pdf}\end{center}
\par\noindent (specifically they considered a final state containing two opposite sign same family leptons, two jets and a neutralino coming from a gluino in the initial state). 
The key idea is that the particle masses can be fully determined -- to within a set of discrete choice-ambiguities -- if sufficiently many events are considered in combination.  In principle, events would need to be considered five-at-a-time.  However to simplify the presentation of the method, the authors made the assumption that the three lightest sparticle masses were already known,  and this allows them to use events in pairs.  The authors noted that it would be possible to extend this method to shorter chains if more than one were present in each event, as the missing momentum constraint would then couple the momenta of the invisible particles and thus couple the constraints on the chains.


Indeed, though couched in a somewhat different language, the so-called ``polynomial'' method of \cite{Cheng:2008mg,Cheng:2009fw} can be thought of as extending the mass relation method to identical pair decays of the form
\par
\begin{center}\includegraphics[scale=0.6,clip]{Mt2ThreeSequential.pdf}.\end{center}\par\noindent This method takes pairs of events and finds the mass hypotheses that are compatible with them.  Perhaps surprisingly (given the large number of particles in the final state) the evidence provided in  \cite{Cheng:2008mg,Cheng:2009fw} suggests that combinatorial ambiguities may not present insurmountable problems for such techniques.  Computer libraries and Mathematica notebooks are freely available \cite{ucdWimpmassLibrary} which provide routines that may be used to determine the masses which are consistent with pairs of events of this type.

\subsection{Matrix element methods, distribution shapes, and combining events}
\label{sec:webberfitsec}

The majority of the mass measurement methods described previously have
been based on conclusions derived from small numbers of events, or
from the local properties of distributions.  For example: invariant
masses were used to measure the mass of single particles decaying to
visible products; local properties of distributions (for example the
location of discrete features, such as a kinematic endpoints) were
used to measure certain relationships between masses; and the
solutions of sets of simultaneous equations derived from a small
number of events were used to find many masses at once under the
assumption that the events were homogeneous.

It should be noted, however, that there are other techniques which
typically become powerful only when looking at much larger numbers of
events.  In the main, these are methods sensitive to the global (or at
least non-local) shapes of differential distributions of observables,
though there are other examples (e.g.~\cite{Webber:2009vm}) which we
will comment on which do not conform to this pattern.  It should be
admitted that the distinction being drawn here\footnote{i.e.~the
distinction between observables based on global properties of
distributions (such as their shapes) and observables based on local
properties (endpoints) or small numbers of events.} is perhaps not as
clear cut as we are suggesting -- one might argue, for example, that
many events are needed to see a kinematic endpoint, or that only by
knowing or using the shape of the distribution near an endpoint can
the endpoint be reliably fitted, and that therefore the use of
kinematic endpoints requires a good understanding of the non-local
properties of distributions and/or large numbers of events.
Nonetheless, we think that the distinction is a useful one to draw in
the sense that, from an experimental perspective, to make use of
``shape'' information one needs a much better understanding of
detector acceptances and efficiencies over the full range of the
differential distribution in question than one does if one is merely
fitting a local property such as a resonance.  Similarly, fitting 
the shape of a ``signal'' distribution over a wide range (a range in
which the background distributions might have very non-trivial
differential distributions of their own) places much more stringent
requirements on the experimenter's understanding of the size and shape
of the underlying backgrounds.\footnote{In contrast a narrow structure, such as a resonance, can often be fitted using a sideband technique with comparatively little understanding of the backgrounds.}


Matrix element methods, also known as likelihood methods or shape
methods, have a long history of making use of all the events in a
sample to constrain a set of model parameters -- not only in particle
physics, but in all areas of the sciences.  In fact, if the
underlying model (for both new physics and relevant backgrounds) is
well understood, and the only remaining question is the determination
of some parameters within that model, then no method can beat the
matrix element or likelihood methods for their ability to extract
information from data~\cite{neymanpearson}. In these methods, the basic idea is that {\em
if} (within the confines of a fully parametrised model) it is
possible to determine the probability with which any given set of
observables is likely to arise, or (equivalently), {\em if} one can
predict the shape of the differential distributions of certain
observables as a function of the parameters of the model in question,
{\em then} it is possible to do whichever of the following is most
desired: (1) to determine which set of model parameters make the data
most likely, (2) to determine confidence intervals for some of the
model parameters, (3) to sample from the posterior distribution of
model parameters induced by the likelihood of the data given an
appropriate prior.  Indeed, there are yet more ways that results of a
shape based analysis could be interpreted or presented -- but the
common feature of each is their dependence on a well understood
likelihood (the probability of the data given a model).  If the
parameters of the model in question are masses (or if the parameters
may be used to derive masses) then these techniques perform model
dependent mass measurements.  For example, \cite{Alwall:2009sv} used \label{sec:alwallme} a matrix element method to place a constraint on the parent and daughter masses in the topology of the type shown in Figure~\ref{fig:mt2sortsoftopologies}(a).  It is interesting to note that the shape of the constraint obtained in \cite{Alwall:2009sv} bears many similarities to that obtained from the corresponding \MTTWO\ analyses.  Possible reasons for this similarity are suggested in \cite{Barr:2009jv}.

One ingredient required by analyses that make use of shapes is,
therefore, an ability to predict the shape of differential
distributions of useful observables.  In some cases, there is no
alternative to using event generators to calculate the shapes of such
distributions by Monte Carlo methods.  There are instances, however,
where the shapes of distributions can be calculated analytically to
provide useful insights into either their nature or into the sets of
circumstances in which the shapes are most useful
\cite{Gjelsten:2004ki,Gjelsten:2005aw,Miller:2005zp,Kraml:2005kb,Smillie:2005ar,Athanasiou:2006ef,Lester:2006yw,Ross:2007rm,Barr:2008ba,Barr:2008hv,Burns:2008cp,Matchev:2009ad,Konar:2009wn}.

It should be noted that the shapes of some of those differential
distributions are sensitive to the spins and couplings of the
particles involved, and that in fact most spin-sensitive analyses rely
on this as their only means of extracting information about those
spins \cite{Barr:2004ze,Barr:2005dz,Smillie:2005ar,Athanasiou:2006ef,Burns:2008cp}.

In practice, matrix element and likelihood methods tend to be applied
only in the mature stages of an analysis.
For example, currently they are being used for top quark mass measurements
\cite{Abazov:2004cs,PhysRevLett.99.182002} and Higgs boson searches
\cite{cdf:9985} at the Tevatron.
Such techniques are less likely to 
perform well in the early stages when there remains some debate
as to the nature of the model being fitted -- and in particular when the
distribution of the background is poorly understood.  As a result,
there is also a large number of methods that are not strictly based
on the statistical likelihood, but which nevertheless try to determine
model parameters from some form of fit over model parameters to some
aspect of the data.

%

Are all methods that look at large homogeneous samples of events based on shapes or differential distributions?  Fortunately not!  For example, though the mass relation method was initially propoposed \cite{Nojiri:2003tu} as a method that should consider pairs of events (thus leading to its discussion earlier in Section~\ref{sec:massrelation}) it has been noted, for example in Section X of \cite{Allanach:2004ub}, that the mass relation method can trivially be extended to work on all events, and indeed becomes less arbitrary in the process.  With similarities to the mass relation method and to the methods of \cite{Cheng:2007xv,Cheng:2008mg,Cheng:2009fw}, another homogeneous event sample technique not based on event shapes has been proposed in \cite{Webber:2009vm} which aims to use linear algebra to efficiently determine unknown masses from events
which are individually under- (but collectively over-) constrained,
by asking us to solve the kinematics of each event in terms of a minimal
number of assumptions, and then minimise a goodness of fit through
variation of these assumptions.  In the example used in
\cite{Webber:2009vm} there are sufficient constraints to permit us to
determine the four-momenta of both of the invisible particles
provided we are prepared to to hypothesise all eight participating masses.  A goodness of fit
function which compares the squares of those LSP momenta to the
hypothesised LSP masses can then be constructed and subsequently
minimised over all possible choices of the set of eight masses.  This
results in an overall ``best fit for the participating masses'' which
uses all events democratically.  Though this is not strictly a likelihood method, one can think of methods in this class as attempting to form a ``heuristic'' likelihood -- by which we mean a function which loosely shares some of the properties of a real likelihood, in particular it is large near ``good'' values of the model parameters, and small elsewhere.

\subsection{Model fitting and adding other observables}

To improve our determination of the masses we may well be willing to make assumptions
about the physics beyond hypothesising topologies and decays.
For example some of the first variables explored in this review (in \secref{sec:threshold})
were those which already made the additional (often implicit) 
assumption that heavy particles are produced near their kinematic threshold, so with $E\approx m$.

Assumptions about the nature of the incoming partons  allow us to turn initial-state radiation to our advantage. QCD radiation adds particles to the final state by an amount which can be calculated, and hence could potentially be exploited to provide information on 
heavy object masses \cite{Papaefstathiou:2009hp} -- though more detailed work is needed to check the practicality of this proposal.

If a particular model is assumed one can combine multiple measurements of 
different topologies to fit for the masses.
The constraints from mass or kinematic edge measurements have been interpreted in the context of supersymmetric models by several groups~\cite{Allanach:2000kt,Lafaye:2004cn,Bechtle:2004pc,Bechtle:2009ty,Roszkowski:2009ye}. If constraints from a future electron-positron collider can be added
\cite{Weiglein:2004hn} then the near-degenerate directions in the LHC-only measurements
can be resolved, leading to improved mass constraints even on those
particles not directly observable at any future $e^+e^-$ machine.

\par

Alternatively, if one is willing to make suppositions about the couplings of 
any new particles (as might be reasonable for example for a supersymmetric model) 
one could further constrain the masses using only LHC data.
The number of events observed is usually a strong function of the mass (both because of 
the parton distribution functions and the explicit dependence of the matrix element on
kinematic variables such as $\hat{s}$). 
If one is willing to entertain hypotheses about the couplings involved, it is possible to interpret the measured cross-sections as constraints in mass space \cite{Lester:2005je,Dreiner:2010gv}, considerably improving the overall mass scale determination even for a modest integrated luminosity of $\sim 1 \ifb$, and with modest uncertainties in the predicted cross sections of order 20\%.

\subsection{Alternative approaches to under-constrained events}\label{sec:maost}

We have already seen that when invisible particles are produced 
it is not always possible to fully reconstruct the event kinematics.  Nonetheless, there are sometimes ways in which it is possible to perform {\em approximate} kinematic reconstructions.

For example, the general point is made in  \cite{Kersting:2009ne,Kang:2009sk} that events in particular corners of kinematic phase space can be ``more reconstructible'' than general events in the bulk.  For example, \cite{Kersting:2009ne} investigates di-chargino production, with each chargino decaying to an opposite sign di-lepton pair and a lighter neutralino.  This is a topology of the form in \autoref{fig:mt2sortsoftopologies}(b).  In general these events are not reconstructible, however the subset of events in which two di-lepton invariant masses (one from each ``side'') are close to their upper kinematic limit are forced to adopt particular kinematic configurations:  in these ``extremal'' events, the decay products of the heavier charginos may be collinear in the chargino rest frame, for example.  If one selects only events near kinematic endpoints, one can therefore make use of this additional kinematic information to render the events sufficiently reconstructible that masses may be determined.  The same trick was applied in Section~20.2.4.1, in the Supersymmetry chapter of \cite{atlasphystdr}.

Another example of performing ``approximate'' reconstruction near kinematic endpoints is the so-called ``MAOS Method''\footnote{``MAOS'' stands for ``\MTTWO\ Assisted On Shell''} of  \cite{Cho:2008tj,Choi:2009hn}.  Here it is noted that for events that are near their \MTTWO\ endpoint (\secref{sec:mttwo}) the missing particles' momenta are constrained to be similar to the values selected by the assignment (or ``splitting'' of \ptmiss) that determines the value of \MTTWO.
It has been demonstrated that by selecting such near-endpoint events, and by using the ``approximate momentum reconstruction'' implied by the splitting selected by the \MTTWO\ minimisation, then it is possible to reconstruct not only the masses of the particles involved, but also place strong constraints on their spins.  This is demonstrated in the the context of a supersymmetric model \cite{Cho:2008tj} and for the determination
of the Higgs boson mass in the channel $h\to W^+W^-\to \ell^+\ell^-\nu\nu$ \cite{Choi:2009hn}.

Not all ``approximate reconstructions'' are motivated by edges of phase space -- an alternative approach is to construct variables from the kinematic configuration which has the greatest likelihood, or from a weighted average of possible  configurations, perhaps weighted by a prior motivated by a Monte Carlo simulation.
This approach has been employed in template-based measurements of the
top quark mass in the di-leptonic channel at the Tevatron \cite{Fatakia:2004wi,Fatakia:2005mz,Adelman:2006rw,Abulencia:2006js,Brandt:2006uc,Meyer:2007zz,Grohsjean:2008uy}.




\section{Conclusions}

The story may be apocryphal, but it has been said that prior to the establishment of the quark model, new particles were being found at such an alarming rate that it was seriously proposed that a Nobel Prize ought to be awarded to the first physicist who {\em couldn't} discover a new particle.   

In the 1950s and early 1960s particle physics may have been expanding into a theoretical vacuum driven by an excess of experimental results.   In the case of mass measurement techniques for the LHC, however, the process seems to have been turned upside down.  The earliest LHC specific techniques were proposed in 1996, or thereabouts~\cite{Paige:1996nx,Hinchliffe:1996iu}, and in the course of the intervening 15 years they have been developed beyond all recognition.  All of this has happened in an almost complete absence of data against which to test these techniques. 
The Tevatron collaborations are owed a debt of thanks both for inventing their own methods, but also for acting as a testbed for some LHC proposals, for example for mass determination in the dileptonic $t\bar{t}$ system.

March of 2010 saw the first collisions at the LHC with centre of mass energies of 7~TeV, and so the long wait for that data is now over.  Those who have invested considerable effort in developing mass measurement techniques are looking on in expectation, waiting to see what the data will bring.  In very little time, experimental collaborations will dash the hopes of phenomenologists the world over by refusing to release {\em any} plots derived from {\em any} of the mass measurement variables which are more complicated than mere invariant masses, as they will be too busy tearing themselves apart in debates over how best to measure the photon reconstruction efficiency in the pseudorapidity range $1.4<|\eta|<1.5$.
The development of mass measurement techniques, which has already seen a period of incredible productivity over the last 15 years, is thus assured a second wind.

While writing this review, nothing was more disheartening than finding the words ``We propose a new variable \ldots'' in one of the abstracts circulated in the daily arXiv digest for hep-ph.   We are pleased to be able to confirm that we, ourselves, have managed to create no new variables during the course of this review.

\begin{acknowledgments}
We are grateful for very helpful discussions with (amongst others)
Johan Alwall, Kiwoon Choi, Christopher Cowden, Ben Gripaios, Rupert Gwenlan, Nick Kersting, Teng Jian Khoo, Ian Woo Kim, Yeong Gyun Kim, Kyoungchul Kong, Jae Sik Lee, Bob McElrath, Konstantin Matchev, David Miller, Mihoko Nojiri, Per Osland, Giacomo Polesello, Mario Serna, Kazuki Sakurai, Dan Tovey and Bryan Webber, and colleagues from the Cambridge Supersymmetry Working Group and the Dalitz Institute (Oxford). We are grateful to Alex Pinder for preparing some of the material for this review. 
AJB acknowledges the support of the Science and Technology Facilities Council,
and thanks Dr Damaris Koch and the Master and Fellows of Peterhouse, Cambridge
for their hospitality.
\end{acknowledgments}

\pagebreak\b
\appendix
\def\thesection{\Alph{section}}
\def\theequation{\thesection.\arabic{equation}}

\section{Some frequently used definitions and formulae}\label{sec:app}
For convenience we include definitions of some of the more frequently-used kinematic variables,
and reproduce some of the most important kinematic endpoint formulae.

\def\myz{{\chi}}
\def\mya{{{l}}}
\def\myb{{{\xi}}}
\def\myc{{{q}}}
\def\msx{{X}}
\def\msone{{{\chi}}}
\def\msslepton{{{l}}}
\def\mstwo{{{\xi}}}
\def\mssquark{{{q}}}
\renewcommand\arraystretch{1.6}
\begin{table}
\begin{center}
\begin{tabular}{c|rcl}
\hline Related edge &
\multicolumn{3}{c} {Kinematic endpoint} \\
\hline
\llEdge & $(\mll^\rmax)^2 $ & $=$ & $ (\myb-\mya)
(\mya -\myz) / \mya  $ \\ 
\llqEdge & $(\mllq^\rmax)^2 $ & $=$ &
$\begin{cases}
( m_\squark-m_\ntlinoOne)^2 \qquad\qquad\text{if $\mya^2 < \myc \myz < \myb^2$ and $ \myb^2 \myz< \myc \mya^2$,}  \\
\max { \left[ { \frac{(\myc-\myb)(\myb-\myz)}{\myb}, \frac{(\myc-\mya)(\mya-\myz)}{\mya}, \frac{(\myc\mya-\myb\myz)(\myb-\mya)}{\myb\mya} } \right] } \qquad\text{otherwise} 
\end{cases}$
\\
%
%
%
 & & & or equivalently \\ 
& & $=$ & $\begin{cases} 
{(\myc-\myb)(\myb-\myz)}/{\myb}  & \text{if $\myb^2 < \myz \myc $,} \\
{(\myc\mya-\myb\myz)(\myb-\mya)}/{\myb\mya}     & \text{if $\mya^2 \myc < \myz \myb^2$,} \\
{(\myc-\mya)(\mya-\myz)}/{\mya} & \text{if $\myc \myz < \mya^2$,} \\
( m_\squark-m_\ntlinoOne)^2 & \text{otherwise.} 
\end{cases}$ \\
\llqThreshold & $(\mllq^\rmin)^2$ & $=$ & $
\bigg[ 2\mya(\myc-\myb)(\myb-\myz)+(\myc+\myb)(\myb-\mya)(\mya-\myz) 
   $ \\
 & &  & $
\qquad   -(\myc-\myb)\sqrt{(\myb+\mya)^2 (\mya+\myz)^2 -16 \myb \mya^2 \myz}  \bigg] / 
(4 \mya \myb) $ \\
%
%
\lqEdgeNear & $(\mlqNear^\rmax)^2 $ & $=$ & $ (\mssquark
-\mstwo) (\mstwo-\msslepton )/\mstwo  $ \\ 
\lqEdgeFar & $(\mlqFar^\rmax)^2 $ & $=$ & $ (\mssquark
-\mstwo) (\msslepton -\msone)/\msslepton   $ \\ 
(just a definition) & $(\mlqEq^\rmax)^2$ & $=$ & $(\mssquark - \mstwo)(\msslepton - \msone)/(2
\msslepton - \msone )$ \\
\lqEdgeHigh & $(\mlqHigh^\rmax)^2$ & $=$ & $\max{\left[{ (\mlqNear^\rmax)^2, (\mlqFar^\rmax)^2 }\right]}$ \\
\lqEdgeLow & $(\mlqLow^\rmax)^2$ & $=$ & $\min{\left[{ (\mlqNear^\rmax)^2,
 (\mlqEq^\rmax)^2 }\right]}$ \\
(alternative form) & 
\multicolumn{3}{l}{$\left((\mlqLow^\rmax)^2,(\mlqHigh^\rmax)^2\right)  =
\begin{cases}
\left( (\mlqNear^\rmax)^2, (\mlqFar^\rmax)^2 \right) & \text{if $2 \mya > \myb + \myz > 2 \sqrt{\myb \myz}$} \\
\left( (\mlqEq^\rmax)^2, (\mlqFar^\rmax)^2 \right) & \text{if $\myb + \myz > 2 \mya > 2 \sqrt{\myb \myz}$}\\
\left( (\mlqEq^\rmax)^2, (\mlqNear^\rmax)^2 \right) & \text{if $\myb + \myz > 2 \sqrt{\myb \myz} > 2 \mya $.}
\end{cases}$}\\
\lqEdgeSum & \multicolumn{3}{l}{$(\mlqPlus^2 + \mlqMinus^2)^\rmax  = (\myc-\myb) (\myb-\myz) /\myb$}\\
\xqEdge  & $(\mxq^\rmax)^2 $ & $=$ & $ \msx+(\mssquark -\mstwo) \left[ \mstwo+\msx-\msone+\sqrt{(\mstwo-\msx-\msone)^2-4 \msx \msone} \right]/ (2 \mstwo)  $ \\
\hline
\end{tabular}
\end{center}
\caption{(Containing results from \cite{Allanach:2000kt,Gjelsten:2004ki,Burns:2009zi,Matchev:2009iw}.) This table lists the absolute kinematic endpoints of invariant mass
distributions formed from decay chains of the type
$\squark \to \ntltwo \to \slepton^\mp \ell^\pm_\mathrm{near} q 
\to \ntlone \ell^\mp_\mathrm{far}\ell^\pm_\mathrm{near} q$ 
for known particle masses. 
The shorthand notation used is: $\myz = m^2_\ntlinoOne$, $\mya = m^2_{\slepton_R}$, $\myb =
m^2_\ntlinoTwo$ and $\myc = m^2_\squark$ and $\msx$ is $m^2_h$ or
$m^2_Z$ as appropriate.
The visible particles are assumed to have negligible mass.
Inversion formulae (i.e.~masses in terms of endpoints) for
certain subsets of the above endpoints are published in
\cite{Gjelsten:2004ki,Burns:2009zi,Matchev:2009iw}.   Note that this table is by no means exhaustive: many other interesting endpoints have been proposed for the $qll$-type chain (including those of \cite{Burns:2009zi,Matchev:2009iw}) and for other types of chain \cite{Agashe:2010gt}. \label{tab:obs}}
\end{table}

\subsection{Summary of simple kinematic variables}\label{sec:app:def}

Bold font symbols indicate Euclidean momentum vectors in three (or two transverse) dimensions.
A subscript $_T$ indicates a quantity built from transverse momentum components. 
\par\vspace{5mm}\noindent
{\bf Invariant mass} (\secref{sec:z}):
\begin{equation} M^2 = \left(\sum_i p_i\right)^2 .\end{equation}
{\bf Transverse energy} (\secref{sec:w})
\begin{equation} e^2 = {m^2 + {\bf p}^2_T}. \label{eq:app:et}\end{equation}
{\bf Effective mass} (\secref{sec:threshold}) -- a typical definition is a sum over the leading $n$ jets:
\begin{equation} \MEFF =  |\Ptmiss|+\sum_{i=1,n}|{\bf p}_{T,i}| \end{equation}
where historically $n$ has always been taken to be ``4'', but more recently $n$ has become dependent on the analysis channel.  Note that $H_T$, the analogue in CMS and at the Tevatron, has a number of different definitions (see equations (\ref{eq:ht2a}),   (\ref{eq:ht2b}) and (\ref{eq:ht2c})), of which the one most frequently used at present is 
\begin{equation} H_T = \sum_{i=1,n}|{\bf p}_{T,i}|  \label{eq:app:ht} \ .\end{equation}
{\bf Transverse mass} (\secref{sec:w}) if $m_{\slashed C}$ is known
\begin{equation} \MT^2 \equiv m_b^2 + m_{\slashed c}^2 + 2 \left( e_b e_\slashed c - {\bf b}_T \cdot \slashed{\bf c}_T \right) \end{equation}
otherwise
\begin{equation} \MT^2 \equiv m_b^2 + \chi^2 + 2 \left( e_b e_\chi - {\bf b}_T \cdot \slashed{\bf c}_T \right) \ \end{equation}
where $\chi$, the hypothesis for $m_{\slashed C}$ is also used in $e_\chi = \sqrt{\chi^2 + {\bf c}_T^2}$.
\par\noindent
{\bf Contransverse mass} (\secref{sec:mct}, \ref{sec:mtovey}):
\begin{equation} \MCT^2 = m_b^2 + m_{b^\prime}^2 
+ 2 ( e_b e_{b^\prime} + {\bf b}_T \cdot{\bf b}_T^\prime ) .\ \end{equation}
{\bf Stransverse mass} (\secref{sec:mttwo} {\it et sequens}):
\begin{equation} \mttwo =      \min_{ \slashed{\bf c}_T + \slashed{\bf c}^\prime_T = \ptmiss }
     \left\{     \max { \left( \MT, \MT' \right) } \right\} .\ \end{equation}
\subsection{Kinematic endpoints of cascade decay chains}\label{sec:app:twobody}
The kinematic endpoints for the decay chain
$\squark \to \ntltwo \to \slepton^\mp \ell^\pm_\mathrm{near} q 
\to \ntlone \ell^\mp_\mathrm{far}\ell^\pm_\mathrm{near} q$ 
(and other chains with the same topology) 
can be found in \autoref{tab:obs}.

\subsection{Some properties of two-body decays and variables related to them}
\noindent
This section summarises some frequently-used results relating to two-body semi-invisible decays of a single particle.  In this section our notation assumes that the decay is labelled
\begin{gather*}
A\to B\slashed{C}
\end{gather*}
and that upstream transverse momentum is defined to be 
\begin{gather*}
\UTM=-{\bf a}_T .
\end{gather*}
\par\noindent
{\bf Transverse momentum}: \par \noindent For any \UTM\ recoil, with the correct invisible particle mass hypothesis $\chi=m_\slashed{C}$ the momentum of each daughter in the parent rest frame is 
\begin{equation}\label{eq:appendix:pstar}
p^* = \frac{\lambda^\half(m_A,m_B,m_\slashed{C})}{2m_A}
\end{equation}
where
\begin{eqnarray}
\lambda(a,b,c) &=& \left(a^2 - (b+c)^2 \right)  \left(a^2-(b-c)^2\right)\nonumber\\
               &=& a^4+b^4+c^4 - 2a^2b^2 - 2 a^2c^2 - 2b^2c^2.
\end{eqnarray}
For $m_B=0$ \eqref{eq:appendix:pstar} simplifies to
\begin{equation}\label{eq:pstarmbzero}
p^* = \frac{m_A^2-m_\slashed{C}^2}{2m_A} \quad : m_B=0
\end{equation}
For single particle production and two-body decay with $\UTM=0$
\begin{equation}
b_T^{\max} = p^* \quad : \UTM=0.
\end{equation}
For fixed $\UTM$ and $m_B=0$,
\begin{equation}
b_T^{\max} = p^*e^y \quad: m_B=0
\end{equation}
where $\sinh y = |\UTM|/m_A$ and $p^*$ for $m_B=0$ is given in \autoref{eq:pstarmbzero}.
For fixed $\UTM$ and $m_B\ne0$,
\begin{equation}
b_T^{\max} = m_B \sinh (y + \eta)
\end{equation}
where $\sinh \eta = p^*/m_B$.
\par\noindent
{\bf Transverse mass}: for any \UTM\ recoil, with the correct invisible particle mass hypothesis 
\begin{equation}
{\MT^{\max}} = m_A \quad : \chi=m_\slashed{C}
\end{equation}
For fixed \UTM, and an arbitrary mass hypothesis $\chi\ne m_\slashed{C}$ and with $m_B$=0, \footnote{After \cite{Burns:2008va}.}
\begin{equation}\label{eq:mtmaxutm}
\left[\MT^{\max}(\chi)\right]^2 = \begin{cases}
\left(p^* e^{- y} + \sqrt{(|\UTM| + p^*e^{- y})^2+\chi^2}\right)^2 - \UTM^2 & \quad : \chi < m_\slashed{C}, m_B=0\\[4mm]
\left(p^* e^{+ y} + \sqrt{(|\UTM| - p^*e^{+ y})^2+\chi^2}\right)^2 - \UTM^2 & \quad : \chi > m_\slashed{C}, m_B=0\\[2mm]
\end{cases}
\end{equation}
where $\sinh y = |\UTM|/m_A$ and $p^*$ for $m_B=0$ is given in \autoref{eq:pstarmbzero}.

\subsection{Endpoints for pairs of semi-invisible decays}\label{sec:app:pairs}
\noindent
This section summarises endpoint formulae for kinematic variables used for pairs of semi-invisible decays.  In this section our notation assumes that the decays are labelled
\begin{gather*}
A\to B\slashed{C} \\
A'\to B'\slashed{C}' 
\end{gather*}
and that the upstream transverse momentum is
\begin{gather*}
\UTM = -({\bf a}_T + {\bf a}^\prime_T).
\end{gather*}
\par
\noindent
{\bf Stransverse mass}: for any \UTM, for the correct invisible particle mass hypothesis,
\begin{equation}
{\MTTWO^{\max}} = m_A \quad : \chi=m_\slashed{C} .
\end{equation}
\par
For fixed \UTM\ and $m_B = 0$ the maximum value of \MTTWO\ for any invisible particle mass hypothesis $\chi$ is given by \autoref{eq:mtmaxutm} but now with $\UTM \mapsto \UTM/2$ \cite{Burns:2008va}:
\begin{equation}\label{eq:mttwomaxutm}
\left[\MTTWO^{\max}(\chi)\right]^2 = \begin{cases}
\left(p^* e^{- y_2} + \sqrt{(|\UTM/2| + p^*e^{- y_2})^2+\chi^2}\right)^2 - \UTM^2/4 & \quad : \chi < m_\slashed{C}, m_B=0\\[4mm]
\left(p^* e^{+ y_2} + \sqrt{(|\UTM/2| - p^*e^{+ y_2})^2+\chi^2}\right)^2 - \UTM^2/4 & \quad : \chi > m_\slashed{C}, m_B=0\\[2mm]
\end{cases}
\end{equation}
where $\sinh y_2 = |\UTM|/2m_A$.
Formulae for ${\MTTWO^{\max}}$ with  $m_B \ne 0 $ are given in \cite{Cho:2007qv} for $\UTM=0$, and in \cite{Burns:2008va} for arbitrary \UTM.

The 1-D decomposed versions of \MTTWO\ (\secref{sec:mttwodecomposed}) constructed from components of momenta parallel to (perpendicular to) \UTM\ \cite{Konar:2009wn} have endpoints:
\begin{eqnarray}
M_{T2\parallel}^{\max} &=& \MTTWO^{\max}\quad \label{eq:mt2paramax}\\
M_{T2\perp}^{\max} &=& p^* + \sqrt{p^{*2}+\tilde{m}_\slashed{C}^2} \quad : m_b=0.\label{eq:mt2perpmax}
\end{eqnarray}
\par\noindent
{\bf Contransverse mass}: for $m_B=0$ and arbitrary \UTM\ \cite{Polesello:2009rn}:
\begin{equation}
\MCT^{\max} = 2p^* e^{y_2}\quad : m_B=0\label{eq:repromctmaxresult}
\end{equation}
and likewise for the projections in the same limit \cite{Matchev:2009ad}:
\begin{eqnarray}
M_{CT\parallel}^{\max} &=& 2p^* e^{y_2}\quad : m_B=0\label{eq:repromctparamaxresult}\\
M_{CT\perp}^{\max} &=& 2p^* \qquad : m_B=0 \label{eq:repromctperpmaxresult}
\end{eqnarray}
where again $\sinh y_2 = |\UTM|/2m_A$ and $p^*$ for $m_B=0$ was defined in \autoref{eq:pstarmbzero}.
Cases with $m_B\ne0$ are considered in \cite{Polesello:2009rn,Matchev:2009ad}.

\bibliography{MassAndSpinMethods}

\providecommand{\href}[2]{#2}\begingroup\raggedright\begin{thebibliography}{10%
0}

\bibitem{Alitti:1990ch}
{\bf UA2} Collaboration, J.~Alitti {\em et.~al.}, {\it A precise determination
  of the {$W$} and {$Z$} masses at the {CERN} $\bar{p}p$ collider},  {\em Phys.
  Lett.} {\bf B241} (1990) 150--164.

\bibitem{Acciarri:1999ft}
{\bf L3} Collaboration, M.~Acciarri {\em et.~al.}, {\it {Measurement of mass
  and width of the $W$ boson at LEP}},  {\em Phys. Lett.} {\bf B454} (1999)
  386--398, [\href{http://xxx.lanl.gov/abs/hep-ex/9909010}{{\tt
  hep-ex/9909010}}].

\bibitem{Abazov:2006bd}
{\bf {D0}} Collaboration, V.~M. Abazov {\em et.~al.}, {\it {Measurement of the
  top quark mass in the lepton + jets final state with the matrix element
  method}},  {\em Phys. Rev.} {\bf D74} (2006) 092005,
  [\href{http://xxx.lanl.gov/abs/hep-ex/0609053}{{\tt hep-ex/0609053}}].

\bibitem{JohanIpmu}
J.~Alwall, ``Measuring sparticles with the matrix element.''
\newblock {IPMU} Focus Week.

\bibitem{Collaboration:2008ub}
{ALEPH}, {CDF}, {D0}, {DELPHI}, {L3}, {OPAL}, and {SLD}, {\it {Precision
  Electroweak Measurements and Constraints on the Standard Model}},
  \href{http://xxx.lanl.gov/abs/0811.4682}{{\tt 0811.4682}}.

\bibitem{Aad:2008zzm}
{\bf ATLAS} Collaboration, G.~Aad {\em et.~al.}, {\it {The ATLAS Experiment at
  the CERN Large Hadron Collider}},  {\em JINST} {\bf 3} (2008) S08003.

\bibitem{:2008zzk}
{\bf CMS} Collaboration, R.~Adolphi {\em et.~al.}, {\it {The CMS experiment at
  the CERN LHC}},  {\em JINST} {\bf 0803} (2008) S08004.

\bibitem{Albrow:1005180}
M.~Albrow {\em et.~al.}, {\it Prospects for diffractive and forward physics at
  the {LHC}},  Tech. Rep. LHCC-G-124. {CERN}-LHCC-2006-039. CMS-Note-2007-002.
  {CERN}-CMS-Note-2007-002. TOTEM-Note-2006-005. {CERN}-TOTEM-Note-2006-005,
  {CERN}, Geneva, Dec, 2006.

\bibitem{Albrow:2008pn}
{\bf FP420 R\&D} Collaboration, M.~G. Albrow {\em et.~al.}, {\it {The FP420
  R\&D Project: Higgs and New Physics with forward protons at the LHC}},  {\em
  JINST} {\bf 4} (2009) T10001, [\href{http://xxx.lanl.gov/abs/0806.0302}{{\tt
  0806.0302}}].

\bibitem{Roberts:1960zz}
A.~Roberts, {\it {A new type of Cerenkov detector for the accurate measurement
  of particle velocity and direction}},  {\em Nucl. Instrum. Meth.} {\bf 9}
  (1960) 55--66.

\bibitem{Richardson:2001df}
P.~Richardson, {\it {Spin correlations in Monte Carlo simulations}},  {\em
  JHEP} {\bf 11} (2001) 029,
  [\href{http://xxx.lanl.gov/abs/hep-ph/0110108}{{\tt hep-ph/0110108}}].

\bibitem{Barr:2004ze}
A.~J. Barr, {\it {Using lepton charge asymmetry to investigate the spin of
  supersymmetric particles at the {LHC}}},  {\em Phys. Lett.} {\bf B596} (2004)
  205--212, [\href{http://xxx.lanl.gov/abs/hep-ph/0405052}{{\tt
  hep-ph/0405052}}].

\bibitem{Goto:2004cpa}
T.~Goto, K.~Kawagoe, and M.~M. Nojiri, {\it {Study of the slepton
  non-universality at the CERN Large Hadron Collider}},  {\em Phys. Rev.} {\bf
  D70} (2004) 075016, [\href{http://xxx.lanl.gov/abs/hep-ph/0406317}{{\tt
  hep-ph/0406317}}].

\bibitem{Goto:2004cp}
T.~Goto, {\it {Neutralino polarization effect in the squark cascade decay at
  LHC}},  \href{http://xxx.lanl.gov/abs/hep-ph/0411360}{{\tt hep-ph/0411360}}.

\bibitem{Smillie:2005ar}
J.~M. Smillie and B.~R. Webber, {\it Distinguishing spins in supersymmetric and
  {Universal Extra Dimension} models at the {Large Hadron Collider}},  {\em
  JHEP} {\bf 10} (2005) 069,
  [\href{http://xxx.lanl.gov/abs/hep-ph/0507170}{{\tt hep-ph/0507170}}].

\bibitem{Battaglia:2005zf}
M.~Battaglia, A.~Datta, A.~De~Roeck, K.~Kong, and K.~T. Matchev, {\it
  {Contrasting supersymmetry and universal extra dimensions at the CLIC
  multi-TeV e+ e- collider}},  {\em JHEP} {\bf 07} (2005) 033,
  [\href{http://xxx.lanl.gov/abs/hep-ph/0502041}{{\tt hep-ph/0502041}}].

\bibitem{Battaglia:2005ma}
M.~Battaglia, A.~K. Datta, A.~De~Roeck, K.~Kong, and K.~T. Matchev, {\it
  {Contrasting supersymmetry and universal extra dimensions at colliders}},
  \href{http://xxx.lanl.gov/abs/hep-ph/0507284}{{\tt hep-ph/0507284}}.

\bibitem{Datta:2005zs}
A.~Datta, K.~Kong, and K.~T. Matchev, {\it {Discrimination of supersymmetry and
  universal extra dimensions at hadron colliders}},  {\em Phys. Rev.} {\bf D72}
  (2005) 096006, [\href{http://xxx.lanl.gov/abs/hep-ph/0509246}{{\tt
  hep-ph/0509246}}].

\bibitem{Meade:2006dw}
P.~Meade and M.~Reece, {\it {Top partners at the {LHC}: Spin and mass
  measurement}},  {\em Phys. Rev.} {\bf D74} (2006) 015010,
  [\href{http://xxx.lanl.gov/abs/hep-ph/0601124}{{\tt hep-ph/0601124}}].

\bibitem{Athanasiou:2006ef}
C.~Athanasiou, C.~G. Lester, J.~M. Smillie, and B.~R. Webber, {\it
  Distinguishing spins in decay chains at the {Large Hadron Collider}},  {\em
  JHEP} {\bf 08} (2006) 055,
  [\href{http://xxx.lanl.gov/abs/hep-ph/0605286}{{\tt hep-ph/0605286}}].

\bibitem{Wang:2006hk}
L.-T. Wang and I.~Yavin, {\it Spin measurements in cascade decays at the
  {LHC}},  {\em JHEP} {\bf 04} (2007) 032,
  [\href{http://xxx.lanl.gov/abs/hep-ph/0605296}{{\tt hep-ph/0605296}}].

\bibitem{Smillie:2006cd}
J.~M. Smillie, {\it Spin correlations in decay chains involving {$W$} bosons},
  {\em Eur. Phys. J.} {\bf C51} (2007) 933--943,
  [\href{http://xxx.lanl.gov/abs/hep-ph/0609296}{{\tt hep-ph/0609296}}].

\bibitem{Alves:2006kn}
A.~Alves, O.~Eboli, and T.~Plehn, {\it Spins in gluino decays},  {\em AIP Conf.
  Proc.} {\bf 903} (2007) 265--268,
  [\href{http://xxx.lanl.gov/abs/hep-ph/0611010}{{\tt hep-ph/0611010}}].

\bibitem{Choi:2006mr}
S.~Y. Choi, K.~Hagiwara, H.~U. Martyn, K.~Mawatari, and P.~M. Zerwas, {\it
  {Spin analysis of supersymmetric particles}},  {\em Eur. Phys. J.} {\bf C51}
  (2007) 753--774, [\href{http://xxx.lanl.gov/abs/hep-ph/0612301}{{\tt
  hep-ph/0612301}}].

\bibitem{Nojiri:2007jm}
M.~M. Nojiri and M.~Takeuchi, {\it {The study of $\tilde{q}_L$ $\tilde{q}_L$
  production at {LHC} in the $l^\pm$ $l^\pm$ channel and sensitivity to other
  models}},  {\em Phys. Rev.} {\bf D76} (2007) 015009,
  [\href{http://xxx.lanl.gov/abs/hep-ph/0701190}{{\tt hep-ph/0701190}}].

\bibitem{Kilic:2007zk}
C.~Kilic, L.-T. Wang, and I.~Yavin, {\it On the existence of angular
  correlations in decays with heavy matter partners},  {\em JHEP} {\bf 05}
  (2007) 052, [\href{http://xxx.lanl.gov/abs/hep-ph/0703085}{{\tt
  hep-ph/0703085}}].

\bibitem{Alves:2007xt}
A.~Alves and O.~Eboli, {\it {Unravelling the sbottom spin at the {CERN}
  {LHC}}},  {\em Phys. Rev.} {\bf D75} (2007) 115013,
  [\href{http://xxx.lanl.gov/abs/0704.0254}{{\tt 0704.0254}}].

\bibitem{Csaki:2007xm}
C.~Csaki, J.~Heinonen, and M.~Perelstein, {\it Testing gluino spin with
  three-body decays},  {\em JHEP} {\bf 10} (2007) 107,
  [\href{http://xxx.lanl.gov/abs/0707.0014}{{\tt 0707.0014}}].

\bibitem{Horsky:2008yi}
R.~Horsky, M.~Kramer, A.~Muck, and P.~M. Zerwas, {\it Squark cascade decays to
  charginos/neutralinos: Gluon radiation},  {\em Phys. Rev.} {\bf D78} (2008)
  035004, [\href{http://xxx.lanl.gov/abs/0803.2603}{{\tt 0803.2603}}].

\bibitem{Burns:2008cp}
M.~Burns, K.~Kong, K.~T. Matchev, and M.~Park, {\it A general method for
  model-independent measurements of particle spins, couplings and mixing angles
  in cascade decays with missing energy at hadron colliders},  {\em JHEP} {\bf
  10} (2008) 081, [\href{http://xxx.lanl.gov/abs/0808.2472}{{\tt 0808.2472}}].

\bibitem{Cho:2008tj}
W.~S. Cho, K.~Choi, Y.~G. Kim, and C.~B. Park, {\it {$M_{T2}$-assisted on-shell
  reconstruction of missing momenta and its application to spin measurement at
  the {LHC}}},  {\em Phys. Rev.} {\bf D79} (2009) 031701,
  [\href{http://xxx.lanl.gov/abs/0810.4853}{{\tt 0810.4853}}].

\bibitem{Gedalia:2009ym}
O.~Gedalia, S.~J. Lee, and G.~Perez, {\it Spin determination via third
  generation cascade decays},  {\em Phys. Rev.} {\bf D80} (2009) 035012,
  [\href{http://xxx.lanl.gov/abs/0901.4438}{{\tt 0901.4438}}].

\bibitem{Kramer:2009kp}
M.~Kramer, E.~Popenda, M.~Spira, and P.~M. Zerwas, {\it {Gluino Polarization at
  the LHC}},  {\em Phys. Rev.} {\bf D80} (2009) 055002,
  [\href{http://xxx.lanl.gov/abs/0902.3795}{{\tt 0902.3795}}].

\bibitem{Ehrenfeld:2009rt}
W.~Ehrenfeld, A.~Freitas, A.~Landwehr, and D.~Wyler, {\it {Distinguishing spins
  in decay chains with photons at the Large Hadron Collider}},  {\em JHEP} {\bf
  07} (2009) 056, [\href{http://xxx.lanl.gov/abs/0904.1293}{{\tt 0904.1293}}].

\bibitem{Barr:2005dz}
A.~J. Barr, {\it {Measuring slepton spin at the {LHC}}},  {\em JHEP} {\bf 02}
  (2006) 042, [\href{http://xxx.lanl.gov/abs/hep-ph/0511115}{{\tt
  hep-ph/0511115}}].

\bibitem{Buckley:2007th}
M.~R. Buckley, H.~Murayama, W.~Klemm, and V.~Rentala, {\it {Discriminating spin
  through quantum interference}},  {\em Phys. Rev.} {\bf D78} (2008) 014028,
  [\href{http://xxx.lanl.gov/abs/0711.0364}{{\tt 0711.0364}}].

\bibitem{Buckley:2008pp}
M.~R. Buckley, B.~Heinemann, W.~Klemm, and H.~Murayama, {\it Quantum
  interference effects among helicities at {LEP-II} and {Tevatron}},  {\em
  Phys. Rev.} {\bf D77} (2008) 113017,
  [\href{http://xxx.lanl.gov/abs/0804.0476}{{\tt 0804.0476}}].

\bibitem{Alves:2008up}
A.~Alves, O.~J.~P. Eboli, M.~C. Gonzalez-Garcia, and J.~K. Mizukoshi, {\it
  {Deciphering the spin of new resonances in Higgsless models}},  {\em Phys.
  Rev.} {\bf D79} (2009) 035009, [\href{http://xxx.lanl.gov/abs/0810.1952}{{\tt
  0810.1952}}].

\bibitem{Buckley:2008eb}
M.~R. Buckley, S.~Y. Choi, K.~Mawatari, and H.~Murayama, {\it Determining spin
  through quantum azimuthal-angle correlations},  {\em Phys. Lett.} {\bf B672}
  (2009) 275--279, [\href{http://xxx.lanl.gov/abs/0811.3030}{{\tt 0811.3030}}].

\bibitem{Boudjema:2009fz}
F.~Boudjema and R.~K. Singh, {\it {A model independent spin analysis of
  fundamental particles using azimuthal asymmetries}},  {\em JHEP} {\bf 07}
  (2009) 028, [\href{http://xxx.lanl.gov/abs/0903.4705}{{\tt 0903.4705}}].

\bibitem{Wang:2008sw}
L.-T. Wang and I.~Yavin, {\it A review of spin determination at the {LHC}},
  {\em Int. J. Mod. Phys.} {\bf A23} (2008) 4647--4668,
  [\href{http://xxx.lanl.gov/abs/0802.2726}{{\tt 0802.2726}}].

\bibitem{Hinchliffe:1996iu}
I.~Hinchliffe, F.~E. Paige, M.~D. Shapiro, J.~Soderqvist, and W.~Yao, {\it
  {Precision {SUSY} measurements at {CERN} {LHC}}},  {\em Phys. Rev.} {\bf D55}
  (1997) 5520--5540, [\href{http://xxx.lanl.gov/abs/hep-ph/9610544}{{\tt
  hep-ph/9610544}}].

\bibitem{Aad:2009wy}
{\bf The ATLAS} Collaboration, G.~Aad {\em et.~al.}, {\it {Expected Performance
  of the ATLAS Experiment - Detector, Trigger and Physics}},
  \href{http://xxx.lanl.gov/abs/0901.0512}{{\tt 0901.0512}}.

\bibitem{Collaboration:1273174}
{\bf {ATLAS}} Collaboration, {The ATLAS Collaboration}, {\it Early
  supersymmetry searches in channels with jets and missing transverse momentum
  with the {ATLAS} detector},  Tech. Rep. ATLAS-COM-CONF-2010-066, {CERN},
  Geneva, Jun, 2010.

\bibitem{Tovey:2000wk}
D.~R. Tovey, {\it {Measuring the {SUSY} mass scale at the {LHC}}},  {\em Phys.
  Lett.} {\bf B498} (2001) 1--10,
  [\href{http://xxx.lanl.gov/abs/hep-ph/0006276}{{\tt hep-ph/0006276}}].

\bibitem{PhysRevLett.88.041801}
{\bf (CDF Collaboration)} Collaboration, T.~{\it et al.}. T.~Affolder, {\it
  Search for gluinos and scalar quarks in $p\bar{p}$ collisions at
  $\sqrt{s}=1.8 tev$ using the missing energy plus multijets signature},  {\em
  Phys. Rev. Lett.} {\bf 88} (Jan, 2002) 041801.

\bibitem{cmsphystdr}
{\bf CMS} Collaboration, {\it {CMS} physics technical design report, volume
  {II}},  2006.
\newblock {{CERN}-LHCC-2006-021}, {CMS-TDR-008-2}.

\bibitem{cms:CMS-PAS-SUS-10-001}
C.~Collaboration, {\it Performance of methods for data-driven background
  estimation in susy searches},  Tech. Rep. CMS-PAS-SUS-10-001, CERN, Geneva,
  July, 2010.

\bibitem{Konar:2008ei}
P.~Konar, K.~Kong, and K.~T. Matchev, {\it {$\surd{\hat{s}}_{min}$ : A global
  inclusive variable for determining the mass scale of new physics in events
  with missing energy at hadron colliders}},  {\em JHEP} {\bf 03} (2009) 085,
  [\href{http://xxx.lanl.gov/abs/0812.1042}{{\tt 0812.1042}}].

\bibitem{Papaefstathiou:2009hp}
A.~Papaefstathiou and B.~Webber, {\it {Effects of QCD radiation on inclusive
  variables for determining the scale of new physics at hadron colliders}},
  {\em JHEP} {\bf 06} (2009) 069,
  [\href{http://xxx.lanl.gov/abs/0903.2013}{{\tt 0903.2013}}].

\bibitem{RevModPhys.57.699}
C.~Rubbia, {\it Experimental observation of the intermediate vector bosons
  {$W^+$, $W^-$, and $Z$}},  {\em Rev. Mod. Phys.} {\bf 57} (Jul, 1985)
  699--722.

\bibitem{Abazov:2009cp}
{\bf {D0}} Collaboration, V.~M. Abazov {\em et.~al.}, {\it {Measurement of the
  $W$ boson mass}},  {\em Phys. Rev. Lett.} {\bf 103} (2009) 141801,
  [\href{http://xxx.lanl.gov/abs/0908.0766}{{\tt 0908.0766}}].

\bibitem{NeervenVermaserenGaemersMT}
W.~L. van Neerven, J.~A.~M. Vermaseren, and K.~Gaemers, {\it {Nationaal
  Instituut voor Kernfysica en Hoge-Energiefysica Report No. NIKHEF-H/82-20a}},
   tech. rep., 1982.

\bibitem{Arnison:1983rp}
{\bf UA1} Collaboration, G.~Arnison {\em et.~al.}, {\it {Experimental
  observation of isolated large transverse energy electrons with associated
  missing energy at $s^{1/2}$ = 540 GeV}},  {\em Phys. Lett.} {\bf B122} (1983)
  103--116.

\bibitem{Banner:1983jy}
{\bf UA2} Collaboration, M.~Banner {\em et.~al.}, {\it {Observation of single
  isolated electrons of high transverse momentum in events with missing
  transverse energy at the {CERN} $\bar{p} p$ collider}},  {\em Phys. Lett.}
  {\bf B122} (1983) 476--485.

\bibitem{PhysRevLett.50.1738}
J.~Smith, W.~L. van Neerven, and J.~A.~M. Vermaseren, {\it Transverse mass and
  width of the {$W$} boson},  {\em Phys. Rev. Lett.} {\bf 50} (May, 1983)
  1738--1740.

\bibitem{PhysRevD.36.295}
V.~Barger, T.~Han, and R.~J.~N. Phillips, {\it Improved transverse-mass
  variable for detecting higgs-boson decays into {$Z$} pairs},  {\em Phys. Rev.
  D} {\bf 36} (Jul, 1987) 295--298.

\bibitem{Cheng:2008hk}
H.-C. Cheng and Z.~Han, {\it Minimal kinematic constraints and {$M_{T2}$}},
  {\em JHEP} {\bf 12} (2008) 063,
  [\href{http://xxx.lanl.gov/abs/0810.5178}{{\tt 0810.5178}}].

\bibitem{Barr:2009jv}
A.~J. Barr, B.~Gripaios, and C.~G. Lester, {\it {Transverse masses and
  kinematic constraints: from the boundary to the crease}},  {\em JHEP} {\bf
  11} (2009) 096, [\href{http://xxx.lanl.gov/abs/0908.3779}{{\tt 0908.3779}}].

\bibitem{Dalitz:1953cp}
R.~H. Dalitz, {\it {On the analysis of $\tau$-meson data and the nature of the
  $\tau$-meson}},  {\em Phil. Mag.} {\bf 44} (1953) 1068--1080.

\bibitem{Amsler:2008zzb}
{\bf Particle Data Group} Collaboration, C.~Amsler {\em et.~al.}, {\it {Review
  of particle physics}},  {\em Phys. Lett.} {\bf B667} (2008) 1.

\bibitem{Paige:1996nx}
F.~E. Paige, {\it Determining {SUSY} particle masses at {LHC}},
  \href{http://xxx.lanl.gov/abs/hep-ph/9609373}{{\tt hep-ph/9609373}}.

\bibitem{atlasphystdr}
{\em {ATLAS} Detector and Physics Peformance TDR}.
\newblock {CERN}, 1999.

\bibitem{Nojiri:1999ki}
M.~M. Nojiri and Y.~Yamada, {\it {Neutralino decays at the {LHC}}},  {\em Phys.
  Rev.} {\bf D60} (1999) 015006,
  [\href{http://xxx.lanl.gov/abs/hep-ph/9902201}{{\tt hep-ph/9902201}}].

\bibitem{Gjelsten:2004ki}
B.~K. Gjelsten, D.~J. Miller, and P.~Osland, {\it {Measurement of {SUSY} masses
  via cascade decays for SPS 1a}},  {\em JHEP} {\bf 12} (2004) 003,
  [\href{http://xxx.lanl.gov/abs/hep-ph/0410303}{{\tt hep-ph/0410303}}].

\bibitem{Birkedal:2005cm}
A.~Birkedal, R.~C. Group, and K.~Matchev, {\it {Slepton mass measurements at
  the LHC}},  \href{http://xxx.lanl.gov/abs/hep-ph/0507002}{{\tt
  hep-ph/0507002}}.

\bibitem{Hinchliffe:2000np}
I.~Hinchliffe and F.~E. Paige, {\it {Lepton flavor violation at the {LHC}}},
  {\em Phys. Rev.} {\bf D63} (2001) 115006,
  [\href{http://xxx.lanl.gov/abs/hep-ph/0010086}{{\tt hep-ph/0010086}}].

\bibitem{Allanach:2008ib}
B.~C. Allanach, J.~P. Conlon, and C.~G. Lester, {\it Measuring smuon-selectron
  mass splitting at the {CERN} {LHC} and patterns of supersymmetry breaking},
  {\em Phys. Rev.} {\bf D77} (2008) 076006,
  [\href{http://xxx.lanl.gov/abs/0801.3666}{{\tt 0801.3666}}].

\bibitem{Bian:2006xx}
G.~Bian, M.~Bisset, N.~Kersting, Y.~Liu, and X.~Wang, {\it {Wedgebox analysis
  of four-lepton events from neutralino pair production at the {LHC}}},  {\em
  Eur. Phys. J.} {\bf C53} (2008) 429--446,
  [\href{http://xxx.lanl.gov/abs/hep-ph/0611316}{{\tt hep-ph/0611316}}].

\bibitem{Bisset:2008hm}
M.~Bisset, R.~Lu, and N.~Kersting, {\it {Improving SUSY Spectrum Determinations
  at the LHC with Wedgebox and Hidden Threshold Techniques}},
  \href{http://xxx.lanl.gov/abs/0806.2492}{{\tt 0806.2492}}.

\bibitem{Hinchliffe:1999zc}
I.~Hinchliffe and F.~E. Paige, {\it {Measurements in SUGRA models with large
  tan(beta) at {LHC}}},  {\em Phys. Rev.} {\bf D61} (2000) 095011,
  [\href{http://xxx.lanl.gov/abs/hep-ph/9907519}{{\tt hep-ph/9907519}}].

\bibitem{Choi:2006mt}
S.~Y. Choi, K.~Hagiwara, Y.~G. Kim, K.~Mawatari, and P.~M. Zerwas, {\it {tau
  polarization in {SUSY} cascade decays}},  {\em Phys. Lett.} {\bf B648} (2007)
  207--212, [\href{http://xxx.lanl.gov/abs/hep-ph/0612237}{{\tt
  hep-ph/0612237}}].

\bibitem{Mawatari:2007mr}
K.~Mawatari, {\it {Tau polarization in {SUSY} cascade decays at {LHC}}},
  \href{http://xxx.lanl.gov/abs/0710.4994}{{\tt 0710.4994}}.

\bibitem{Godbole:2008it}
R.~M. Godbole, M.~Guchait, and D.~P. Roy, {\it {Using Tau Polarization to probe
  the Stau Co-annihilation Region of mSUGRA Model at LHC}},  {\em Phys. Rev.}
  {\bf D79} (2009) 095015, [\href{http://xxx.lanl.gov/abs/0807.2390}{{\tt
  0807.2390}}].

\bibitem{Bachacou:1999zb}
H.~Bachacou, I.~Hinchliffe, and F.~E. Paige, {\it {Measurements of masses in
  SUGRA models at {CERN} {LHC}}},  {\em Phys. Rev.} {\bf D62} (2000) 015009,
  [\href{http://xxx.lanl.gov/abs/hep-ph/9907518}{{\tt hep-ph/9907518}}].

\bibitem{Lester:2001zx}
C.~G. Lester, {\em Model independent sparticle mass measurements at {ATLAS}}.
\newblock PhD thesis, University of Cambridge, St John's College, Cambridge,
  UK, 2001.
\newblock Presented on 12 Dec 2001, {{CERN}-THESIS-2004-003}
  \url{http://cdsweb.cern.ch/record/705139}.

\bibitem{Allanach:2000kt}
B.~C. Allanach, C.~G. Lester, M.~A. Parker, and B.~R. Webber, {\it {Measuring
  sparticle masses in non-universal string inspired models at the {LHC}}},
  {\em JHEP} {\bf 09} (2000) 004,
  [\href{http://xxx.lanl.gov/abs/hep-ph/0007009}{{\tt hep-ph/0007009}}].

\bibitem{phystdr}
{\em {ATLAS} Detector and Physics Peformance TDR}.
\newblock CERN, 1999.

\bibitem{Matchev:2009iw}
K.~T. Matchev, F.~Moortgat, L.~Pape, and M.~Park, {\it {Precise reconstruction
  of sparticle masses without ambiguities}},  {\em JHEP} {\bf 08} (2009) 104,
  [\href{http://xxx.lanl.gov/abs/0906.2417}{{\tt 0906.2417}}].

\bibitem{Gjelsten:2005aw}
B.~K. Gjelsten, D.~J. Miller, and P.~Osland, {\it {Measurement of the gluino
  mass via cascade decays for SPS 1a}},  {\em JHEP} {\bf 06} (2005) 015,
  [\href{http://xxx.lanl.gov/abs/hep-ph/0501033}{{\tt hep-ph/0501033}}].

\bibitem{Miller:2005zp}
D.~J. Miller, P.~Osland, and A.~R. Raklev, {\it {Invariant mass distributions
  in cascade decays}},  {\em JHEP} {\bf 03} (2006) 034,
  [\href{http://xxx.lanl.gov/abs/hep-ph/0510356}{{\tt hep-ph/0510356}}].

\bibitem{Burns:2009zi}
M.~Burns, K.~T. Matchev, and M.~Park, {\it {Using kinematic boundary lines for
  particle mass measurements and disambiguation in {SUSY}-like events with
  missing energy}},  {\em JHEP} {\bf 05} (2009) 094,
  [\href{http://xxx.lanl.gov/abs/0903.4371}{{\tt 0903.4371}}].

\bibitem{Costanzo:2009mq}
D.~Costanzo and D.~R. Tovey, {\it {Supersymmetric particle mass measurement
  with invariant mass correlations}},  {\em JHEP} {\bf 04} (2009) 084,
  [\href{http://xxx.lanl.gov/abs/0902.2331}{{\tt 0902.2331}}].

\bibitem{Lester:2006yw}
C.~G. Lester, {\it {Constrained invariant mass distributions in cascade decays:
  The shape of the '$m_{qll}$ threshold' and similar distributions}},  {\em
  Phys. Lett.} {\bf B655} (2007) 39--44,
  [\href{http://xxx.lanl.gov/abs/hep-ph/0603171}{{\tt hep-ph/0603171}}].

\bibitem{freedmanlester}
R.~Freedman and C.~G. Lester.
\newblock Student project, {ATL-COM-PHYS-2007-086}
  \url{http://cdsweb.cern.ch/record/1067733}.

\bibitem{Alves:2006df}
A.~Alves, O.~Eboli, and T.~Plehn, {\it {It's a gluino}},  {\em Phys. Rev.} {\bf
  D74} (2006) 095010, [\href{http://xxx.lanl.gov/abs/hep-ph/0605067}{{\tt
  hep-ph/0605067}}].

\bibitem{Agashe:2010gt}
K.~Agashe, D.~Kim, M.~Toharia, and D.~G.~E. Walker, {\it {Distinguishing Dark
  Matter Stabilization Symmetries Using Multiple Kinematic Edges and Cusps}},
  \href{http://xxx.lanl.gov/abs/1003.0899}{{\tt 1003.0899}}.

\bibitem{Matchev:2009ad}
K.~T. Matchev and M.~Park, {\it A general method for determining the masses of
  semi-invisibly decaying particles at hadron colliders},
  \href{http://xxx.lanl.gov/abs/0910.1584}{{\tt 0910.1584}}.

\bibitem{Konar:2009wn}
P.~Konar, K.~Kong, K.~T. Matchev, and M.~Park, {\it {Superpartner mass
  measurements with 1D decomposed $M_{T2}$}},
  \href{http://xxx.lanl.gov/abs/0910.3679}{{\tt 0910.3679}}.

\bibitem{Lester:2006cf}
C.~G. Lester, M.~A. Parker, and M.~J. White, {\it {Three body kinematic
  endpoints in {SUSY} models with non- universal Higgs masses}},  {\em JHEP}
  {\bf 10} (2007) 051, [\href{http://xxx.lanl.gov/abs/hep-ph/0609298}{{\tt
  hep-ph/0609298}}].

\bibitem{Barr:2008ba}
A.~J. Barr, G.~G. Ross, and M.~Serna, {\it The precision determination of
  invisible-particle masses at the {LHC}},  {\em Phys. Rev.} {\bf D78} (2008)
  056006, [\href{http://xxx.lanl.gov/abs/0806.3224}{{\tt 0806.3224}}].

\bibitem{Serna:2008zk}
M.~Serna, {\it {A short comparison between $m_{T2}$ and $m_{CT}$}},  {\em JHEP}
  {\bf 06} (2008) 004, [\href{http://xxx.lanl.gov/abs/0804.3344}{{\tt
  0804.3344}}].

\bibitem{Allanach:2001xz}
B.~C. Allanach {\em et.~al.}, {\it Measuring supersymmetric particle masses at
  the {LHC} in scenarios with baryon-number {R}-parity violating couplings},
  {\em JHEP} {\bf 03} (2001) 048,
  [\href{http://xxx.lanl.gov/abs/hep-ph/0102173}{{\tt hep-ph/0102173}}].

\bibitem{Butterworth:2009qa}
J.~M. Butterworth, J.~R. Ellis, A.~R. Raklev, and G.~P. Salam, {\it
  {Discovering baryon-number violating neutralino decays at the LHC}},  {\em
  Phys. Rev. Lett.} {\bf 103} (2009) 241803,
  [\href{http://xxx.lanl.gov/abs/0906.0728}{{\tt 0906.0728}}].

\bibitem{Butterworth:2008iy}
J.~M. Butterworth, A.~R. Davison, M.~Rubin, and G.~P. Salam, {\it {Jet
  substructure as a new Higgs search channel at the LHC}},  {\em Phys. Rev.
  Lett.} {\bf 100} (2008) 242001,
  [\href{http://xxx.lanl.gov/abs/0802.2470}{{\tt 0802.2470}}].

\bibitem{Plehn:2009rk}
T.~Plehn, G.~P. Salam, and M.~Spannowsky, {\it {Fat Jets for a Light Higgs}},
  {\em Phys. Rev. Lett.} {\bf 104} (2010) 111801,
  [\href{http://xxx.lanl.gov/abs/0910.5472}{{\tt 0910.5472}}].

\bibitem{ATL-PHYS-PUB-2009-081}
{\it Reconstruction of high mass $t\overline{t}$ resonances in the lepton+jets
  channel},  Tech. Rep. ATL-PHYS-PUB-2009-081. ATL-COM-PHYS-2009-255, CERN,
  Geneva, May, 2009.

\bibitem{Nisati:1997gb}
A.~Nisati, S.~Petrarca, and G.~Salvini, {\it {On the possible detection of
  massive stable exotic particles at the {LHC}}},  {\em Mod. Phys. Lett.} {\bf
  A12} (1997) 2213--2222, [\href{http://xxx.lanl.gov/abs/hep-ph/9707376}{{\tt
  hep-ph/9707376}}].

\bibitem{atl-muon-99-006}
G.~Polesello and A.~Rimoldi, {\it Reconstruction of quasi-stable charged
  sleptons in the {ATLAS} muon spectrometer}, . {ATL-MUON-99-006}.

\bibitem{Kazana:687147}
M.~Kazana, G.~Wrochna, and P.~Zalewski, {\it Study of the {NLSP} from the
  {GMSB} models in the {CMS} detector at the {LHC}},  Tech. Rep.
  CMS-CR-1999-019, CERN, Geneva, Jul, 1999.

\bibitem{Ambrosanio:2000ik}
S.~Ambrosanio, B.~Mele, S.~Petrarca, G.~Polesello, and A.~Rimoldi, {\it
  {Measuring the {SUSY} breaking scale at the {LHC} in the slepton NLSP
  scenario of GMSB models}},  {\em JHEP} {\bf 01} (2001) 014,
  [\href{http://xxx.lanl.gov/abs/hep-ph/0010081}{{\tt hep-ph/0010081}}].

\bibitem{Ambrosanio:2000zu}
S.~Ambrosanio {\em et.~al.}, {\it {SUSY long-lived massive particles: Detection
  and physics at the LHC}},  \href{http://xxx.lanl.gov/abs/hep-ph/0012192}{{\tt
  hep-ph/0012192}}.

\bibitem{Allanach:2001sd}
B.~C. Allanach, C.~M. Harris, M.~A. Parker, P.~Richardson, and B.~R. Webber,
  {\it {Detecting exotic heavy leptons at the Large Hadron Collider}},  {\em
  JHEP} {\bf 08} (2001) 051,
  [\href{http://xxx.lanl.gov/abs/hep-ph/0108097}{{\tt hep-ph/0108097}}].

\bibitem{Ellis:2006vu}
J.~R. Ellis, A.~R. Raklev, and O.~K. Oye, {\it {Gravitino dark matter scenarios
  with massive metastable charged sparticles at the {LHC}}},  {\em JHEP} {\bf
  10} (2006) 061, [\href{http://xxx.lanl.gov/abs/hep-ph/0607261}{{\tt
  hep-ph/0607261}}].

\bibitem{Raklev:2009mg}
A.~R. Raklev, {\it {Massive Metastable Charged (S)Particles at the LHC}},
  \href{http://xxx.lanl.gov/abs/0908.0315}{{\tt 0908.0315}}.

\bibitem{Kilian:2004uj}
W.~Kilian, T.~Plehn, P.~Richardson, and E.~Schmidt, {\it {Split supersymmetry
  at colliders}},  {\em Eur. Phys. J.} {\bf C39} (2005) 229--243,
  [\href{http://xxx.lanl.gov/abs/hep-ph/0408088}{{\tt hep-ph/0408088}}].

\bibitem{Hewett:2004nw}
J.~L. Hewett, B.~Lillie, M.~Masip, and T.~G. Rizzo, {\it {Signatures of
  long-lived gluinos in split supersymmetry}},  {\em JHEP} {\bf 09} (2004) 070,
  [\href{http://xxx.lanl.gov/abs/hep-ph/0408248}{{\tt hep-ph/0408248}}].

\bibitem{Kraan:2005ji}
A.~C. Kraan, J.~B. Hansen, and P.~Nevski, {\it {Discovery potential of
  R-hadrons with the {ATLAS} detector}},  {\em Eur. Phys. J.} {\bf C49} (2007)
  623--640, [\href{http://xxx.lanl.gov/abs/hep-ex/0511014}{{\tt
  hep-ex/0511014}}].

\bibitem{atl-phys-pub-2006-005}
S.~Hellman, D.~Milstead, and M.~Ramstedt, {\it A strategy to detect and
  identify gluino {R}-hadrons with the {ATLAS} experiment at the {LHC}}, .
  ATL-PHYS-PUB-2006-005.

\bibitem{atl-phys-pub-2005-022}
S.~Tarem, S.~Bressler, E.~Duchovni, and L.~Levinson, {\it Can {ATLAS} avoid
  missing the long lived stau?}, . {ATL-PHYS-PUB-2005-022}.

\bibitem{Tarem:2009zz}
S.~Tarem, S.~Bressler, H.~Nomoto, and A.~Di~Mattia, {\it {Trigger and
  reconstruction for heavy long-lived charged particles with the {ATLAS}
  detector}},  {\em Eur. Phys. J.} {\bf C62} (2009) 281--292.

\bibitem{Fairbairn:2006gg}
M.~Fairbairn {\em et.~al.}, {\it {Stable massive particles at colliders}},
  {\em Phys. Rept.} {\bf 438} (2007) 1--63,
  [\href{http://xxx.lanl.gov/abs/hep-ph/0611040}{{\tt hep-ph/0611040}}].

\bibitem{deCampos:2007bn}
F.~de~Campos {\em et.~al.}, {\it {Probing bilinear R-parity violating
  supergravity at the {LHC}}},  {\em JHEP} {\bf 05} (2008) 048,
  [\href{http://xxx.lanl.gov/abs/0712.2156}{{\tt 0712.2156}}].

\bibitem{Allanach:2001if}
B.~C. Allanach, A.~J. Barr, M.~A. Parker, P.~Richardson, and B.~R. Webber, {\it
  Extracting the flavour structure of a baryon-number {R}-parity violating
  coupling at the {LHC}},  {\em JHEP} {\bf 09} (2001) 021,
  [\href{http://xxx.lanl.gov/abs/http://arXiv.org/abs/hep-ph/0106304}{{\tt
  http://arXiv.org/abs/hep-ph/0106304}}].

\bibitem{Barr:2002ex}
A.~J. Barr, C.~G. Lester, M.~A. Parker, B.~C. Allanach, and P.~Richardson, {\it
  {Discovering anomaly-mediated supersymmetry at the {LHC}}},  {\em JHEP} {\bf
  03} (2003) 045, [\href{http://xxx.lanl.gov/abs/hep-ph/0208214}{{\tt
  hep-ph/0208214}}].

\bibitem{Kawagoe:2003jv}
K.~Kawagoe, T.~Kobayashi, M.~M. Nojiri, and A.~Ochi, {\it {Study of the gauge
  mediation signal with non-pointing photons at the {CERN} {LHC}}},  {\em Phys.
  Rev.} {\bf D69} (2004) 035003,
  [\href{http://xxx.lanl.gov/abs/hep-ph/0309031}{{\tt hep-ph/0309031}}].

\bibitem{Han:2009ss}
T.~Han, I.-W. Kim, and J.~Song, {\it {Kinematic Cusps: Determining the Missing
  Particle Mass at the LHC}},  \href{http://xxx.lanl.gov/abs/0906.5009}{{\tt
  0906.5009}}.

\bibitem{Barr:2009mx}
A.~J. Barr, B.~Gripaios, and C.~G. Lester, {\it {Measuring the Higgs boson mass
  in dileptonic W-boson decays at hadron colliders}},  {\em JHEP} {\bf 07}
  (2009) 072, [\href{http://xxx.lanl.gov/abs/0902.4864}{{\tt 0902.4864}}].

\bibitem{Huang:2008qd}
P.~Huang, N.~Kersting, and H.~H. Yang, {\it Model-independent {SUSY} masses
  from 4-lepton kinematic invariants at the {LHC}},  {\em Phys. Rev.} {\bf D77}
  (2008) 075011, [\href{http://xxx.lanl.gov/abs/0801.0041}{{\tt 0801.0041}}].

\bibitem{Huang:2008ae}
P.~Huang, N.~Kersting, and H.~H. Yang, {\it {Hidden Thresholds: A Technique for
  Reconstructing New Physics Masses at Hadron Colliders}},
  \href{http://xxx.lanl.gov/abs/0802.0022}{{\tt 0802.0022}}.

\bibitem{Choi:2009hn}
K.~Choi, S.~Choi, J.~S. Lee, and C.~B. Park, {\it {Reconstructing the Higgs
  boson in dileptonic W decays at hadron collider}},  {\em Phys. Rev.} {\bf
  D80} (2009) 073010, [\href{http://xxx.lanl.gov/abs/0908.0079}{{\tt
  0908.0079}}].

\bibitem{Tovey:2008ui}
D.~R. Tovey, {\it {On measuring the masses of pair-produced semi-invisibly
  decaying particles at hadron colliders}},  {\em JHEP} {\bf 04} (2008) 034,
  [\href{http://xxx.lanl.gov/abs/0802.2879}{{\tt 0802.2879}}].

\bibitem{Dimopoulos:1981zb}
S.~Dimopoulos and H.~Georgi, {\it {Softly Broken Supersymmetry and SU(5)}},
  {\em Nucl. Phys.} {\bf B193} (1981) 150.

\bibitem{Cheng:2002ab}
H.-C. Cheng, K.~T. Matchev, and M.~Schmaltz, {\it {Bosonic supersymmetry?
  Getting fooled at the {CERN} {LHC}}},  {\em Phys. Rev.} {\bf D66} (2002)
  056006, [\href{http://xxx.lanl.gov/abs/hep-ph/0205314}{{\tt
  hep-ph/0205314}}].

\bibitem{Lester:1999tx}
C.~G. Lester and D.~J. Summers, {\it {Measuring masses of semiinvisibly
  decaying particles pair produced at hadron colliders}},  {\em Phys. Lett.}
  {\bf B463} (1999) 99--103,
  [\href{http://xxx.lanl.gov/abs/hep-ph/9906349}{{\tt hep-ph/9906349}}].

\bibitem{oxbridgeStransverseMassLibrary}
A.~J. Barr and C.~G. Lester, ``Oxbridge stransverse mass library.''
\newblock \url{http://www.hep.phy.cam.ac.uk/~lester/mt2/index.html}.

\bibitem{zenuhanStransverseMassLibrary}
H.-C. Cheng and Z.~Han, ``{UCD} stransverse mass library.''
\newblock \url{http://daneel.physics.ucdavis.edu/~zhenyuhan/mt2.html}.

\bibitem{ucdWimpmassLibrary}
H.-C. Cheng, D.~Engelhardt, J.~F. Gunion, Z.~Han, G.~Marandella, and
  B.~McElrath, ``{\tt WIMPMASS} library.''
\newblock
  \url{http://particle.physics.ucdavis.edu/hefti/projects/doku.php?id=wimpmass%
}.

\bibitem{minuit2Library}
``Minuit2 minimization library.''
\newblock \url{http://seal.web.cern.ch/seal/MathLibs/Minuit2/html/}.

\bibitem{James:1975dr}
F.~James and M.~Roos, {\it {Minuit: A System for Function Minimization and
  Analysis of the Parameter Errors and Correlations}},  {\em Comput. Phys.
  Commun.} {\bf 10} (1975) 343--367.

\bibitem{Barr:2003rg}
A.~Barr, C.~Lester, and P.~Stephens, {\it {m(T2) : The Truth behind the
  glamour}},  {\em J. Phys.} {\bf G29} (2003) 2343--2363,
  [\href{http://xxx.lanl.gov/abs/hep-ph/0304226}{{\tt hep-ph/0304226}}].

\bibitem{cdftop}
``Simultaneous template-based top quark mass measurement in the lepton+jets and
  dilepton channels using $3.2 \mathrm{fb}^{-1}$ of {CDF} data (the dilepton
  channel uses a new variable $m_{T2}$.).''
\newblock {CDF} public note 9679.

\bibitem{Cho:2008cu}
W.~S. Cho, K.~Choi, Y.~G. Kim, and C.~B. Park, {\it {Measuring the top quark
  mass with $m_{T2}$ at the {LHC}}},  {\em Phys. Rev.} {\bf D78} (2008) 034019,
  [\href{http://xxx.lanl.gov/abs/0804.2185}{{\tt 0804.2185}}].

\bibitem{Baumgart:2006pa}
M.~Baumgart, T.~Hartman, C.~Kilic, and L.-T. Wang, {\it Discovery and
  measurement of sleptons, {Binos}, and {Winos} with a {$Z^\prime$}},  {\em
  JHEP} {\bf 11} (2007) 084,
  [\href{http://xxx.lanl.gov/abs/hep-ph/0608172}{{\tt hep-ph/0608172}}].

\bibitem{Lytken:2003rw}
E.~Lytken, {\it {Prospects for slepton searches with ATLAS}}, .

\bibitem{Barr:2009wu}
A.~J. Barr and C.~Gwenlan, {\it {The race for supersymmetry: using $M_{T2}$ for
  discovery}},  {\em Phys. Rev.} {\bf D80} (2009) 074007,
  [\href{http://xxx.lanl.gov/abs/0907.2713}{{\tt 0907.2713}}].

\bibitem{Cho:2007qv}
W.~S. Cho, K.~Choi, Y.~G. Kim, and C.~B. Park, {\it Transverse mass for pairs
  of gluinos},  {\em Phys. Rev. Lett.} {\bf 100} (2008) 171801,
  [\href{http://xxx.lanl.gov/abs/0709.0288}{{\tt 0709.0288}}]. \copyright\
  (2008) by the American Physical Society.

\bibitem{Gripaios:2007is}
B.~Gripaios, {\it Transverse observables and mass determination at hadron
  colliders},  {\em JHEP} {\bf 02} (2008) 053,
  [\href{http://xxx.lanl.gov/abs/0709.2740}{{\tt 0709.2740}}].

\bibitem{Barr:2007hy}
A.~J. Barr, B.~Gripaios, and C.~G. Lester, {\it Weighing {WIMPs} with kinks at
  colliders: Invisible particle mass measurements from endpoints},  {\em JHEP}
  {\bf 02} (2008) 014, [\href{http://xxx.lanl.gov/abs/0711.4008}{{\tt
  0711.4008}}].

\bibitem{Cho:2007dh}
W.~S. Cho, K.~Choi, Y.~G. Kim, and C.~B. Park, {\it {Measuring superparticle
  masses at hadron collider using the transverse mass kink}},  {\em JHEP} {\bf
  02} (2008) 035, [\href{http://xxx.lanl.gov/abs/0711.4526}{{\tt 0711.4526}}].

\bibitem{Cohen:2010wv}
T.~Cohen, E.~Kuflik, and K.~M. Zurek, {\it {Extracting the Dark Matter Mass
  from Single Stage Cascade Decays at the LHC}},
  \href{http://xxx.lanl.gov/abs/1003.2204}{{\tt 1003.2204}}.

\bibitem{Burns:2008va}
M.~Burns, K.~Kong, K.~T. Matchev, and M.~Park, {\it Using subsystem $m_{T2}$
  for complete mass determinations in decay chains with missing energy at
  hadron colliders},  {\em JHEP} {\bf 03} (2009) 143,
  [\href{http://xxx.lanl.gov/abs/0810.5576}{{\tt 0810.5576}}].

\bibitem{Cheng:2007xv}
H.-C. Cheng, J.~F. Gunion, Z.~Han, G.~Marandella, and B.~McElrath, {\it Mass
  determination in {SUSY}-like events with missing energy},  {\em JHEP} {\bf
  12} (2007) 076, [\href{http://xxx.lanl.gov/abs/0707.0030}{{\tt 0707.0030}}].

\bibitem{Hinchliffe:1998ys}
I.~Hinchliffe and F.~E. Paige, {\it {Measurements in gauge mediated {SUSY}
  breaking models at {LHC}}},  {\em Phys. Rev.} {\bf D60} (1999) 095002,
  [\href{http://xxx.lanl.gov/abs/hep-ph/9812233}{{\tt hep-ph/9812233}}].

\bibitem{Cho:2009ve}
W.~S. Cho, J.~E. Kim, and J.-H. Kim, {\it {Shining on buried new particles}},
  \href{http://xxx.lanl.gov/abs/0912.2354}{{\tt 0912.2354}}.

\bibitem{Polesello:2009rn}
G.~Polesello and D.~R. Tovey, {\it {Supersymmetric particle mass measurement
  with the boost-corrected contransverse mass}},  {\em JHEP} {\bf 03} (2010)
  030, [\href{http://xxx.lanl.gov/abs/0910.0174}{{\tt 0910.0174}}].

\bibitem{Barr:2010ii}
A.~J. Barr, C.~Gwenlan, C.~G. Lester, and C.~J.~S. Young, {\it {A comment on
  'Amplification of endpoint structure for new particle mass measurement at the
  LHC'}},  \href{http://xxx.lanl.gov/abs/1006.2568}{{\tt 1006.2568}}.

\bibitem{Chang:2009dh}
S.~Chang and A.~de~Gouvea, {\it {Neutrino alternatives for missing energy
  events at colliders}},  {\em Phys. Rev.} {\bf D80} (2009) 015008,
  [\href{http://xxx.lanl.gov/abs/0901.4796}{{\tt 0901.4796}}].

\bibitem{Nojiri:2008vq}
M.~M. Nojiri, K.~Sakurai, Y.~Shimizu, and M.~Takeuchi, {\it {Handling jets +
  missing $E_T$ channel using inclusive $m_{T2}$}},  {\em JHEP} {\bf 10} (2008)
  100, [\href{http://xxx.lanl.gov/abs/0808.1094}{{\tt 0808.1094}}].

\bibitem{Konar:2009qr}
P.~Konar, K.~Kong, K.~T. Matchev, and M.~Park, {\it Dark matter particle
  spectroscopy at the {LHC}: Generalizing $m_{T2}$ to asymmetric event
  topologies},  \href{http://xxx.lanl.gov/abs/0911.4126}{{\tt 0911.4126}}.

\bibitem{Nojiri:2008hy}
M.~M. Nojiri, Y.~Shimizu, S.~Okada, and K.~Kawagoe, {\it {Inclusive transverse
  mass analysis for squark and gluino mass determination}},  {\em JHEP} {\bf
  06} (2008) 035, [\href{http://xxx.lanl.gov/abs/0802.2412}{{\tt 0802.2412}}].

\bibitem{Barnett:1993ea}
R.~M. Barnett, J.~F. Gunion, and H.~E. Haber, {\it {Discovering supersymmetry
  with like sign dileptons}},  {\em Phys. Lett.} {\bf B315} (1993) 349--354,
  [\href{http://xxx.lanl.gov/abs/hep-ph/9306204}{{\tt hep-ph/9306204}}].

\bibitem{Baer:1995nq}
H.~Baer, C.-h. Chen, F.~Paige, and X.~Tata, {\it {Signals for minimal
  supergravity at the {CERN} large hadron collider: Multi-jet plus missing
  energy channel}},  {\em Phys. Rev.} {\bf D52} (1995) 2746--2759,
  [\href{http://xxx.lanl.gov/abs/hep-ph/9503271}{{\tt hep-ph/9503271}}].

\bibitem{Lester:2007fq}
C.~Lester and A.~Barr, {\it {$M_{TGen}$} : Mass scale measurements in
  pair-production at colliders},  {\em JHEP} {\bf 12} (2007) 102,
  [\href{http://xxx.lanl.gov/abs/0708.1028}{{\tt 0708.1028}}].

\bibitem{MatchevUnpublished}
A.~Barr, P.~Konar, K.~Kong, C.~Lester, K.~Matchev, and M.~Park unpublished.

\bibitem{Plehn:2008ae}
T.~Plehn and T.~M.~P. Tait, {\it {Seeking Sgluons}},  {\em J. Phys.} {\bf G36}
  (2009) 075001, [\href{http://xxx.lanl.gov/abs/0810.3919}{{\tt 0810.3919}}].

\bibitem{Alwall:2009zu}
J.~Alwall, K.~Hiramatsu, M.~M. Nojiri, and Y.~Shimizu, {\it {Novel
  reconstruction technique for New Physics processes with initial state
  radiation}},  {\em Phys. Rev. Lett.} {\bf 103} (2009) 151802,
  [\href{http://xxx.lanl.gov/abs/0905.1201}{{\tt 0905.1201}}].

\bibitem{Ross:2007rm}
G.~G. Ross and M.~Serna, {\it Mass determination of new states at hadron
  colliders},  {\em Phys. Lett.} {\bf B665} (2008) 212--218,
  [\href{http://xxx.lanl.gov/abs/0712.0943}{{\tt 0712.0943}}].

\bibitem{Barr:2008hv}
A.~J. Barr, A.~Pinder, and M.~Serna, {\it {Precision Determination of
  Invisible-Particle Masses at the {CERN} {LHC}: II}},  {\em Phys. Rev.} {\bf
  D79} (2009) 074005, [\href{http://xxx.lanl.gov/abs/0811.2138}{{\tt
  0811.2138}}].

\bibitem{Nojiri:2007pq}
M.~M. Nojiri, G.~Polesello, and D.~R. Tovey, {\it {A hybrid method for
  determining {SUSY} particle masses at the {LHC} with fully identified cascade
  decays}},  {\em JHEP} {\bf 05} (2008) 014,
  [\href{http://xxx.lanl.gov/abs/0712.2718}{{\tt 0712.2718}}].

\bibitem{Kim:2009si}
I.-W. Kim, {\it Algebraic singularity method for mass measurement with missing
  energy},  {\em Phys. Rev. Lett.} {\bf 104} (2010) 081601,
  [\href{http://xxx.lanl.gov/abs/0910.1149}{{\tt 0910.1149}}].

\bibitem{Nojiri:2003tu}
M.~M. Nojiri, G.~Polesello, and D.~R. Tovey, {\it {Proposal for a new
  reconstruction technique for SUSY processes at the LHC}},
  \href{http://xxx.lanl.gov/abs/hep-ph/0312317}{{\tt hep-ph/0312317}}.

\bibitem{Kawagoe:2004rz}
K.~Kawagoe, M.~M. Nojiri, and G.~Polesello, {\it {A new {SUSY} mass
  reconstruction method at the {CERN} {LHC}}},  {\em Phys. Rev.} {\bf D71}
  (2005) 035008, [\href{http://xxx.lanl.gov/abs/hep-ph/0410160}{{\tt
  hep-ph/0410160}}].

\bibitem{Cheng:2008mg}
H.-C. Cheng, D.~Engelhardt, J.~F. Gunion, Z.~Han, and B.~McElrath, {\it
  Accurate mass determinations in decay chains with missing energy},  {\em
  Phys. Rev. Lett.} {\bf 100} (2008) 252001,
  [\href{http://xxx.lanl.gov/abs/0802.4290}{{\tt 0802.4290}}].

\bibitem{Cheng:2009fw}
H.-C. Cheng, J.~F. Gunion, Z.~Han, and B.~McElrath, {\it {Accurate Mass
  Determinations in Decay Chains with Missing Energy: II}},  {\em Phys. Rev.}
  {\bf D80} (2009) 035020, [\href{http://xxx.lanl.gov/abs/0905.1344}{{\tt
  0905.1344}}].

\bibitem{Alwall:2009sv}
J.~Alwall, A.~Freitas, and O.~Mattelaer, {\it Measuring sparticles with the
  matrix element},  {\em AIP Conf. Proc.} {\bf 1200} (2010) 442--445,
  [\href{http://xxx.lanl.gov/abs/0910.2522}{{\tt 0910.2522}}].

\bibitem{Webber:2009vm}
B.~Webber, {\it {Mass determination in sequential particle decay chains}},
  {\em JHEP} {\bf 09} (2009) 124,
  [\href{http://xxx.lanl.gov/abs/0907.5307}{{\tt 0907.5307}}].

\bibitem{neymanpearson}
J.~Neyman and E.~S. Pearson, {\it On the problem of the most efficient tests of
  statistical hypotheses},  {\em Phil. Trans. Royal Soc. A} {\bf 231} (1933).

\bibitem{Kraml:2005kb}
S.~Kraml and A.~R. Raklev, {\it {Same-sign top quarks as signature of light
  stops at the {LHC}}},  {\em Phys. Rev.} {\bf D73} (2006) 075002,
  [\href{http://xxx.lanl.gov/abs/hep-ph/0512284}{{\tt hep-ph/0512284}}].

\bibitem{Abazov:2004cs}
{\bf D0} Collaboration, V.~M. Abazov {\em et.~al.}, {\it {A precision
  measurement of the mass of the top quark}},  {\em Nature} {\bf 429} (2004)
  638--642, [\href{http://xxx.lanl.gov/abs/hep-ex/0406031}{{\tt
  hep-ex/0406031}}].

\bibitem{PhysRevLett.99.182002}
T.~Aaltonen, A.~Abulencia, J.~Adelman, T.~Affolder, T.~Akimoto, M.~G. Albrow,
  S.~Amerio, D.~Amidei, A.~Anastassov, K.~Anikeev, A.~Annovi, J.~Antos,
  M.~Aoki, G.~Apollinari, T.~Arisawa, A.~Artikov, W.~Ashmanskas, A.~Attal,
  A.~Aurisano, F.~Azfar, P.~Azzi-Bacchetta, P.~Azzurri, N.~Bacchetta,
  W.~Badgett, A.~Barbaro-Galtieri, V.~E. Barnes, and B.~A. Barnett, {\it
  Precise measurement of the top-quark mass in the $lepton+jets$ topology at
  cdf ii},  {\em Phys. Rev. Lett.} {\bf 99} (Oct, 2007) 182002.

\bibitem{cdf:9985}
{\bf CDF} Collaboration, T.~C. collaboration, ``Search for standard model higgs
  boson production in association with a w boson using matrix element
  techniques with 4.3 fb-1 of cdf data.'' CDF note 9985, October, 2009.

\bibitem{Allanach:2004ub}
{\bf Beyond the Standard Model Working Group} Collaboration, B.~C. Allanach
  {\em et.~al.}, {\it {Les Houches 'Physics at TeV Colliders 2003' Beyond the
  Standard Model Working Group: Summary report}},
  \href{http://xxx.lanl.gov/abs/hep-ph/0402295}{{\tt hep-ph/0402295}}.

\bibitem{Lafaye:2004cn}
R.~Lafaye, T.~Plehn, and D.~Zerwas, {\it {SFITTER: SUSY parameter analysis at
  LHC and LC}},  \href{http://xxx.lanl.gov/abs/hep-ph/0404282}{{\tt
  hep-ph/0404282}}.

\bibitem{Bechtle:2004pc}
P.~Bechtle, K.~Desch, and P.~Wienemann, {\it {Fittino, a program for
  determining MSSM parameters from collider observables using an iterative
  method}},  {\em Comput. Phys. Commun.} {\bf 174} (2006) 47--70,
  [\href{http://xxx.lanl.gov/abs/hep-ph/0412012}{{\tt hep-ph/0412012}}].

\bibitem{Bechtle:2009ty}
P.~Bechtle, K.~Desch, M.~Uhlenbrock, and P.~Wienemann, {\it {Constraining SUSY
  models with Fittino using measurements before, with and beyond the LHC}},
  {\em Eur. Phys. J.} {\bf C66} (2010) 215--259,
  [\href{http://xxx.lanl.gov/abs/0907.2589}{{\tt 0907.2589}}].

\bibitem{Roszkowski:2009ye}
L.~Roszkowski, R.~Ruiz~de Austri, and R.~Trotta, {\it {Efficient reconstruction
  of CMSSM parameters from LHC data - A case study}},
  \href{http://xxx.lanl.gov/abs/0907.0594}{{\tt 0907.0594}}.

\bibitem{Weiglein:2004hn}
{\bf {LHC}/LC Study Group} Collaboration, G.~Weiglein {\em et.~al.}, {\it
  {Physics interplay of the {LHC} and the ILC}},  {\em Phys. Rept.} {\bf 426}
  (2006) 47--358, [\href{http://xxx.lanl.gov/abs/hep-ph/0410364}{{\tt
  hep-ph/0410364}}].

\bibitem{Lester:2005je}
C.~G. Lester, M.~A. Parker, and .~White, Martin~J., {\it Determining {SUSY}
  model parameters and masses at the {LHC} using cross-sections, kinematic
  edges and other observables},  {\em JHEP} {\bf 01} (2006) 080,
  [\href{http://xxx.lanl.gov/abs/hep-ph/0508143}{{\tt hep-ph/0508143}}].

\bibitem{Dreiner:2010gv}
H.~K. Dreiner, M.~Kramer, J.~M. Lindert, and B.~O'Leary, {\it {SUSY parameter
  determination at the LHC using cross sections and kinematic edges}},
  \href{http://xxx.lanl.gov/abs/1003.2648}{{\tt 1003.2648}}.

\bibitem{Kersting:2009ne}
N.~Kersting, {\it {A Simple Mass Reconstruction Technique for SUSY particles at
  the LHC}},  {\em Phys. Rev.} {\bf D79} (2009) 095018,
  [\href{http://xxx.lanl.gov/abs/0901.2765}{{\tt 0901.2765}}].

\bibitem{Kang:2009sk}
Z.~Kang, N.~Kersting, S.~Kraml, A.~R. Raklev, and M.~J. White, {\it {Neutralino
  Reconstruction at the LHC from Decay-frame Kinematics}},
  \href{http://xxx.lanl.gov/abs/0908.1550}{{\tt 0908.1550}}.

\bibitem{Fatakia:2004wi}
{\bf D0} Collaboration, S.~N. Fatakia, {\it {A method of extracting the mass of
  the top quark in the di-lepton channel using the D0 detector}},
  \href{http://xxx.lanl.gov/abs/hep-ex/0409013}{{\tt hep-ex/0409013}}.

\bibitem{Fatakia:2005mz}
S.~N. Fatakia, {\it A measurement of the mass of the top quark in the di-lepton
  channels using the {D0} detector at fermilab}, . {FERMILAB}-THESIS-2005-15.

\bibitem{Adelman:2006rw}
{\bf CDF} Collaboration, J.~A. Adelman {\em et.~al.}, {\it {Measurement of the
  top quark mass using the template method in the lepton plus jets channel with
  in situ $W \to jj$ calibration at CDF-II}}, . {FERMILAB}-FN-0794-E.

\bibitem{Abulencia:2006js}
{\bf CDF} Collaboration, A.~Abulencia {\em et.~al.}, {\it {Measurement of the
  top quark mass using template methods on dilepton events in proton antiproton
  collisions at $\sqrt{s} =$ 1.96~TeV}},  {\em Phys. Rev.} {\bf D73} (2006)
  112006, [\href{http://xxx.lanl.gov/abs/hep-ex/0602008}{{\tt
  hep-ex/0602008}}].

\bibitem{Brandt:2006uc}
O.~Brandt, {\it {Measurement of the mass of the top quark in dilepton final
  states with the {D0} detector}}, . {FERMILAB}-MASTERS-2006-03.

\bibitem{Meyer:2007zz}
J.~M. Meyer, {\it {Measurement of the Top Quark Mass using Dilepton Events and
  a Neutrino Weighting Algorithm with the D0 Experiment at the Tevatron (Run
  II)}}, . FERMILAB-THESIS-2007-65.

\bibitem{Grohsjean:2008uy}
{\bf {D0}} Collaboration, A.~Grohsjean, {\it {Measurements of the Top Quark
  Mass in the Dilepton Decay Channel at the {D0} Experiment}},
  \href{http://xxx.lanl.gov/abs/0810.3711}{{\tt 0810.3711}}.

\end{thebibliography}\endgroup
\end{document}